\documentclass{aa}

\usepackage{natbib,twoopt}
\bibpunct{(}{)}{;}{a}{}{,} 
\usepackage{amssymb}
\usepackage[varg]{txfonts}
\usepackage[tight]{subfigure}

\usepackage{hyperref}
\usepackage{graphicx}
\usepackage{color}

\newcommand{\nbody}{$N$-body}
\newcommand{\msun}{M_{\sun}}
\newcommand{\mcl}{M_{\mathrm{ecl}}}
\newcommand{\mmax}{m_{\mathrm{max}}}

\newcommand{\fej}{f_{\mathrm{ej,O}}}

\newcommand{\rh}{r_{\mathrm{h}}}
\newcommand{\rhi}{r_{\mathrm{h}}{\left(0\right)}}
\newcommand{\rhoh}{\rho_{\mathrm{h}}}
\newcommand{\kms}{\ensuremath{\mathrm{km\,s^{-1}}}}
\newcommand{\uden}{\ensuremath{\msun\,\mathrm{pc}^{-3}}}

\newcommand{\amass}{\langle m \rangle}
\newcommand{\mms}{m_{\mathrm{massive}}}

\newcommand{\trh}{t_{\mathrm{rh}}}
\newcommand{\tms}{t_{\mathrm{ms}}}
\newcommand{\tej}{\tau_{\mathrm{ej}}}
\newcommand{\vcl}{\sigma_{\mathrm{cl}}}
\newcommand{\ai}{\langle\alpha_{\mathrm{i}}\rangle}
\newcommand{\aej}{\alpha_{\mathrm{ej}}}
\newcommand{\ain}{\langle\alpha_{\mathrm{in}}\rangle}
\newcommand{\alpin}{\alpha_{\mathrm{in}}}

\begin{document}
\label{firstpage}

\title{Dynamical ejections of massive stars from young star clusters under diverse initial conditions}

\author{Seungkyung Oh\inst{\ref{inst1},\ref{inst2}}\fnmsep\thanks{Member of the International Max Planck Research School
(IMPRS) for Astronomy and Astrophysics at the Universities of Bonn and Cologne}
 \and
 Pavel Kroupa\inst{\ref{inst1}}}
\institute{Helmholtz-Institut f\"ur Strahlen- und Kernphysik (HISKP), University of Bonn, 
Nussallee 14-16, 53115 Bonn, Germany \\ \email{skoh@astro.uni-bonn.de} \label{inst1}
\and
Argelander Institut f\"ur Astronomie, University of Bonn, Auf dem H\"ugel 71, 53121 Bonn, Germany \label{inst2}} 

\date{Received 1 February 2016 / Accepted 23 March 2016}

\titlerunning{Dynamical ejection of massive stars}
\authorrunning{S. Oh \& P. Kroupa}

\abstract{
We study the effects that initial conditions of star clusters and their massive star population have  
on dynamical ejections of massive stars from star clusters up to an age of 3\,Myr. 
We use a large set of direct \nbody\ calculations for moderately massive star clusters 
($\mcl \approx 10^{3.5}\,\msun$). 
We vary the initial conditions of the calculations, such as the initial half-mass radius of the clusters, 
initial binary populations for massive stars and initial mass segregation. 
We find that the initial density is the most influential parameter for the ejection fraction 
of the massive systems. 
The clusters with an initial half-mass radius $\rhi$ of 0.1 (0.3)\,pc can eject up to 50\% (30\%) 
of their O-star systems on average, while initially larger ($\rhi = 0.8$\,pc) clusters, that is,  
lower density clusters, eject hardly any OB stars (at most ${\approx}4.5$\%).
When the binaries are composed of two stars of similar mass, the ejections are most effective.  
Most of the models show that the average ejection fraction decreases with decreasing stellar mass. 
For clusters that are efficient at ejecting O stars, the mass function of the ejected stars is 
 top-heavy compared to the given initial mass function (IMF), 
while the mass function of stars that remain in the cluster becomes slightly steeper 
(top-light) than the IMF. The top-light mass functions of stars in 3\,Myr old clusters 
in our \nbody\ models agree well with the mean mass function of young intermediate-mass 
clusters in M31, as reported previously.  
This implies that the IMF of the observed young clusters is the canonical IMF. 
 We show that the multiplicity fraction of the ejected massive stars 
can be as high as ${\approx}60$\%, that massive high-order multiple systems can be dynamically ejected, 
and that high-order multiples become common especially in the cluster.  
We also discuss binary populations of the ejected massive systems. 
Clusters that are initially not mass-segregated begin ejecting massive stars after a time delay 
that is  caused by mass segregation.  
When a large kinematic survey of  massive field stars becomes available, 
for instance through \emph{Gaia}, our results may be used to constrain the birth configuration 
of massive stars in star clusters. 
The results presented here, however, already show that the birth mass-ratio distribution for O-star primaries  
must be near uniform for mass ratios $q\gtrsim 0.1$.
}

\keywords{methods: numerical -- stars: kinematics and dynamics -- stars: massive -- 
open clusters and associations: general -- galaxies: star clusters: general}

\maketitle

\section{Introduction}
\label{ejsec:intro}

Massive runaways \citep{Bl61} are massive stars 
that move with a high peculiar velocity (${\gtrsim}30\,\kms$) or 
that are located at large distances from the Galactic plane.  
They have been observed in the Galaxy \citep[and references therein]{Bl61,GB86,Hoogerwerf01,Get12} 
and nearby galaxies \citep{Get12}.
At high redshift, massive runaways may have played an important role in reionising the Universe \citep{CK12}.

\citet{Bl61} proposed the binary supernova hypothesis to explain the origin of runaways, 
in which the initially less massive star in a binary that is composed of two massive stars obtains a high velocity when the more massive star explodes as a supernova.
\citet{PRA67} proposed another mechanism for the origin of runaways, namely the dynamical ejection through energetic few-body interactions. When a close encounter between stars involves a binary, binding energy can
be transformed into kinetic energy and a runaway can thus be produced. 
\citet{LD90} showed with small-number \nbody\ calculations that the most efficient way of producing 
massive runaways is a close binary--binary encounter.
Dynamical ejections of massive stars have been studied numerically
with three/four-body scattering experiments \citep[e.g.][]{Leonard95,GPE04,GGP09,GG11} 
and with full \nbody\ calculations of real-sized young star clusters \citep[e.g.][]{FP11,BKO12,PS12,OKP15}.  
These two scenarios are considered as the main processes for populating massive stars 
outside of clusters, especially runaways, from star clusters. 
In reality both processes will play a role \citep{Hoogerwerf01,Tetzlaff11}. 
The combination, the two-step ejection process, is a key mechanism to provide massive field stars 
that cannot be traced back to their birth cluster, which leads to a misinterpretation of such stars 
as being formed in isolation \citep{PK10}. 

In the supernova scenario, the ejection requires a supernova explosion, implying that 
the ejection will occur only after a few Myr from the birth of the stars.  
But a dynamical ejection can occur any time in a stellar lifetime and even during the formation time, 
that is, at ages younger than a Myr. 
There are very young star clusters that are too young to have had a supernova, but which have runaway OB stars around them, 
for example, NGC~6611 \citep{GB08}, NGC~6357 \citep{Get11}, NGC~3603 \citep{RL12,RL13,Gvaramadze13}, 
and Westerlund~2 \citep[cf. \citealt{Hur15}]{RL11}. 
This implies  that the massive stars are ejected from their natal cluster at an early age. 
Those field O stars that can be traced back to a star cluster or an association 
appear to have left their birth place at a very early age \citep{SR08}.  
Very massive (${\gtrsim}100\,\msun$) runaways \citep[e.g. 30 Dor 016,][]{Evans10} and 
very massive  single stars and binaries in apparent isolation 
(e.g. VFTS 682, \citealt{Bestenlehner11}; R144, \citealt{Sana13}) in the Large Magellanic Cloud are most 
likely the outcome of dynamical ejections \citep{BKO12,OKB14}.  
These observations therefore imply that the dynamical ejection process is very significant 
for the massive star population within a cluster (e.g. loss of massive stars in the cluster) 
and outside of a cluster (e.g. origin of the field massive stars).

The dynamical ejection process is sensitive to the initial conditions of massive star populations 
and to the properties of their birth cluster. The runaways can therefore be used as constraints to probe 
the initial configuration of massive stars in star clusters, under the assumption that dynamical ejection 
is the dominant process in producing the massive runaways. 
By assuming that OB runaways are dynamically ejected from star clusters, \citet{CP92} deduced 
the birth configuration of OB stars using observed properties of OB runaways,  
such as their proportion and the high ratio of O to B stars in the runaway population. 
They suggested that massive stars form in compact groups of binaries with mass ratios 
biased toward unity and containing a deficit of low-mass stars. This configuration would correspond 
to the mass-segregated core of a dense young star cluster.
The authors provided analytical results and neglected the cluster potential.

If all the field O stars, not only the runaways, 
have formed in clusters and then populated the Galactic field 
dominantly by the dynamical ejection process, studying the properties of the dynamically ejected population 
may help to constrain the initial configuration of massive stars in their birth cluster. 
It is therefore our aim to study how the properties of the ejected massive stars vary with 
various initial conditions using direct \nbody\ calculations that include the entire cluster.

Using the theoretical young cluster library presented in \citet{OK12} and \citet{OKP15}, 
we here study the dynamical ejection of massive stars from young star clusters with
diverse initial conditions.  
The paper is structured as follows. Initial conditions of our \nbody\ models are briefly described in Sect.~\ref{ejsec:nbody}. 
Section~\ref{ejsec:fej} describes the fraction of ejected OB stars in different models.  
Properties of ejected stars produced in the \nbody\ models, such as the velocity distribution
and mass function, are presented in Sect.~\ref{ejsec:prop}. In Sect.~\ref{ejsec:multi}, 
multiplicity fractions and binary populations of ejected massive systems are illustrated. 
The discussion and summary follow in Sects.~\ref{ejsec:dis} and \ref{ejsec:sum}.

\section{$N$-body models}
\label{ejsec:nbody}

\begin{table*}
\caption{ \label{ejtab:models} List of the \nbody\ models and their initial conditions.}
\centering
\begin{tabular}{@{}lcccccc@{}}
\hline\hline
Model & $\rhi$ (pc) & IMS & $f_{\mathrm{bin}}$ & IPD & Pairing method & $e$ dist.\\
\hline
MS1OP  & 0.1 & Y & 1 & Kroupa & OP & thermal\\
MS3OP\_SPC & 0.3 & Y & 1 & Sana et al. & OP & $e=0$\\
MS3OP\_SP & 0.3 & Y & 1 & Sana et al. & OP &thermal\\
MS3UQ\_SP & 0.3 & Y & 1 & Sana et al. & uniform $q$-dist. &thermal\\
MS3OP  & 0.3 & Y & 1 & Kroupa & OP &thermal\\
MS3RP  & 0.3 & Y & 1 & Kroupa & RP &thermal\\
MS3S   & 0.3 & Y & 0 & - & - & - \\
NMS3OP & 0.3 & N & 1 & Kroupa & OP &thermal\\
NMS3RP & 0.3 & N & 1 & Kroupa & RP &thermal\\
NMS3S  & 0.3 & N & 0 & - & - & - \\ 
MS8OP  & 0.8 & Y & 1 & Kroupa & OP &thermal\\
MS8RP  & 0.8 & Y & 1 & Kroupa & RP &thermal\\
MS8S   & 0.8 & Y & 0 & - & - & - \\
NMS8OP & 0.8 & N & 1 & Kroupa & OP &thermal\\
NMS8RP & 0.8 & N & 1 & Kroupa & RP &thermal\\
NMS8S  & 0.8 & N & 0 & - & - & - \\
\hline
\end{tabular}
\tablefoot{Initial half-mass radius, $\rhi$, is listed in Col.~2. 
 Column 3 denotes initial mass segregation (IMS), N standing for the initially unsegregated 
 cluster model and Y for the initially mass-segregated one. The initial binary fraction,
  $f_{\mathrm{bin}}$, is listed in Col.~4.
 Initial period distributions (IPD) applied to massive binaries are listed in Col.~5, where 
 Sana et al. and Kroupa refer to the distributions introduced in \citet[][Eq.~(\ref{eq:P_Sana})]{Set12} and \citet[][Eq.~(\ref{eq:P_Kroupa})]{PK95b}. Column~6 is the pairing method for massive binaries 
 (primary mass $m_{1}\geq 5\,\msun$, for details see Sect.~\ref{ejsec:nbody}). 
 The last column gives the initial eccentricity distribution for massive binaries.  In binary-rich clusters, all low-mass binaries have the period distribution of \citet{PK95b} and a thermal eccentricity distribution, and their component masses are randomly paired for consistency with observational data \citep{PK95a,PK95b,MK11}.
 Each model is computed $N_{\mathrm{run}}=100$ times with different random number seeds 
 to obtain representative statistics.}
\end{table*}

We adopted \nbody\ models of $\mcl=10^{3.5}\,\msun$ clusters  
from the theoretical young star cluster library of \citet{OK12} and its extension in \citet{OKP15}. 
With this mass, model clusters are the most efficient at ejecting O-star systems \citep{OKP15} 
and they have the largest number of ejected O-star systems per set of initial conditions 
because more runs (100 realisations for each set of initial conditions) 
were performed than for the clusters with higher masses (e.g. 10 runs for $10^4\,\msun$ models). 
The library contains a wide variation of the initial conditions for star clusters, such as 
the initial half-mass radius, initial mass segregation, binary fraction, and two pairing 
methods for massive binaries. Thus the cluster mass we chose here suits the aim of studying 
how the ejection fraction and the properties
of the ejected massive stars depend on the initial conditions. 

Common initial conditions for all models are briefly described as follows \citep[see][for more details]{OK12,OKP15}.
Stellar masses are randomly drawn from the two-part power-law canonical initial mass function (IMF),
\begin{equation}
\xi(m) \propto m^{-\alpha_{i}},
\label{eq:imf}
\end{equation} 
with power-law indices of $\alpha_1 = 1.3$ for $0.08 \leq m/\msun < 0.5$ 
and of $\alpha_2 = 2.3$ (the Salpeter index) for $m/\msun \geq 0.5$, 
where $m$ is the stellar mass \citep{KP01,Ket13}. 
The lower mass limit for stellar masses adopted here is 0.08 $\msun$. 
The upper stellar mass limit of the IMF, $\mmax$, is derived from the relation between 
maximum stellar mass and cluster mass  \citep{WK06,WKB10,WKP13}, e.g. $\mmax\approx 79.2\,\msun$ for
the $10^{3.5}\,\msun$ cluster.
Initial positions and velocities of stars (centre of mass in the case of a binary system) 
are generated according to the Plummer density profile with the assumption 
that the clusters are initially in virial equilibrium. 
In initially mass-segregated models (with a name starting with MS), 
the more massive star is more closely bound to the cluster \citep{BDK08}.  
Mainly two different initial half-mass radii of $\rhi=0.3$ and $0.8$\,pc are used, but we also include a model 
with $\rhi=0.1$\,pc for comparison. The $0.3$\,pc model-sets are consistent with the radius-mass relation 
of very young embedded clusters \citep{MK12}. 

The typically large angular momenta of star-forming cloud cores (\citealt{Goodman93}; Fig.~13 in \citealt{PK95b})  
imply that most, if not all, stars form in a binary multiple system \citep{Get07,DK13,Reipurth14},  
as shown in observations, for instance, the high binary fractions for massive stars in clusters \citep{Set14,Cet12} 
and for protostars \citep{Det07,Chen13}. 
We therefore include primordial binaries in most of our models for a more realistic initial set-up 
of the young star clusters, especially for the massive star population in them. 
In this study, we assume a binary fraction of 100\% for the initially binary-rich models. 
Modelling a cluster with primordial binaries requires distributions of initial binary parameters
such as eccentricity, $e$, period, $P$ in days, and mass ratio, 
$q=m_{2}/m_{1}$ where $m_{1}$ and $m_{2}$ are the masses of the binary components and $m_{1} \ge m_{2}$.

The models in \citet{OK12} adopted Eq.~(8) of \citet{PK95b}, 
\begin{equation}\label{eq:P_Kroupa}
f\left(\log_{10}P\right) = 2.5\frac{\log_{10}P -1}{45 + \left(\log_{10}P-1\right)^{2}},
\end{equation}
 as the initial period distribution for all binaries \citep{MK11,Marks14,Leigh15}. 
This period distribution was deduced from studies of solar-mass binaries that have been intensively studied \citep{DM91}, 
 and is also applied to the massive binaries in the library since the period distribution of 
massive binaries is poorly constrained in observation and theory. 
However, recent observational studies show a high fraction of short-period binaries for massive 
(e.g. O star) binaries \citep{Set12,Ket14}. 
\citet{OKP15} additionally performed cluster models with the \citet{Set12} period distribution 
for massive binaries with a primary mass ${\geq}5\,\msun$ \citep[Eq.~(3) in][]{OKP15},
\begin{equation}\label{eq:P_Sana}
f\left(\log_{10}P\right)= 0.23\times \left(\log_{10}P\right)^{-0.55},
\end{equation} 
with a range $0.15 \leq\log_{10} (P/\mathrm{days})\leq6.7$.
These models are indicated with \_SP in their names, for example MS3UQ\_SP, MS3OP\_SP, and MS3OP\_SPC.

For lower mass binaries ($m_{1}< 5\,\msun$), masses of binary components are randomly 
chosen from the IMF. For massive binaries ($m_{1}\geq 5\,\msun$), we use three different pairing methods. 
The first is random pairing (RP) as for lower mass binaries, the second is ordered pairing \citep[OP,][]{OK12},  
which produces $q$ biased towards unity by sorting stars with decreasing mass after generating the mass-array from the IMF and then pairing them in the sorted order. 
The last is a uniform mass-ratio distribution (UQ, the MS3UQ\_SP model) in which the companion is chosen from 
the stellar masses already drawn from the IMF with the closest mass ratio to a value drawn 
from a uniform distribution for $q\geq0.1$. The uniform distribution is adopted to imprint the observed 
mass-ratio distribution of O-star binaries \citep{Set12,Ket14}. The very important point we need to emphasize
 here is that our procedure of first drawing $N$ stars from the IMF with a combined mass of $10^{3.5}\,\msun$ 
and only then pairing stars from this array to binaries is essential so as to not change the shape of the stellar IMF.

In the model libraries, the eccentricities are generated from the thermal distribution, 
$f(e)=2e$ \citep{Kr08}. While the observed eccentricity distribution of low-mass and solar-type binaries 
is well accounted for by the thermal distribution, that of the high-mass counterpart is not well described with a simple functional form. 
As a simple test, we additionally include calculations that are the same as the MS3OP\_SP models, 
but with all massive binaries (with primary star ${\geq}5\,\msun$) being initially in a circular orbit (MS3OP\_SPC).  

We include single-star cluster models, that is, those with an initial binary fraction of 0, for comparison. 
These models are indicated in the model names with the suffix S. 

Each cluster is calculated with the direct \nbody\ code \textsc{nbody6} \citep{Aa03} up to 5\,Myr 
and stellar evolution is taken into account with the stellar evolution library \citep{HPT00,HTP02} 
that is implemented in the code. Throughout this paper, however, we use the snapshots at 3\,Myr because after this time, stellar evolution significantly changes the masses of the most massive stars in the clusters, and the first supernova occurs at around 4\,Myr .

Table~\ref{ejtab:models} lists the initial conditions for each of the models. 
The wide range of initial conditions will help us to study how properties of 
the ejected OB stars change with cluster initial condition 
and to constrain the initial configuration of massive stars in a star cluster.
Sixteen models are considered in total and each model contains 100 realisations 
with a different random seed number. The results are therefore either averages or the compilation 
of 100 runs. 

\section{Effects of the initial conditions on ejection fractions}
\label{ejsec:fej}
In this section we discuss how each initial condition affects the resulting dynamical ejection fractions, especially those of massive stars. 
Before discussing the results, we define terms and quantities used in this paper for clarity. 
We refer throughout to a star more massive than $17.5\,\msun$ as an O star, 
to one with a mass between $3$ and $17.5\,\msun$ as a B star, and to one with a mass between $1.65$ and $3\,\msun$ as an A star \citep{Ad04}. The mass is determined from the snapshots at 3\,Myr. Although we switched 
on stellar evolution, their masses change little from the initial values unless stars collide with other stars. 
We call a star massive when its mass exceeds $5\,\msun$ (all O and early-B stars) in this study. 
Throughout this paper, we use the word system for either a single star or a multiple system 
(see Sect.~\ref{ejsec:findmultiple} for the procedure for searching multiples). The spectral type of the system 
is defined to be that of the primary star (the most massive component) in the case of a multiple system.

From snapshots at 3\,Myr, we regard a system as an escapee when it is located farther 
than $3\times\rh(3\,\mathrm{Myr})$ from the cluster centre and is moving faster than the escape velocity at its distance. 
The averaged $\rh(3\,\mathrm{Myr})$ is approximately 0.5, 0.55--0.7, and 0.85--1.0\,pc (depending on models) for clusters with $\rhi=0.1$, 0.3, and 0.8\,pc, respectively \citep{OK12,OKP15}. 
We assume that the escapees are dynamically ejected since the dynamical ejection process 
is the only way to move massive stars outwards at a young age before the first supernova explosion. 
A cluster within an extreme environment, such as the Galactic centre, may lose its massive stars
through tidal stripping by a strong tidal field if the cluster forms with a size similar 
to its tidal radius and without primordial mass segregation 
\citep[e.g.][but the observation indicates that those stars are possibly ejected from young star clusters 
through energetic encounters with other stars, \citealt{Det15}]{HSH14}. 
However, it is worth mentioning that lower mass escapees are not necessarily ejected. 
Slowly moving stars may be the outcome of evaporation that is driven by two-body relaxation for these escapees. 

\begin{figure*}
 \centering
 \resizebox{0.93\hsize}{!}{\includegraphics{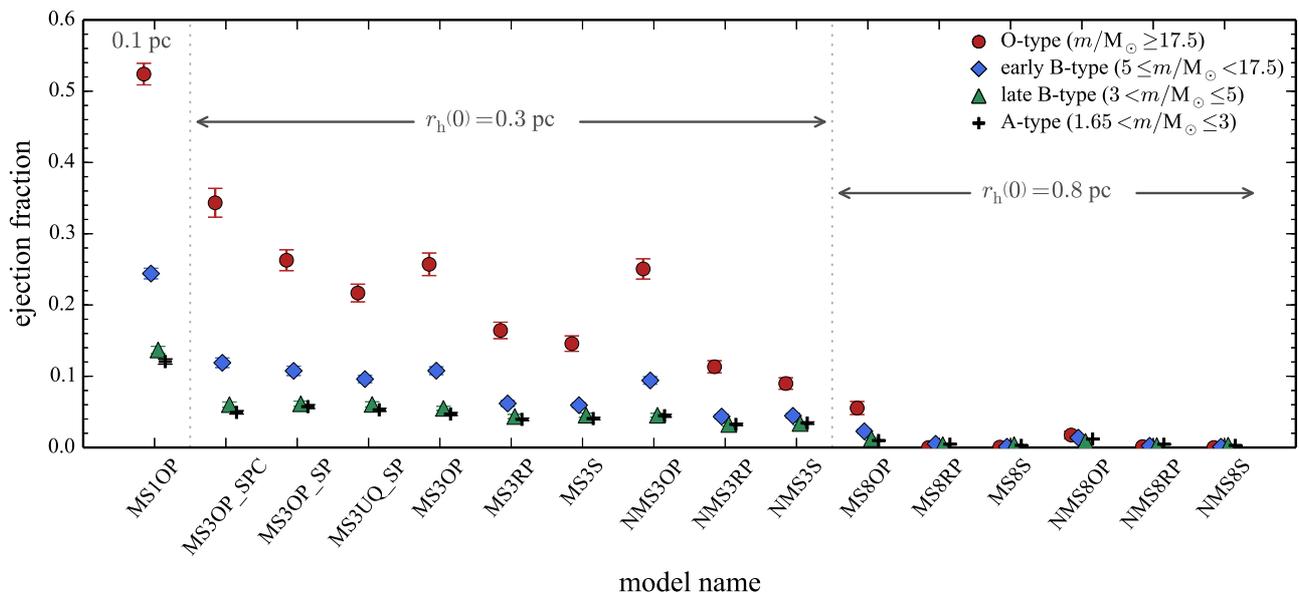}}
 \caption{
   Averaged ejection fractions (Eq.~(\ref{eq:fej})), $\langle f_{\mathrm{ej,ST}}\rangle$, 
   of systems for four different primary-mass groups at 3\,Myr. 
    Model names are indicated on the x-axis (see Table~\ref{ejtab:models}). 
    The error bars are the standard deviation of the mean. 
  }
  \label{ejfig:fej}
\end{figure*}

\begin{figure*}
 \centering
 \resizebox{0.93\hsize}{!}{\includegraphics{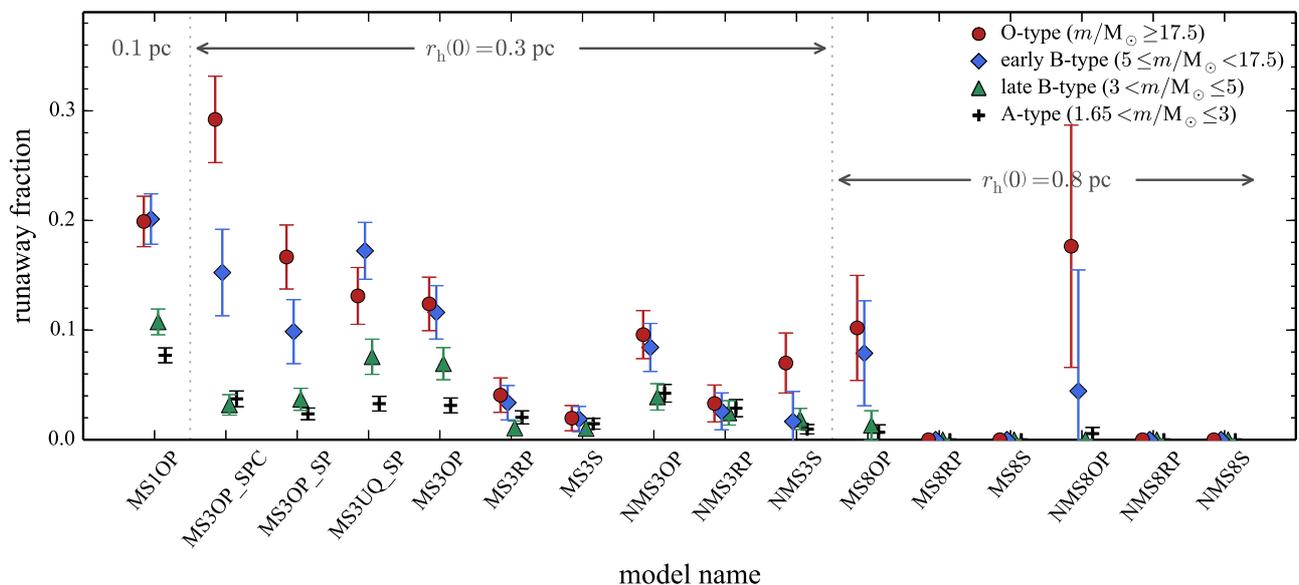}} 
 \caption{Runaway fraction among the ejected systems, the number of runaways divided by the number of ejected stars (Eq.~(\ref{eq:fr})). 
 The fraction is derived from all runs for each model. The error bars indicate Poisson uncertainties. }
 \label{ejfig:fr} 
\end{figure*}

The ejection fraction of a certain spectral type (ST) star,\footnote{We use the term spectral type to distinguish groups with 
different stellar masses.} $f_{\mathrm{ej,ST}}$, is defined as
\begin{equation}\label{eq:fej}
  f_{\mathrm{ej,ST}} = \frac{N_{\mathrm{ej,ST}}}{N_{\mathrm{sys,ST}}},
\end{equation}
where $N_{\mathrm{ej,ST}}$ and $N_{\mathrm{sys,ST}}$ are the number of the ejected systems and 
the total number of systems for a certain stellar type in a calculation, respectively.
Throughout the study, we use the average of ejection fractions from 100 realisations of each model. 
An ejected system moving with a velocity greater than $30\,\kms$ is referred to as a runaway.  
The runaway fraction is the ratio of the number of the runaway systems, $N_{\mathrm{r,ST}}$,  
to the number of ejected systems. We use the runaway fraction relative to all ejected systems of each model, 
\begin{equation}\label{eq:fr}
  f_\mathrm{r,ST}= \frac{ \sum N_{\mathrm{r,ST}}}{ \sum N_{\mathrm{ej,ST}}}.
\end{equation}

In the following subsections, we discuss the results depending on the initial conditions, such as initial size, mass segregation, and binary populations.
Figures~\ref{ejfig:fej} and \ref{ejfig:fr} show the averaged ejection fractions and the runaway fractions of different stellar mass groups  for all models.
\begin{figure*}
 \centering
 \resizebox{0.93\hsize}{!}{\includegraphics[width=170mm]{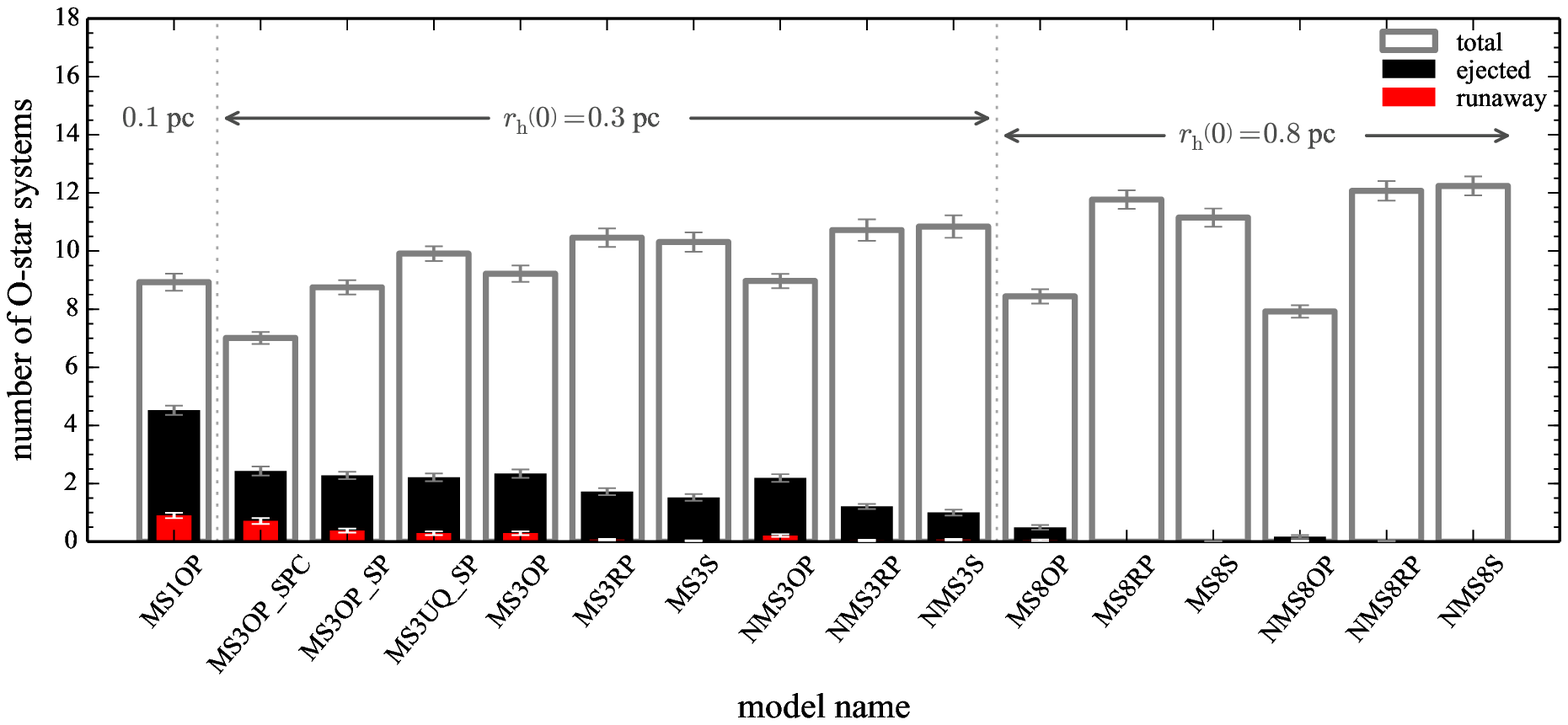}}
 \caption{Averaged numbers, $\bar{N}$, of total (open grey), ejected (black), and runaway (red bars) 
 O-star systems in the models at 3\,Myr. The error bars are the standard deviation of the mean. 
 The total number varies because of the different binary fraction and/or the different mass-ratio distribution at 3\,Myr.}
 \label{ejfig:no} 
\end{figure*}

\subsection{Initial size (density) of cluster}
\label{ejsec:icsize}
The close encounter rate is proportional to the cluster density.  
Since we chose the same cluster mass for all models, it is expected that the ejection is more efficient 
with decreasing initial size of the cluster in our models.

The highest ejection fraction appears in the most compact cluster in our model library (MS1OP) 
with an average O-star ejection fraction of about 52\%, which is about twice higher than for the model with $\rhi=0.3$\,pc (MS3OP). 
Clusters with $\rhi = 0.3$\,pc ($\rhoh \approx 1.4\times10^{4}\,\uden$)  
eject ${\approx}9$--$34$\% of O-type systems depending on the initial conditions (Fig.~\ref{ejfig:fej}).  
However, clusters with $\rhi = 0.8$\,pc ($\rhoh \approx 740\,\uden$) 
eject hardly any massive stars. Only the binary-rich clusters with OP massive binaries eject  
a few percent of OB stars (${\lesssim}3.5$\% for O-type and ${<}1$\% 
for B-type systems). None of the single-star clusters or RP binary-rich clusters 
eject OB stars.  It is therefore very unlikely that this or larger clusters with 
the mass studied here populate massive stars into the field through dynamical ejections.
Most of the massive stars in the field should form in compact star clusters ($\rhi\lesssim 0.5$\,pc), 
which eject massive stars efficiently if they have a dynamically ejected origin. 
But if the cluster is too compact ($\rhi\lesssim 0.1$\,pc), the ejection of O-star system is 
so efficient that it may lead to a higher fraction of O stars relative to all young stars 
in the field than in the cluster.

The determining factor for the number of ejected O-star systems is the initial cluster size. 
For the same initial half-mass radius, the averaged number of the ejected O-star systems is similar 
and approximately independent of the other initial conditions, 
although the ejection fraction varies as a result of the different number of total O-star systems,  
which depends on the initial binary fraction, the mass ratio of O-star binaries, and the stellar merger rate (Fig.~\ref{ejfig:no}).

Although the dependence of the O-star runaway fraction on the model properties is similar 
to that of the O-star ejection fraction, the runaway fraction does not seem to be strongly 
dependent on the initial size (i.e. density) of the cluster. 
The ejection velocity of a system attained after an energetic close encounter 
is mainly determined by configurations of the systems that are involved in the encounter. 
For example, for a single--binary encounter, the ejection velocity $v_{\mathrm{ej}}$ is 
of the order of the orbital velocity of the binary \citep{Heggie80,GB86,PS12}, 
\begin{equation}
 v_{\mathrm{ej}} \propto \left(\dfrac{m}{a}\right)^{1/2}, \label{eq:vej}
\end{equation} 
where $a$ and $m$ are the semi-major axis of the binary and the typical mass of the stars 
engaging in the encounter, respectively. 
The ejection velocity is higher when a tighter binary is involved \citep[e.g.][]{GGP09}.
It is expected that the denser cluster (i.e. the cluster with the higher velocity dispersion) 
is more efficient in producing runaways among ejected systems 
if the initial density is the only difference between clusters. 
The reason is that binaries with an orbital velocity similar to the velocity dispersion of the cluster 
are dynamically most active,  and a higher velocity is required for escapees 
from a denser cluster because of its deeper gravitational potential 
\citep[see also Fig.~8 in][for the runaway fraction as a function of cluster mass for clusters with the same initial half-mass radius]{OKP15}. 
Based on comparing MS1OP and MS3OP (the two models differ only in size, i.e. density) in Fig.~\ref{ejfig:fr}, 
the initially denser cluster has a higher runaway fraction. 
However, clusters with a lower density (a larger size) can produce a higher fraction of runaways among ejected systems 
than denser (more compact) clusters if a higher fraction of energetic (e.g. close) binaries are present in the less dense clusters.  
For example, the model with short period (Eq.~(\ref{eq:P_Sana})) and 
only circular massive binaries (MS3OP\_SPC) produces runaways 
more efficiently than the smaller cluster model that has the Kroupa period distribution (Eq.~(\ref{eq:P_Kroupa})) 
and the thermal eccentricity distribution (MS1OP).         
An initial binary population (especially, with which clusters maintain a high fraction of short-period binaries after an early dynamical evolution)  
can therefore be more important than the initial density for efficiency with which runaways are produced among ejected stars (see also Sect.~\ref{ejsec:icibp}).
The high O-star runaway fractions in some models that have a low ejection fraction  
(e.g. the MS3S, MS8OP, and NMS8OP models) 
arise from the small number of the ejected O-star systems.

\subsection{Initial mass-segregation} 
\label{ejsec:icims}
For cluster models with $\rhi=0.3$\,pc, the final ejection fraction 
is not strongly sensitive to whether the cluster is initially mass-segregated when
a massive binary is initially composed of two massive stars. 
The ejection fraction of OB stars for the MS3OP model, for instance, is only $1$--$2$\% higher than 
 that of the NMS3OP model.  

However, differences are shown when the massive binaries are randomly paired or 
when there is no initial binary. The MS clusters eject OB stars almost twice as often as the NMS clusters. 
These initial conditions are unrealistic, however, because many massive stars in
young star clusters are binaries and the companions are very likely massive stars.

For clusters with $\rhi=0.8$\,pc, the MS8OP model is more efficient at ejecting massive stars 
than the NMS8OP model, even though the massive binaries are composed of two massive stars.
This is due to the larger size of the clusters, which leads to a significantly longer 
mass-segregation timescale even if massive stars are paired with other massive stars. 
Such large initially not mass-segregated clusters therefore need time for the massive stars 
to segregate to the core. Ejections of massive stars arise from there on.

In realistic models with characteristic radii of a few tenths of a pc and 
a high proportion of short-period binaries with a similar spectral-type companion, 
it would be difficult to determine whether the cluster was initially mass-segregated 
by only considering the ejection fraction of massive stars 
because the initial mass segregation is not a strong factor for the ejection fraction at 3\,Myr.  
But other properties, such as the time distribution when
the ejections occur, may show differences between initially mass-segregated and not mass-segregated clusters. 
It is expected that the MS clusters would start ejecting massive stars earlier 
than the NMS clusters would because the dense core of massive stars already exists in 
the centre of the MS clusters at the beginning of the cluster evolution.
This is discussed in Sect.~\ref{ejsec:tej}.

\subsection{Initial binary population}
\label{ejsec:icibp}
We varied four parameters for the initial binary populations of massive systems.
The first was the binary fraction. 
The dynamical ejection process most likely involves at least one binary, regardless of
whether it was primordial or dynamically formed.  
During the energetic close encounter between a binary and a single star or binary, 
 the binding energy of the binary transforms into kinetic energy and a star may be ejected. 
Binaries have a larger cross-section,  
which increases the probability of energetic encounters. The dynamical formation of binaries is inefficient. 
Thus a high initial binary fraction enhances the interaction rate, resulting in a significantly 
higher ejection fraction.
Including primordial binaries plays a central role in ejecting massive stars from the cluster 
because single-star clusters eject OB stars slightly later than binary-rich clusters. 
This is a result of the time that first needs to pass for binary systems to form dynamically, as is evident 
in the example presented here (clusters with $\rhi=0.3$\,pc).  
  
The second parameter we varied was the pairing method (i.e. mass-ratio distribution) for the massive binaries. 
None of the single-star clusters with $\rhi=0.8$\,pc ejects OB star systems by 3\,Myr (Fig.~\ref{ejfig:fej}).
The situation does not change when the initial binary fraction is 100\%,  
but the massive binaries are paired with companions randomly chosen from the whole IMF.    
For clusters with $\rhi=0.3$\,pc, both single-star and RP binary-rich clusters
eject approximately $4$--$6$\% of O-star systems and ${\approx}1.5$--$2.3$\% of B-star systems by 3\,Myr. 
The ejection fractions are similar for both cluster models, meaning that the difference is smaller than $1$\%.  
This is because most massive binaries paired randomly behave like single stars  
because their mass ratios are very low. The mass ratio of the system 
would be $q \approx 10^{-3}$ in the most extreme case, for instance, if the most massive star 
(${\approx}80\,\msun$) is paired to the least massive star (${\approx}0.08\,\msun$) in the cluster.

When massive binaries initially have a massive companion (OP and UQ models), however, clusters 
eject massive stars more efficiently. 
For initially mass-segregated clusters with $\rhi=0.3$\,pc, for instance, 
ejection fractions of massive systems for the OP cluster models 
are more than twice as high (differences of up to $16$\% for O-star systems and $6$\% for early-B star systems) 
as those of the single-star or RP clusters.  
Runaway fractions for the OP models are also higher than those of the single-star or RP models (Fig.~\ref{ejfig:fr}). 
For the OP models, the massive binaries involved in the close encounters initially have a higher 
binding energy as a result of the higher masses of binary components, 
which increases the ejection velocity (Eq.~(\ref{eq:vej})),  
compared to the massive binaries for the single-star or RP models.
Not only do they show higher 
ejection and runaway fractions, but the ejection of massive systems from these clusters occurs 
when they are younger than 1\,Myr (see Sect.~\ref{ejsec:tej}). 
We also note that the OP model (MS3OP\_SP) in which O-star binaries are initially paired with an O star 
shows higher ejection and runaway fractions of O-star systems than the UQ model (MS3UQ\_SP) in which
 a large portion of O-star binaries initially have a B-star companion. 
This difference in companions of O-star binaries may result in a higher runaway fraction of B-star systems  
for the UQ model than for the OP model (Fig.~\ref{ejfig:fr}). 
 For the UQ model B stars can engage in close encounters with O stars more often than in the OP model. 
Therefore, the UQ model has a higher probability of producing B-star runaways 
since the least massive star among interacting stars gains the highest velocity.

The third parameter we varied was the initial period distribution function. Given two initial period distribution functions 
considered for the massive binaries in this study (Eqs.~(\ref{eq:P_Kroupa}) and (\ref{eq:P_Sana})), 
the distribution is not a significant factor in determining the ejection efficiency of O-star systems. 
For example, the models with the two different initial period distributions (e.g. MS3OP\_SP and MS3OP) 
show similar results for the O-star ejection fraction and the number of the ejected O-star systems (Figs.~\ref{ejfig:fej} and \ref{ejfig:no}). 
However, the model with the short-period massive binaries (MS3OP\_SP) produces more runaways 
than the model with the softer Kroupa period distribution (MS3OP)  (Fig.~\ref{ejfig:fr}).   

The last parameter we changed was the initial eccentricity distribution function. 
For the initial eccentricity distribution of massive binaries, all binary-rich models but one assumed the thermal distribution, 
which well reproduces the observed eccentricity distribution for low-mass binaries \citep{PK95b}.  
Since massive short-period binaries are mostly found in circular orbits \citep[e.g. $e=0$ for $P\leq2$\,days,][]{Set12}, 
we included a model (MS3OP\_SPC) for which all massive binaries were in circular orbits.
The two models in which only the initial eccentricity distribution differs (all systems having initial $e=0$, MS3OP\_SPC, 
versus the thermal distribution, MS3OP\_SP) show a similar average number of the ejected O-star systems at 3\,Myr (black bars in Fig.~\ref{ejfig:no}). 
However, the number of total O-star systems (open grey bars in Fig.~\ref{ejfig:no}) are different for the two models. 
For the MS3OP\_SPC model, the binaries with a short period in a circular orbit hardly break up or merge at all, 
keeping the number close to the initial number of O-star systems, which is half of the number 
of individual O stars. For the MS3OP\_SP model in contrast, a large portion of massive binaries, 
composed of similar-mass components, merge and form a single massive star as a result of their short period and high eccentricity. 
Thus more O stars are formed by merging of less massive stars during the evolution of the clusters.   
With the increase of the total number of O stars by collisions in the MS3OP\_SP model and the smaller 
total number of O-star systems in the MS3OP\_SPC model, the ejection efficiency of the latter model 
is higher than that of the former model, although both models result in the same average number of the ejected O-star systems. 
Differences also appear in the numbers of O-star runaways (Fig.~\ref{ejfig:no}) and the runaway fractions (Fig.~\ref{ejfig:fr}) 
of the two models. With initially only circular binaries, more O-star runaways are produced than in the model with the thermal initial eccentricity distribution.   
The reason may be that the eccentric orbits are mostly near their apo-centre where orbital velocities are lower 
and thus produce lower slingshot velocities than circular orbits with the same semi-major axis.
This means that the initial eccentricity also plays a role in the ejection efficiency 
and in the properties of the ejected systems (Sect.~\ref{ejsec:prop}), especially 
when massive binaries are initially highly biased towards short periods 
(e.g. an initial period distribution given by Eq.~(\ref{eq:P_Sana})).

\section{Properties of the ejected massive systems}
\label{ejsec:prop}
In this section, we consider the properties (e.g. velocities, the mass function, and multiplicities) 
of the ejected OB stars and study how they vary with different initial conditions. 
Here, we only examine the results from the clusters with $\rhi = 0.3$\,pc 
since the clusters with $\rhi = 0.8$\,pc eject hardly any OB stars (most models do not eject OB stars, Fig.~\ref{ejfig:fej}). 
For the $\rhi=0.1$\,pc models, the ejection fraction is high,  
although these clusters are too small to be realistic, and therefore only one set of models 
was calculated for the effect of cluster size (density) on the dynamical ejection.  

\subsection{Ratio of ejection and runaway fractions for O- to B-star systems}
\label{ejsec:fejra}

As already mentioned in the previous section, the ejection fraction generally increases with increasing stellar mass.
For models with $\rhi\leq 0.3$\,pc, the ratios of the ejection fraction of the O-type to 
that of early-B type systems  
are ${\approx}2$--$2.9$, while the ratios of early-B to late-B type systems are ${<}2.0$. 
This tendency indicates that the difference of the ejection efficiency decreases with decreasing stellar mass 
and the difference almost disappears in the low-mass (${<}5\,\msun$) regime.  
The reason for this decrease of the ejection fraction with the stellar mass would be 
that O-star systems quickly sink into the cluster core and are ejected at the earliest time when 
the cluster core is in the densest phase.

In the observational studies, the fraction of runaways for the different spectral types have been presented.
For example, \citet{Bl61} reported the frequency of runaways to be $21$, $2.5$, and $1.5$\% 
for stars with spectral types of O5--O9.5, B0--B0.5, and  B1--B5, respectively. 
\citet{GB86} collected the runaway fractions of O- and B-type stars from several publications and listed them 
as $10$--$25$\% and $2$\%, respectively. 
This means that the ratio of the O- to B-star runaway fraction is between $5$--$10$, depending on the literature.
This decrease in runaway fraction with decreasing stellar mass extends to lower masses. 
The frequency of high-velocity A stars relative to the total number is $0.5$--$1.0$\% \citep{GB86}. 
\citet{Stetson81,Stetson83} found evidence for a young, high-velocity A-star population that accounts 
for $0.1$--$0.2$\% of the A stars in the vicinity of the Sun.
 
In our models, the sharp decrease of the ejection fraction with decreasing stellar mass appears as well. 
When only the ejected systems are considered, the runaway fractions 
of O-star and early-B star systems are almost the same for a few models, although the runway fractions 
generally decrease with decreasing stellar mass (Fig.~\ref{ejfig:fr}).   
In the MS3UQ\_SP model, which is the most realistic model in the library, the runaway fraction 
of the ejected systems for B-star systems is higher than that for O-star systems. 
If we consider entire populations of each mass group for the runaway fraction 
(i.e. the systems that are ejected and those that remain in clusters, for which we replace the denominator 
of Eq.~(\ref{eq:fr}) by $\Sigma N_\mathrm{sys,ST}$), 
the sharp decrease of the runaway fraction with decreasing group mass becomes pronounced. 
For the MS3OP\_SPC model, for example, the runaway fraction relative to the entire O-star systems 
is  ${\approx}10$\%,  while the runaway fraction of early-B type systems (primary mass $5\leq m_{1}/\msun<17.5$) 
relative to the entire early-B type systems is  ${\approx}1.8$\%. In the MS3UQ\_SP model, 
the fractions are $2.8$\% and $1.7$\% for O-star and early-B star systems, respectively. 
In all models the runaway fractions of O-star systems are higher than those of B-star systems.
Although the model values are slightly lower than those of the observed ones, 
they agree with the observational finding  that the runaway fraction increases 
with increasing stellar mass when the fraction of runaways of all systems, whether ejected or not, is considered. 
But this tendency is not a unique feature of the dynamical ejection process. 
It also appears in the binary supernova scenario \citep{ELT11}.  
This means that if the observed samples are a mixture of outcomes of the two processes, 
the supernova scenario will enhance the tendency that appears in our models.
We note that the runaway fractions reported in the literature are space frequencies 
or derived from given star catalogues that may include selection biases and contamination. 
The difference of our theoretical results compared to the observation might also be caused by the different initial configurations 
that the O- and B-star systems may have. In our calculations, 
the distribution function for periods, eccentricities, and mass ratios of 
initial binary population are the same for both O-star and early-B star systems, 
and in reality this may not be the case.

\subsection{Velocities of the ejected massive star systems}
\label{ejsec:vel}

\begin{figure}
  \resizebox{\hsize}{!}{\includegraphics{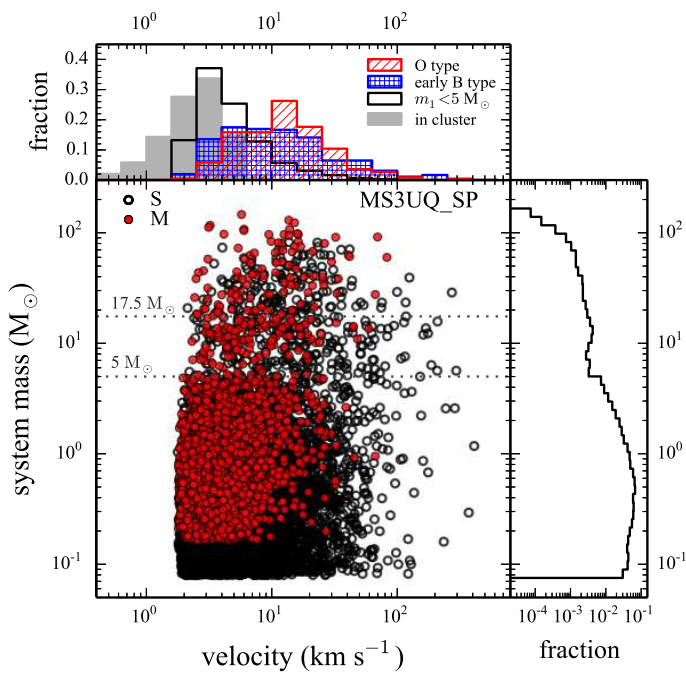}}
  \caption{System mass versus velocity of all escapees from the most realistic model (MS3UQ\_SP) of our
   models. Filled red and open black circles are multiple systems and single stars, respectively. 
   The \textit{top panel} presents the histograms of velocities  of ejected systems for three different (primary) mass groups 
   and all systems that remain in the clusters. The \textit{lower right panel} is the histogram of the system mass for all ejected systems.}
  \label{ejfig:mv}
\end{figure}

\begin{figure}
  \resizebox{\hsize}{!}{\includegraphics{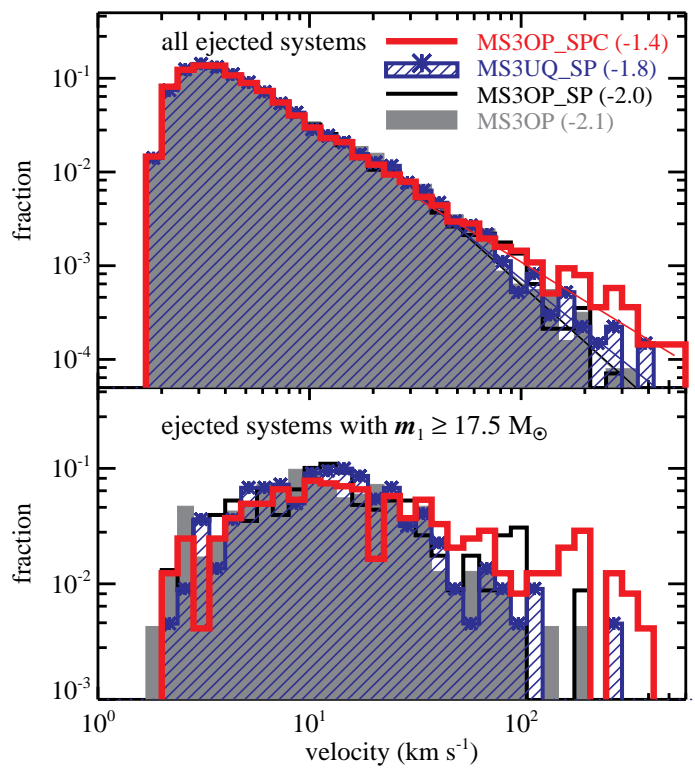}}
  \caption{Velocity distributions of the all ejected systems (\textit{top panel}) and 
  of the ejected O-star systems (\textit{bottom panel}) in four models, 
  MS3OP\_SPC, MS3UQ\_SP, MS3OP\_SP, and MS3OP. The solid lines in the \textit{upper panel} are a linear 
  fit for the systems with a velocity ${\geq}30\,\kms$.   }
  \label{ejfig:vfit}
\end{figure}

In Fig.~\ref{ejfig:mv} we plot the system masses and velocities of all ejected systems 
from the MS3UQ\_SP model at 3\,Myr.  
The histograms in the upper panel of Fig.~\ref{ejfig:mv} 
show velocity distributions of all ejected systems with different stellar masses and 
the distribution of those that remain in the star cluster.  
Systems that remain in the cluster have a velocity lower than 
$10\,\kms$, with a peak near $3$--$4\,\kms$, which is the velocity dispersion 
of a cluster with a mass of $10^{3.5}\,\msun$ and a size 
of $\rh=0.3$\,pc. O-star systems ejected from the cluster show a velocity peak close 
to the escape velocity at the cluster centre, for example about $11\,\kms$ for the $10^{3.5}\,\msun$ 
cluster with $\rhi=0.3$\,pc and about $19\,\kms$ for the $10^{3.5}\,\msun$ cluster with 
$\rhi=0.1$\,pc (see the MS1OP model in Fig.~\ref{aejfig:mv}).
Because massive stars are ejected from the cluster centre, ejected massive
stars require the velocity to be higher than the central escape velocity of the cluster. 
This velocity peak therefore varies with the initial condition of a cluster.
With the same initial half-mass radius, the peak will move to a higher velocity with increasing cluster mass. 

The velocity distribution also varies with the stellar mass of the ejected systems.
The top panel of Fig.~\ref{ejfig:mv} exhibits the velocity distributions of the ejected systems 
for different stellar masses.  
The more massive systems tend to have a distribution biased towards higher velocities 
(see also Fig.~\ref{aejfig:mv} for other models), although in Fig.~\ref{ejfig:mv} the difference 
between the velocity distribution of ejected O-star and of ejected  early-B star systems is marginal. 
\citet{BKO12} also showed that the average velocity of ejected stars increases with stellar mass. 
The lowest velocity of ejected systems slightly increases with system mass 
in the main panel of Fig.~\ref{ejfig:mv}. 
This may be because fewer massive systems are ejected compared to their lower mass counterparts,  
but it may also caused by the effect of the stars being in the cluster. 
As mentioned above, massive stars are ejected from the cluster centre where the gravitational potential 
of the cluster is deepest, therefore a higher velocity is required to escape the cluster, 
while less massive stars are most likely ejected at a location farther from the cluster centre 
where the escape velocity is lower.
Furthermore, massive systems ejected with a low velocity can fall back to the cluster, failing 
to escape the cluster, by dynamical mass-segregation processes if they move away from the cluster centre 
slowly enough to interact with low-mass stars in the outer part of the cluster.  

However,  the relatively less massive systems have the highest velocity
because the least massive of the interacting stars generally gains the highest velocity 
by the energy exchange. This can explain the higher runaway fraction of B-star systems compared 
to O-star systems for the MS3UQ\_SP model in which B stars are most likely engaged 
because a large portion of O-star binaries initially have a B-star companion 
(Fig.~\ref{ejfig:fr}, see also Sect.~\ref{ejsec:icibp}).
For the same reason, the multiple systems have lower velocities than single stars 
\citep[the main panel of Fig.~\ref{ejfig:mv}, see also][]{LD90}. 
An ejection of a binary through a single--binary or a binary--binary encounter accompanies 
an ejection of single star(s),  
and the single star very likely attains a higher velocity because its mass is lower than 
the mass of the ejected binary. 
The velocity distribution of the low-mass group ($m_{1} < 5\,\msun$) overlaps with 
that of stars that remain in the cluster at the peak of the distributions. 
The reason may be that the low-mass group includes low-mass star systems 
that evaporate from the cluster in addition to the ejected systems.

Figure~\ref{ejfig:vfit} exhibits the velocity distribution of ejected systems for four different models 
that are efficient at ejecting O-star systems.
In the lower panel of the figure, we show that the models with a high proportion of 
short-period massive binaries produce more high-velocity stars.  
The model with only circular massive binaries (MS3OP\_SPC) is interesting. It produces a long high-velocity tail 
of ejected O-star systems (Fig.~\ref{aejfig:mv}).  Based on 100 runs of the model, seven single O stars 
have a velocity exceeding $200\,\kms$, mostly close to  $300\,\kms$, and the fastest one has a mass 
of $26.6\,\msun$ and a velocity $381.5\,\kms$. 
Furthermore, two single massive ($m\approx9.5$ and $12\,\msun$) stars in the model move faster 
than $400\,\kms$ ($v\approx462$ and $474\,\kms$, respectively).
This indicates that with realistic massive binary populations, an energetic close encounter in a star cluster 
can be the origin of the high velocity of the runaway red supergiant star that has recently been discovered in M31 
\citep[J004330.06+405258.4, with an inferred initial mass of $12$--$15\,\msun$ and 
an estimated peculiar velocity of  $400$--$450\,\kms$,][]{EM15}.
The production of such a high-velocity massive star in the MS3OP\_SPC model is most likely due to energetic 
(short-period) massive binaries in the model engaging in close encounters that produce high-velocity stars,  
while in the MS3OP\_SP and MS3UQ\_SP models such systems disappear from a cluster by merging 
as a result of their high initial eccentricities (i.e. short peri-centre distances) at the early stage of the evolution.

The solid lines in the upper panel of Fig.~\ref{ejfig:vfit} are the linear fit to all ejected systems 
with $v\geq 30\,\kms$ and their slopes are indicated next to the model names. 
Our results show that the velocity distribution depends on the massive binary population, 
especially at high velocity ($v>70\,\kms$); this agrees with the results of \citet{PS12}. 
Our study shows in particular that differences solely in the massive binary populations 
result in different velocity distributions of the dynamically ejected systems, 
even though the low-mass binaries, which constitute the majority of cluster members, 
are set to have identical initial binary populations for the four models shown in Fig.~\ref{ejfig:vfit}. 
Furthermore, our results suggest that the velocity distribution depends not only on the initial period \citep[separation in][]{PS12} distribution, but also on the initial mass ratios and eccentricities of massive binaries.
Since the binary populations and cluster properties (size and mass) adopted in this study differ from those in \citet{PS12},  we do not reproduce the same velocity distribution (i.e. a power-law slope) as those in their paper.
The slopes we derived from our models, $-1.4$--$-2.0$, are shallower than those in the high-velocity 
regime in \citet{PS12}, $-2.5$--$-3.4$.   
We were unable to separately fit the velocity distribution in the high-velocity regime ($v > 80\,\kms$), 
which was done in \citet{PS12}, because most of our models do not produce a sufficient number of such high-velocity stars for the analysis 
because our cluster mass is lower and we used fewer runs than they.

The highest velocity of the escapees from the cluster is dependent on the cluster model.
None of the single-star or RP clusters ejects OB stars with velocities higher than 
$100\,\kms$. In the OP models, however, some of the escapees exceed $100\,\kms$.
Especially for the MS3OP\_SPC model in which massive binaries are initially on a circular orbit, 
several O stars have a velocity exceeding $200\,\kms$.  
This clearly shows that the initial population of massive binaries is important in producing high-velocity massive stars.

\subsection{Age of the clusters at the time of the ejections}
\label{ejsec:tej}
\begin{figure}
 \resizebox{\hsize}{!}{\includegraphics{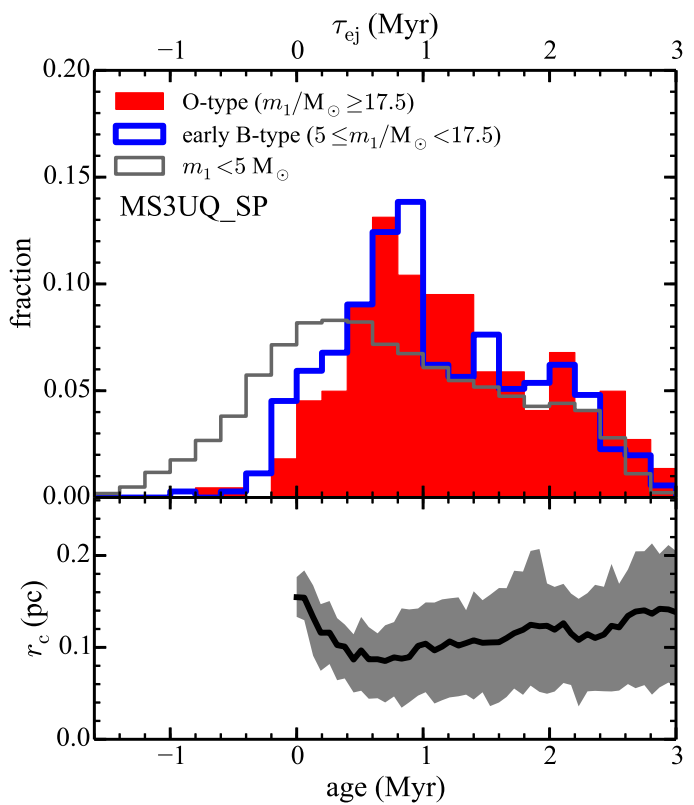}}
 \caption{\textit{Top} : distribution of estimated ages of the cluster when the massive systems 
 are ejected from the cluster, $\tej$, for three different (primary star) mass groups in the MS3UQ\_SP model. 
 All ejected systems from 100 runs for each model are counted.
 \textit{Bottom} : averaged core radius, $r_{\mathrm{c}}$, of the same model 
 as a function of time (black line). The lower and upper boundary of the grey shaded area are 
 the 17th and 84th percentile, i.e., 68\% of clusters have a core radius ranging within the area.  
 The secondary maximum is shown in the \textit{top panel} near 2.1\,Myr when the core radius shrinks again 
 as a result of the Spitzer instability.
}
 \label{ejfig:tauej1}
\end{figure}

\begin{figure}
 \resizebox{\hsize}{!}{\includegraphics{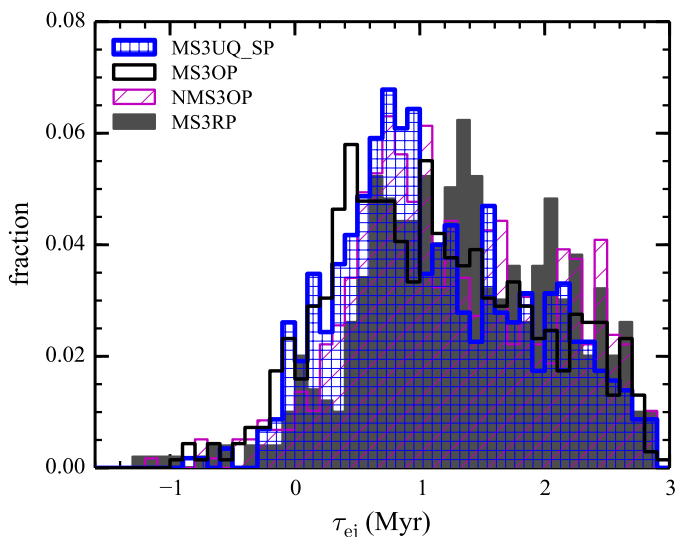}}
 \caption{Distribution of $\tej$ of the ejected massive systems 
 for four different models, MS3UQ\_SP, MS3OP, NMS3OP, and MS3RP. 
 All ejected massive systems from 100 runs for each model are counted.}
 \label{ejfig:tauej2}
\end{figure}

It is not possible to determine the exact moment at which the close encounters that have resulted in ejections 
of the systems have occurred for individual ejected systems in our \nbody\ models because of the 
complexity and the time resolution of the output. 
Here we calculate observationally relevant estimates for when systems are ejected from a cluster by calculating 
the approximate travelling time, $r_{\mathrm{sys}}/v_{\mathrm{sys}}$, where $r_{\mathrm{sys}}$ and $v_{\mathrm{sys}}$ 
are the present distance and velocity of the system relative to the cluster centre.
From a snapshot at $t$ (in this study $t=3$\,Myr), the cluster age when a system has been 
ejected, $\tej$, can be deduced from
\begin{equation}\label{eq:tej}
\tej= t - \frac{r_{\mathrm{sys}}}{v_{\mathrm{sys}}},
\end{equation} 
 under the assumption that the system was ejected from the cluster centre 
and did not experience any further interactions with other systems after it had been ejected. 
This is a reasonable assumption since $r_{\mathrm{sys}}\gg \rh(t)$ and $v_{\mathrm{sys}} \gg \vcl(t)$ 
the velocity dispersion in the cluster.

The distribution of $\tej$ for the ejected systems of the MS3UQ\_SP model at 3\,Myr is shown in Fig.~\ref{ejfig:tauej1}. 
The peak of the $\tej$ distribution, at which the ejections occur most often, 
varies from model to model. In the most energetic case, the MS3OP model 
(the OP cluster with initial mass segregation), the peak appears around 0.5\,Myr 
after the beginning of the calculations, while single-star clusters without initial mass 
segregation eject massive stars most efficiently after 1\,Myr.  

As stars move away from the cluster, the cluster potential can slow the ejected stars,
especially those with low-velocities. This can result in negative values of $\tej$ for the systems that  
have been ejected at the very early stage of cluster evolution. The systems with negative $\tej$ 
indeed have a low velocity (${<}10\,\kms$, see Fig.~\ref{aejfig:tejv}).

The $\tej$ distributions of the massive systems are expected to be related with 
the evolution of the cluster core. The core radius is as given by \textsc{nbody6} 
(Eq.~(15.4) in \citealt{Aa03}; see also \citealt{CH85}). 
Figure~\ref{ejfig:tauej1} shows that  the number of ejection increases as the cluster core shrinks 
and that the maxima of the $\tej$ distributions appear approximately when the core radius is the smallest. 
This shows that the number of massive systems in the core decreases and  the core expands. 

Figure~\ref{ejfig:tauej2} shows that the $\tej$ distribution is dependent on 
the initial configuration of the cluster. 
Evident dilation (the peak at later than 1\,Myr) is present in the model in which massive binaries 
have a companion randomly chosen from the IMF (MS3RP, grey filled histogram 
in Fig.~\ref{ejfig:tauej2}) compared to the models in which massive binaries preferentially 
have a massive companion (the other three models in the same figure).  
In the random-pairing models the massive primaries exchange their low-mass companions 
for massive companions, which causes a time delay until ejections become energetically possible. 
The peak of the $\tej$ distribution occurs later in the 
initially not mass-segregated cluster (e.g. NMS3OP, magenta hatched histogram in the figure) 
than in the mass-segregated one (e.g. MS3OP, black histogram in the same figure).  
The ejection time delay in the not mass-segregated clusters compared 
to the mass-segregated ones is expected to be the timescale at which dynamical mass segregation 
first creates a core of massive stars. 
The dynamical mass-segregation time, $\tms$, is expressed as \citep{Spitzer87}
\begin{equation}\label{eq:tms}
\tms = \frac{\amass}{\mms}\trh,
\end{equation}
where $\amass\approx 0.57\,\msun$, $\mms$, and $\trh$ are the average mass of the stars, 
the mass of the massive system, and the relaxation time at the half-mass radius, respectively. 
The half-mass relaxation time can be estimated as \citep{BT08} 
\begin{equation}\label{eq:trh}
\trh = \frac{0.78\,\mathrm{Gyr}}{\ln(\lambda N)}\frac{1\,\msun}{\amass}\left(\frac{\mcl}{10^5\,\msun}\right)^{1/2}\left(\frac{\rh}{1\,\mathrm{pc}}\right)^{3/2},
\end{equation}  
where $\lambda\approx 0.1$ \citep{GH94}.  
 For a $20\,\msun$ star in a $10^{3.5}\,\msun$ cluster 
with $\rh=0.3$\,pc, the equation gives $\tms$ of about 0.18\,Myr.
The difference between peak locations of the MS3OP (black) and NMS3OP 
(magenta histogram in Fig.~\ref{ejfig:tauej2}) models is ${\approx}0.3$\,Myr.  
For clusters with $\rh=0.8$\,pc, the $\tms\approx 0.78$\,Myr 
for the star with the same mass and is consistent in the difference between the peaks of 
the MS8OP and the NMS8OP models (${\approx}1$\,Myr). 

The ejections require the formation of a core of massive stars. Even though a cluster 
is initially mass-segregated, the cluster needs a time to build a dynamically compact core of massive stars. 
Both MS3OP and MS8OP models are initially mass-segregated, but thanks to 
its shorter mass-segregation timescale, the MS3OP model ejects massive stars 
more efficiently at an earlier time than the MS8OP model (see panels in Fig.~\ref{aejfig:tej}).

\subsection{Present-day mass function of the ejected systems}
\label{ejsec:mf}
\begin{figure} 
   \resizebox{\hsize}{!}{\includegraphics{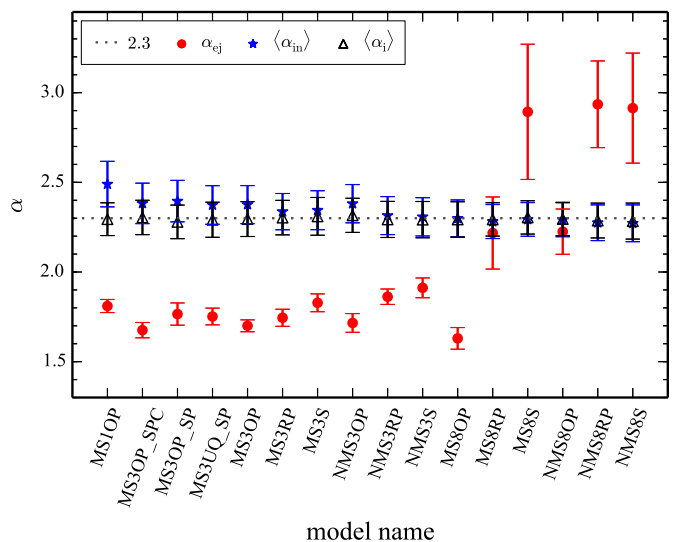}} 
   \caption{Mass function slopes for all (individual) stars with mass ${\geq}2\,\msun$.
    Red filled circles are the present-day mass functions for the ejected stars ($\aej$). 
    Blue stars are the averaged present-day mass function slopes of the stars 
    that remain in the clusters at 3\,Myr ($\ain$), while black open
    triangles are those at 0\,Myr (i.e. IMF, $\ai$). The canonical value of the 
    upper IMF, $\alpha_{2}=2.3$ (Eq.~(\ref{eq:imf})), is indicated with a dotted grey line.}  \label{ejfig:mfall}
\end{figure}

The ejection fraction drops with decreasing stellar mass, as shown in Sect.~\ref{ejsec:fej} and in \citet{BKO12}. 
It is therefore expected that the present-day mass function (MF) of the ejected stars 
is top-heavy compared to the IMF.
In this section we investigate the slope of the present-day mass function in the high-mass regime 
($m\geq2\,\msun$)\footnote{We chose the lower mass limit of $2\,\msun$ for MF following \citet{Weisz15} 
to compare our models to their results.}
of the ejected stars. For comparison, the mass function of stars that remain in the cluster is also presented. 
We counted all individual stars with $m\geq\,2\msun$, that is, for a multiple system 
each component was counted separately, and only stars more massive than $2\,\msun$ were taken into account.
 In reality, a fraction of the observed systems are possibly unresolved multiple systems.   
These systems may result in an observed MF deviating from the true stellar MF. 
However, \citet{Weidner09}, who studied the effect of unresolved multiple systems on the mass function 
of massive stars, found the difference between a slope of the observed MF and of the MF of all stars 
to be small, at most 0.1. The MFs of all stars can therefore be comparable with the observed MF.
The slopes were obtained with the method of \citet{MU05}, in which bin sizes are varied for 
the same number of stars in each. 
We used ten bins for each fitting, and the number of stars in each bin was the same  
(5--370 stars per bin for the MFs of the ejected stars, ${\approx}20$ stars per bin for the IMF 
and the MFs of stars that remain in their host cluster).  For the ejected stars, we used the stars 
from all 100 runs for each model because of their small number per cluster,  especially for models 
that eject hardly any massive stars. We calculated the MF power-law index $\alpha$ and its uncertainty $\sigma(\alpha)$ with 
a nonlinear least-squares fit\footnote{We used \textsc{curve\_fit} in python (\textsc{scipy} package). 
\citet{MU05} used \textsc{curvefit} in the standard IDL distribution. Both \textsc{curve\_fit} and 
\textsc{curvefit} are nonlinear least-squares fitting routines based on a gradient-expansion algorithm \citep{NR86,Bevington92}. } 
\citep[for details see][]{MU05}. For the slope of stars that remain in clusters and  for the initial slope, 
we computed the values from the individual runs and then averaged them.

In Fig.~\ref{ejfig:mfall} we show the MF slopes for stars with a mass ${\ge}2\,\msun$  
that are ejected and for those that remain in the clusters at 3\,Myr.  
In addition, we present the fitted IMF slopes. For all models, the average values of the fitted IMF slope, $\ai$, 
very well reproduce the adopted canonical value of the IMF, $\alpha_{2}=2.3$, with uncertainties smaller than 0.1.
The mass functions of the ejected stars are top-heavy with $1.7 \lesssim \alpha \lesssim 1.8$ 
for the clusters that actively eject massive stars ($\fej\gtrsim0.2$, see Fig.~\ref{ejfig:fej}). 
As we showed in Fig.~\ref{ejfig:fej}, 
these clusters in general eject more massive stars more efficiently. It is therefore naturally expected that 
the mass function of the ejected stars is top-heavy. \citet{OKP15} showed that the mass function 
of ejected O stars is top-heavy compared to the canonical IMF.

For the clusters with a higher ejection fraction of massive stars, 
a higher efficiency in ejecting more massive stars causes the mass function of stars that remain
in the cluster to become steeper ($2.37<\ain<2.50$, Fig.~\ref{ejfig:mfall}), or top-light, at 3\,Myr. 
This is similar to the mass function slope $2.45$ found for stars with 
the same lower mass limit ($2\,\msun$) in young (${\lesssim}25$\,Myr) intermediate-mass 
($10^3$--$10^4\,\msun$) star clusters in M31 \citep{Weisz15}. 
It also confirms the findings by \citet{PK06} that the Orion nebular cluster (ONC) is deficient in massive stars.   
This may imply that these star clusters have ejected massive stars and their IMFs may not be deviating 
from the canonical IMF. 
Without significant (close to none) ejection of stars, however, the average slope of 
the present-day mass function of stars in the cluster becomes slightly shallower 
than the initial values because of stellar evolution 
(see most of the model clusters with $\rhi=0.8$\,pc in Fig.~\ref{ejfig:mfall}).

It should be noted that this tendency can be obtained only with many  clusters, 
not from an individual cluster. 
Too few stars are ejected from a single cluster to statistically derive 
the mass function of the ejected stars. 
Furthermore, it would be almost impossible to find all the stars that a cluster has ejected in observations without 
the precise kinematic knowledge of all stars in the field and all nearby star clusters \citep[cf.][on the ONC]{PK06}.  
It may therefore be difficult to determine whether the steeper mass function observed from 
a single cluster is due to evolutionary effects or stochastic effects in the IMF.

\begin{figure*}
\centering   
 \subfigure[system MF for systems with a primary mass ${\geq}2\,\msun$]{
     \includegraphics[width=8.9cm]{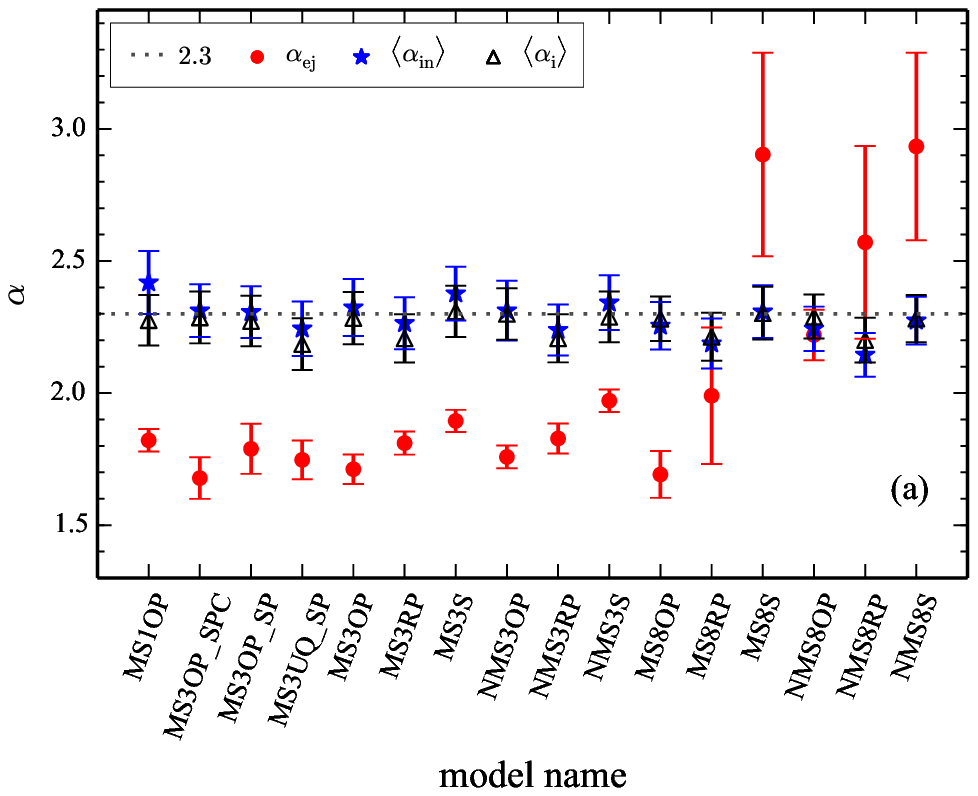}
     \label{ejfig:mfsysa}
     }  
  \subfigure[system MF for systems with a system mass ${\geq}2\,\msun$]{
     \includegraphics[width=8.9cm]{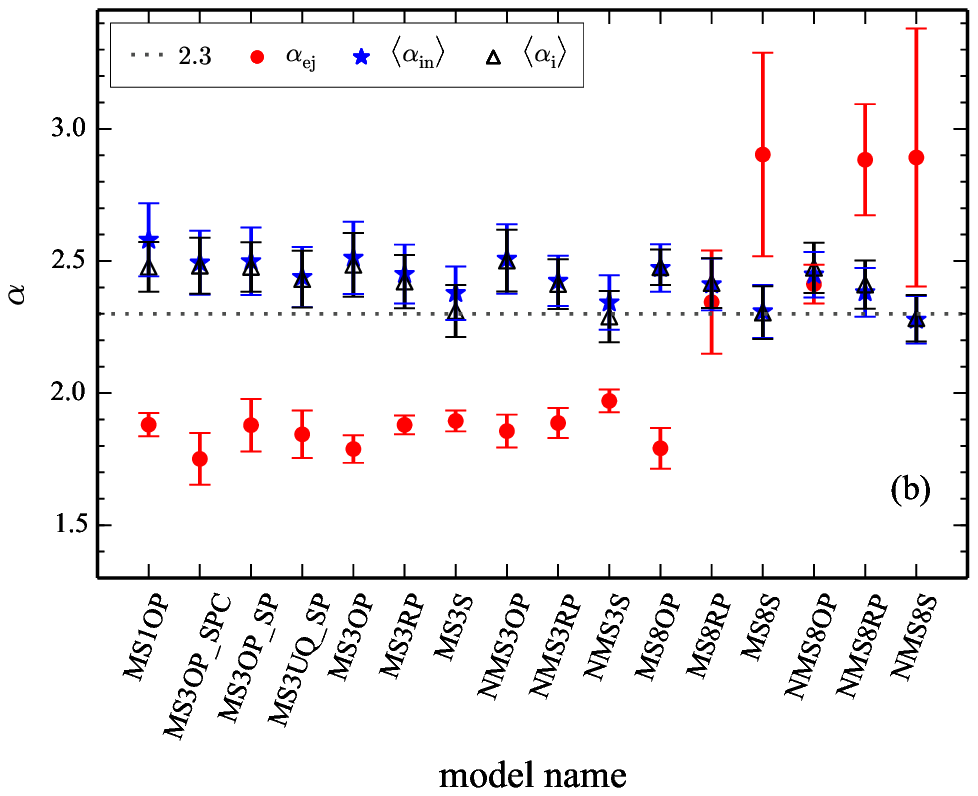}
     \label{ejfig:mfsysb}
     }       
    \caption{System mass function slopes for systems \textbf{a}) with a primary mass ${\geq}2\,\msun$  
    and \textbf{b}) with a system mass ${\geq}2\,\msun$. Symbols are the same as in Fig.~\ref{ejfig:mfall}.}\label{ejfig:mfsys}
\end{figure*} 

\begin{figure}
\resizebox{\hsize}{!}{\includegraphics{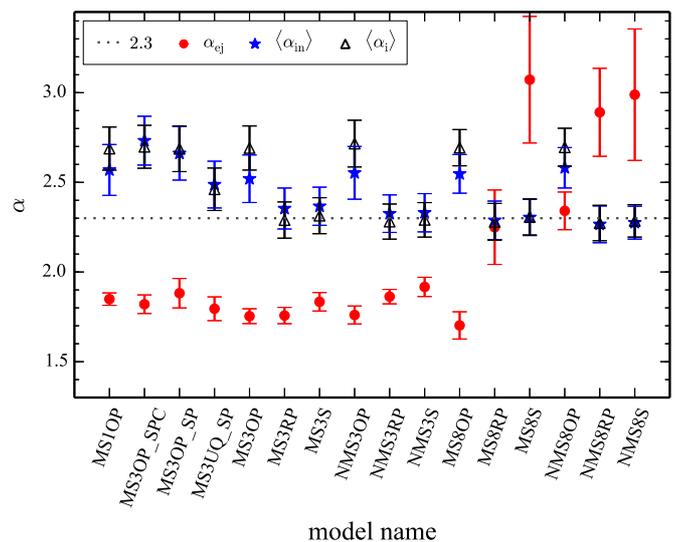}}
 \caption{Primary mass function slopes for systems with a primary mass ${\geq}2\,\msun$. 
 Symbols are the same as in Fig.~\ref{ejfig:mfall}.}    \label{ejfig:mfp}
\end{figure}

In addition to the MF of all stars, we derived the system MF for systems with a primary mass  
${\geq}2\,\msun$  and with a system mass  ${\geq}2\,\msun$ (Fig.~\ref{ejfig:mfsys}) 
as well as the primary MF for systems with a primary mass ${\geq}2\,\msun$ (Fig.~\ref{ejfig:mfp}) 
in the same way as described above to investigate the effect of multiplicity on these mass functions. 
For the ejected systems, the multiplicity does not significantly affect the system MFs and the primary MF. 
They are highly top-heavy for clusters that are efficient at ejecting massive systems and bottom-heavy 
for the clusters that eject hardly any massive stars, as shown in the individual stellar MF of the ejected stars in Fig.~\ref{ejfig:mfall}. 
This is due to their relatively low multiplicity fraction (Sect.~\ref{ejsec:multi}). 
However, the system MF and the primary MF for the systems that remain in the cluster, 
which have a higher multiplicity fraction than the ejected systems (Fig.~\ref{ejfig:fbin}), 
are different from the MF of all stars in some models. 
The system MFs show that $\ain$ becomes larger than the derived IMF $\ai$, a similar trend as in 
 the MF of individual stars, regardless of how we sample the systems,  
 that is, whether we use a primary or a system mass. 
However, the values of the MF slopes are different when different criteria are used to sample the systems.  
For systems with a primary mass ${\geq}2\,\msun$ (Fig.~\ref{ejfig:mfsysa}), 
the slopes of the derived IMF for system masses are slightly lower than the canonical IMF value of $2.3$ 
(especially for the RP models), but the difference is negligible, while for systems with a system mass 
${\geq}2\,\msun$ (Fig.~\ref{ejfig:mfsysb}),  $\ain$ and $\ai$ for binary-rich clusters are 
higher than the canonical IMF value \citep[e.g. $2.4\leq \ai \leq 2.5$, see also][]{Weidner09}.
On the other hand, $\ain$ of the primary MF becomes smaller than the derived IMF 
for most of the models in which massive stars are initially paired with another massive star (OP and UQ models).
These models also show that the derived IMF of primary masses is very different from the canonical value (Fig.~\ref{ejfig:mfp}).
This is mainly due to our choice for the lower limit of $2\,\msun$ when, in these models, 
two different pairing methods are used for systems with $m_{1}<5\,\msun$ and with $m_{1} \geq 5\,\msun$.
When $5\,\msun$ is chosen as the lower mass limit of the MFs, 
the primary MFs show indices similar to the MFs of all stars more massive than $5\,\msun$, 
while the system MFs are quite different.

This indicates that neither the primary MF nor the system MF is a good indicator for the MF of all stars 
because they can significantly deviate from the MF of all stars depending on the assumed parameters 
such as a criterion for sampling systems (see Fig.~\ref{ejfig:mfsys}), a lower mass limit, and an initial mass-ratio distribution of binaries. 
Fortunately, however, our MF for all stars can be used for a comparison with the observed MF 
even with the high proportion of unresolved multiples 
since the effect of unresolved multiple systems on the observed MF is small \citep{Weidner09}, 
as mentioned above. 
Again, the observed steeper MF for massive stars found in \citet{Weisz15} agrees well 
with our MF for all stars that remain in the cluster for models that are efficient at ejecting massive stars. 
This suggests that the origin of these observed steeper MF slopes for massive stars in young star clusters 
may be the preferential loss of massive stars by dynamical ejections, even though the IMF has the canonical slope 
(in agreement with the conclusion drawn by \citealt{PK06} on the IMF in the ONC).

\section{Dynamically ejected massive multiple systems}
\label{ejsec:multi}
It has been shown that binary systems involving massive stars, 
even very massive ones \citep[${\gtrsim}100\ \msun$,][]{OKB14}, can be ejected from a star cluster in theory 
 \citep[e.g.][]{FP11,BKO12,OKP15} and in observations \citep[e.g.][]{GB86,Sana13}, 
 and that some are even runaways.  
Here we show that higher order multiple systems (e.g. a system with $N \geq3$ components) can be dynamically ejected as well.  
We present multiplicity fractions among the ejected massive systems in the following subsections.  
We only discuss the initially mass-segregated cluster models with $\rhi=0.3$\,pc in this section. 
Unsegregated cluster models would show only little different values compared to those of the segregated models, 
and their trend of the binary fraction related with the other initial conditions is similar as for the mass-segregated models. 
The larger sized models eject only a few systems and are therefore inappropriate to study the multiplicity fraction 
of the ejected systems.

\subsection{Finding all multiple systems}
\label{ejsec:findmultiple}
Binary systems are found in a snapshot by searching a closest companion for each star and calculating whether 
the binding energy is negative. After this, the identified binaries are replaced by their centre-of-mass system 
and the search is repeated to find triple and quadruple systems. See also \citet{PK95a} for further details.

\subsection{Multiplicity fraction} 
\label{ejsec:hiarch}

\begin{figure}[tb]
    \resizebox{\hsize}{!}{\includegraphics{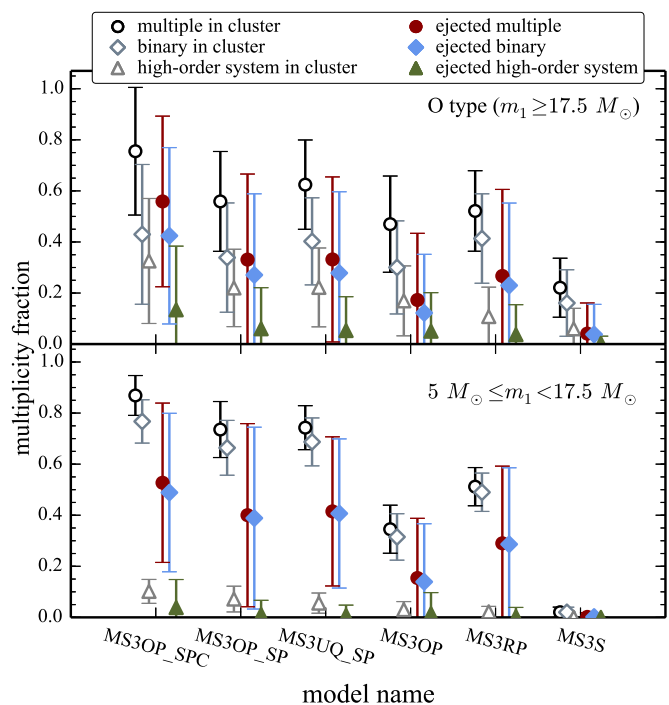}}
    \caption{Averaged multiplicity fractions of O-star systems (\textit{top}) and less massive systems (\textit{bottom}) 
    at 3\,Myr. Open and filled symbols are the values of the systems that remain in their host cluster and those of the
    ejected systems, respectively. 
    The error bars are standard deviations of the values from the individual clusters. \label{ejfig:fbin}}
\end{figure}

Considering that higher order multiple systems form during few-body interactions,
it is not surprising that some of them are ejected, although the incidence is very rare. 
The multiplicity fraction is defined as 
\begin{equation}
f_{\mathrm{multi}}= \frac{N_\mathrm{multi}}{N_\mathrm{sys}},
\end{equation}
where $N_\mathrm{multi}$ and $ N_\mathrm{sys}$ are the number of multiple (binary, triple, quadruple, etc.) systems and all systems, respectively.
The binary fraction ($f_{\mathrm{bin}}$) can be described as $f_{\mathrm{multi}}$ by using 
the number of binary systems ($N_\mathrm{bin}$) in place of $N_\mathrm{multi}$ 
(similarly for $f_{\mathrm{trip}}$, $f_{\mathrm{quad}}$, etc.).

In Fig.~\ref{ejfig:fbin} we show the multiplicity, binary, and higher order multiple fractions of massive systems 
 for systems that remain in their host cluster and for dynamically ejected systems.   
The multiplicity fractions vary with the initial conditions. 
However, all the models in the figure show that the multiplicity fractions of ejected massive systems are generally lower than those of systems that remain in a cluster. 
This is because the ejection efficiency is lower for binaries because  on average they have higher system masses than single stars. Binary or multiple systems are also usually decomposed during the ejecting encounter. 
For the ejected systems, the O-star systems have a higher high-order multiplie fraction than less massive systems. This is because the O-star systems undergo significant dynamical processing, including captures into binaries, in the cluster centre, and because they have a high binding mass. 

For O-star systems that remain in their host cluster, high-order multiple fractions are significantly high, even comparable to the binary fraction,
while most multiple systems of the ejected systems are binaries. 
The fraction of the ejected high-order multiple O-star systems is small, 
but their occurrence implies that the energetic interactions 
that eject O-star systems can be more complicated than just binary-binary scattering. 
We note that the companions of many high-order massive multiple 
systems are generally low-mass stars. 
Even though most of the massive star systems are centrally concentrated, many low-mass stars 
are also present in the central part of the cluster, which means that the probability 
of a massive system to capture a low-mass system is higher 
than to capture another massive system. 
This is also favourable energetically because capturing a low-mass star requires a smaller amount of 
energy that has to be absorbed.

For less massive multiple systems, binary systems dominate mostly (the lower panel of Fig.~\ref{ejfig:fbin}) compared to O-star systems. The high-order multiple fraction is significantly lower than the binary fraction 
because of the rare occurrences of their formation by dynamical interactions in the majority of clusters.  
Especially for the ejected systems, the multiplicity fraction is almost equal to their binary fraction for 
all models shown in Fig.~\ref{ejfig:fbin}. 

\begin{figure*}[bt]
\resizebox{\hsize}{!}{\includegraphics{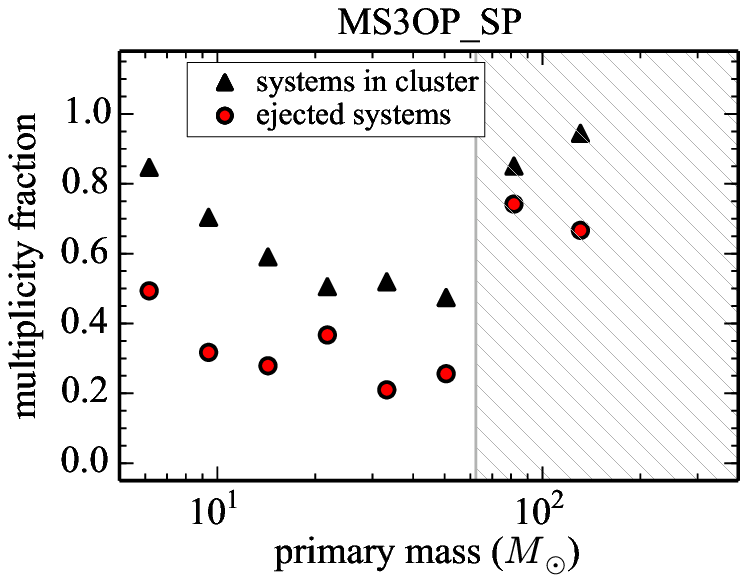}
                      \includegraphics{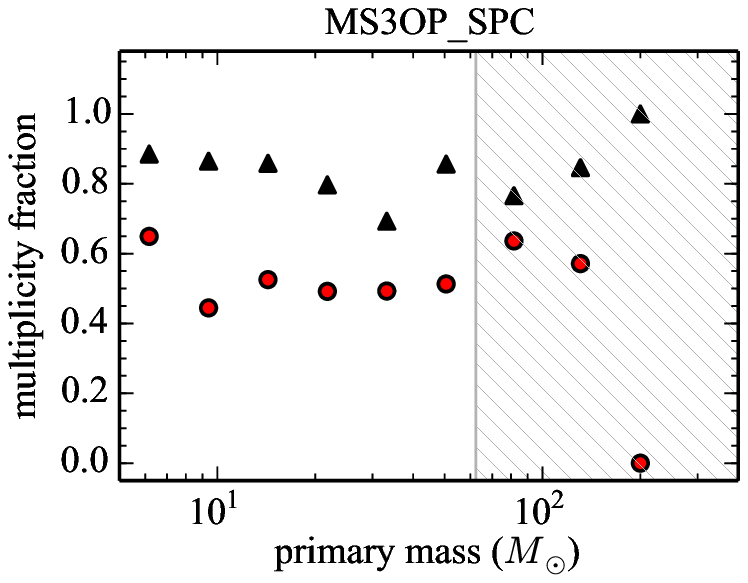}
                      \includegraphics{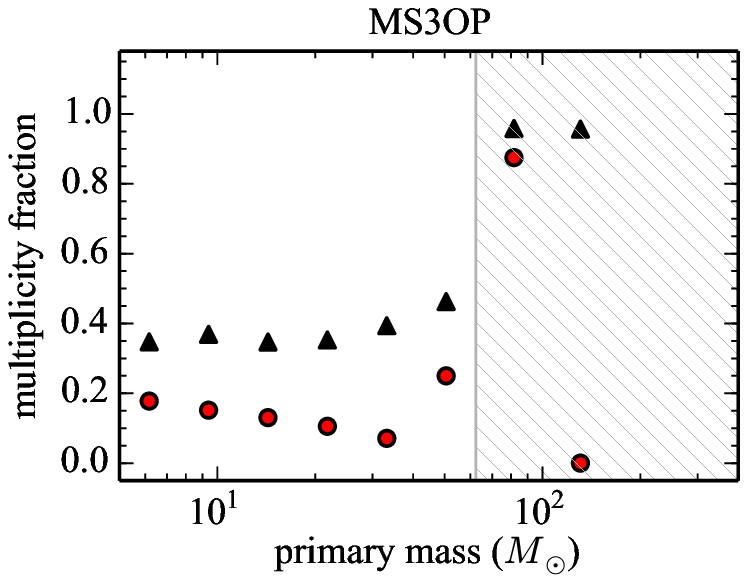}}
\caption{Averaged multiplicity fraction as a function of primary mass of systems for primary mass ${\geq}5\,\msun$ 
  from three models (MS3OP\_SP, MS3OP\_SPC, and MS3OP). 
  Black triangles and red circles are the multiplicity fractions of systems that remain in the cluster and 
  of the dynamically ejected systems, respectively. Model names are indicated at the top of subfigures.
  Grey hatched area indicates where a primary mass is higher than $62.4\,\msun$, that is, the evolved mass 
  of a $\mmax$ star ($79.2\,\msun$ at $t=0$\,Myr) at 3\,Myr, evolved by single stellar evolution. 
 Primary stars of all systems inside the area are therefore stellar mergers. 
 Technically, these stars are blue stragglers that lie above the main sequence \citep[cf.][]{Dalessandro13}.\label{ejfig:fmufm}}
\end{figure*}

The ejected less massive systems generally show a slightly higher multiplicity fraction than O-star systems. 
In Fig.~\ref{ejfig:fmufm} the multiplicity fraction of the ejected systems decreases with increasing primary mass 
in all three models, although the trend varies with the model for the systems that remain in the cluster. 
We note that the multiplicity fractions of systems at the high-mass end (grey shaded area in Fig.~\ref{ejfig:fmufm}) 
are high,  particularly for the systems that remain in their host cluster. 
These systems contain primaries that are merger products and do not follow the general trends shown in lower mass counterparts. 
Because they are the most massive systems in the clusters, 
the systems are generally situated close to the centre of their cluster, 
where the probability of capturing other cluster members is high. 
An observer would see these stars as blue stragglers (because of their merger nature), 
but above the main sequence because of their multiplicity nature \citep[cf.][]{Dalessandro13}.

The differences in the multiplicity fractions for massive systems that remain in their host cluster 
 indicate that the main mechanism for the removal of natal massive binary systems depends on the initial binary population. 
For the SP models, the multiplicity fraction is higher for less massive ($5\leq m_{1}/\msun < 17.5$) systems 
than for O-star systems. The reason is that in these models the main mechanism 
for reducing the binary fraction is merging of the two stars in the system, which has a higher probability 
to occur in a binary system with more massive (i.e. larger) components.  
This is more clearly shown in the left panel (MS3OP\_SP) of Fig.~\ref{ejfig:fmufm}, where the multiplicity fraction 
decreases with increasing primary mass, especially for systems that remain in their host cluster. 
Furthermore, the merger products naturally have a higher mass than the primary mass of the progenitor system, 
which leads to an increase of the number of single stars in the high-mass regime. 
Moreover, the significantly lower value of the multiplicity fraction in the MS3OP\_SP model compared 
to the MS3OP\_SPC model means that it is more likely for the short-period
binaries  with eccentric orbits to merge than it is for those with circular orbits. 
Likewise, the slightly smaller multiplicity fraction of O-star systems in the MS3OP\_SP models 
than in the MS3UQ\_SP models is probably due to the larger components of the initial binary systems, 
which more frequently lead to mergers.  
The MS3OP\_SPC model in Fig.~\ref{ejfig:fmufm} shows generally higher multiplicity fractions 
than the other models and little dependency of the multiplicity fraction on primary mass. 
The effect of pairing more massive companions in the MS3OP\_SP model is less prominent among 
less massive ($5\leq m_{1}/\msun < 17.5$) systems. The multiplicity fractions of  
the MS3OP\_SP and MS3UQ\_SP models are almost the same in the lower panel of Fig.~\ref{ejfig:fbin}. 
 This again confirms that the merging of binaries is the main mechanism that removes 
 initial binary systems for the models with the \citet{Set12} period distribution. 

However, for the models with the Kroupa period distribution (Eq.~(\ref{eq:P_Kroupa})), 
the main mechanism of binary removal is the disruption of binary systems through dynamical interactions 
with other cluster member systems. 
The less massive systems are more vulnerable to a binary disruption because their 
 absolute binding energy is lower than that of the more massive systems. 
For these models, the multiplicity fraction of less massive systems is therefore lower
than those of O-star systems.  In the right panel (MS3OP) of Fig.~\ref{ejfig:fmufm} we show 
that the multiplicity fraction of systems that remain in their birth cluster (weakly) 
increases with increasing primary mass. 
The following mechanism naturally explains why the multiplicity fraction in the MS3OP model is lower 
than that in the MS3RP model (Fig.~\ref{ejfig:fbin}). 
While in the former model two single massive stars generally emerge when a massive binary system is ionised,  
only one single massive star appears in the latter.  The difference between the two models increases 
for less massive systems because they are more prone to be ionised by the dynamical interactions 
than O-star systems, which is a result of their lower binding energy.

\subsection{Binary populations}
\label{ejsec:binpop}
\begin{figure*}[tb]
 \centering
 \includegraphics[width=15cm]{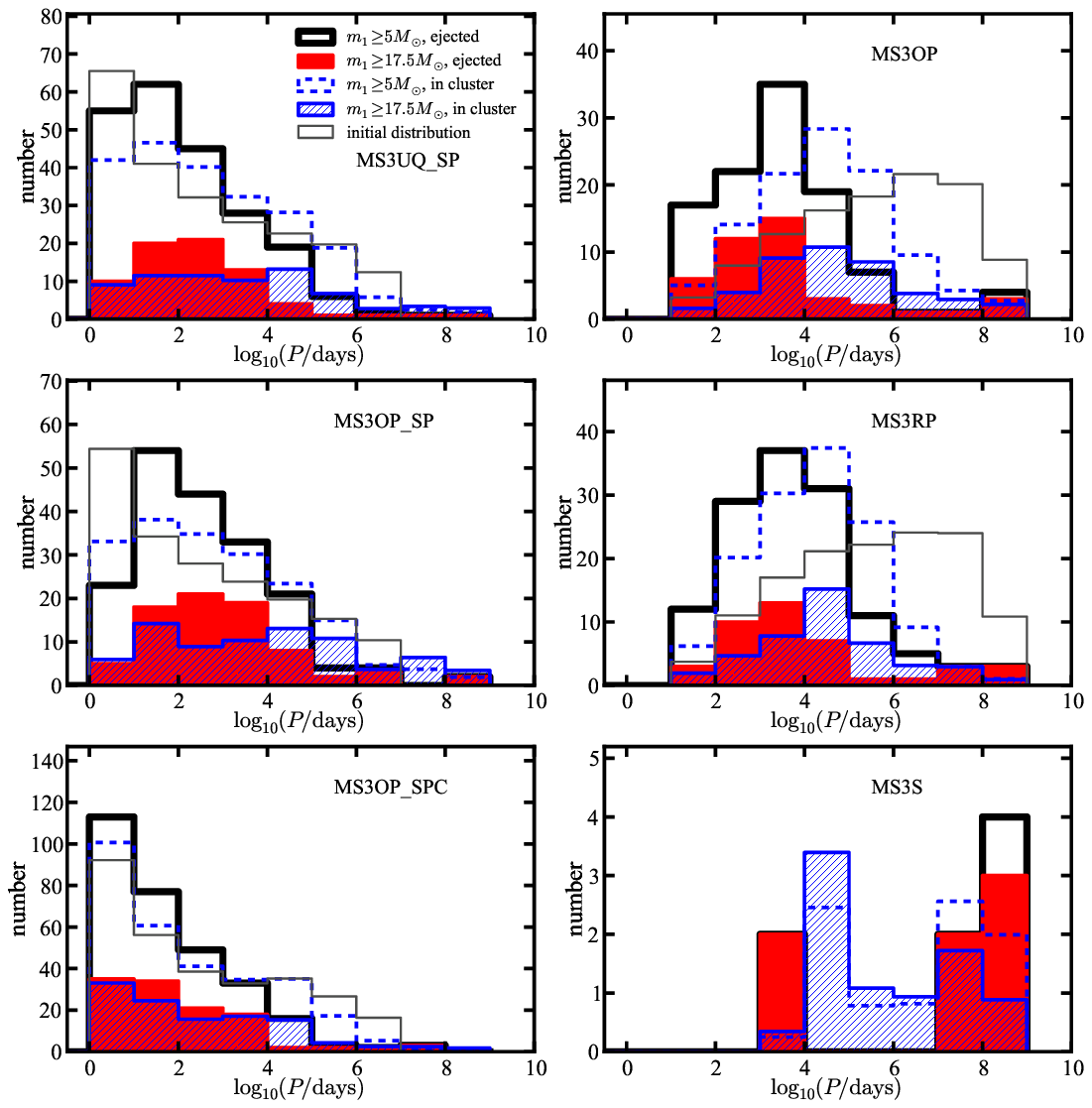}
    \caption{Period distribution  (number of binaries in a period bin) of the ejected massive  binary systems 
    for initially mass-segregated clusters with $\rhi=0.3$\,pc. Figures in the \textit{left column} are the clusters 
    with an initial period distribution according to \citet{Set12} for massive binaries (Eq.~(\ref{eq:P_Sana})), 
    while those in the \textit{right column} are the clusters with the \citet{PK95b} period distribution (Eq.~(\ref{eq:P_Kroupa})).
    The grey solid line indicates the initial distribution. Initial distribution and the distributions 
    of systems that remain in their birth cluster are scaled to the total number of the ejected binary systems. }
    \label{ejfig:binpd}
\end{figure*}

\begin{figure*}[htbp]
 \centering
 \includegraphics[width=15cm]{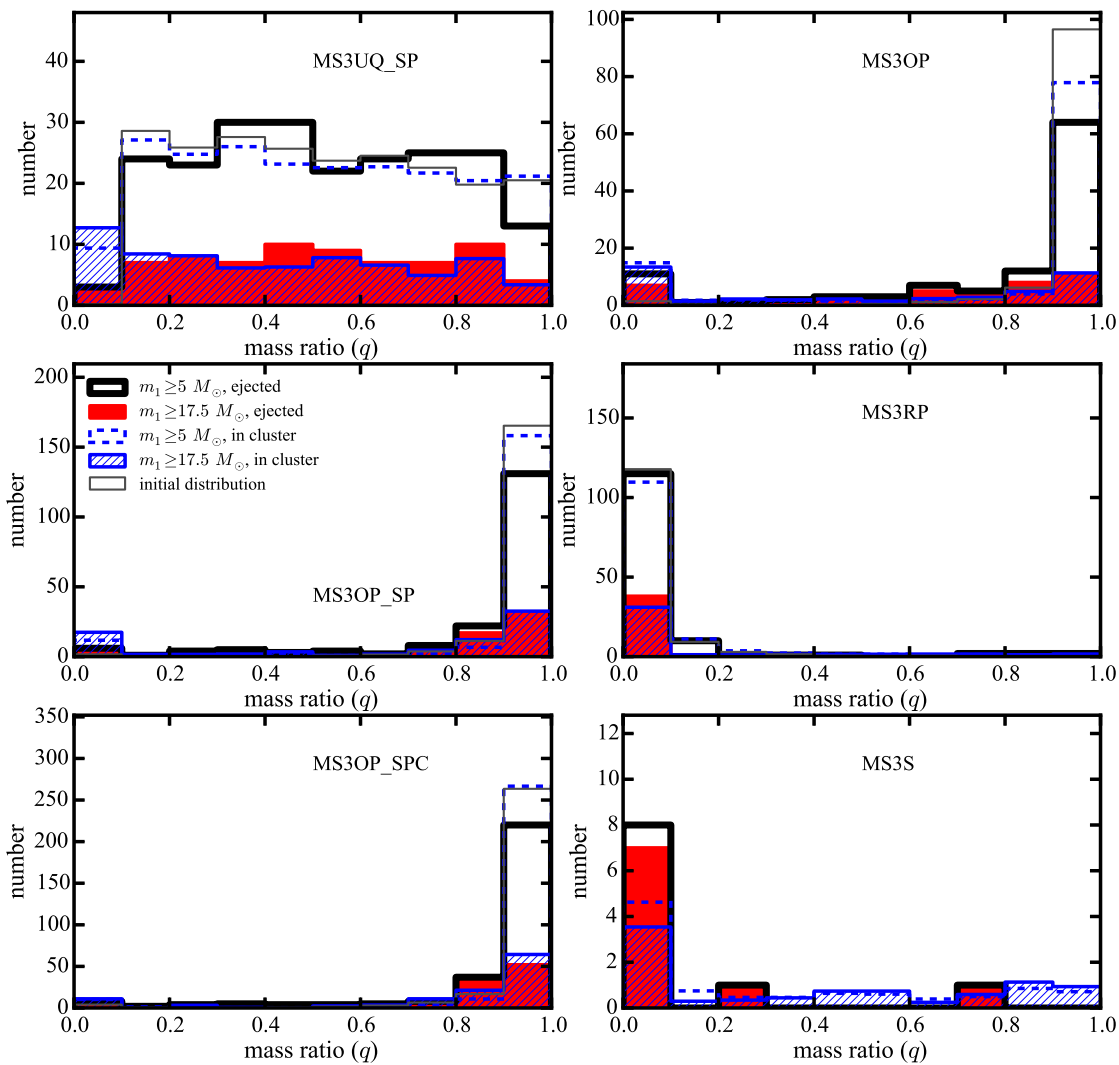}
 \caption{Same as Fig.~\ref{ejfig:binpd}, but for the mass-ratio distributions. }
 \label{ejfig:binmr}
\end{figure*}

\begin{figure*}[htbp]
 \centering
 \includegraphics[width=15cm]{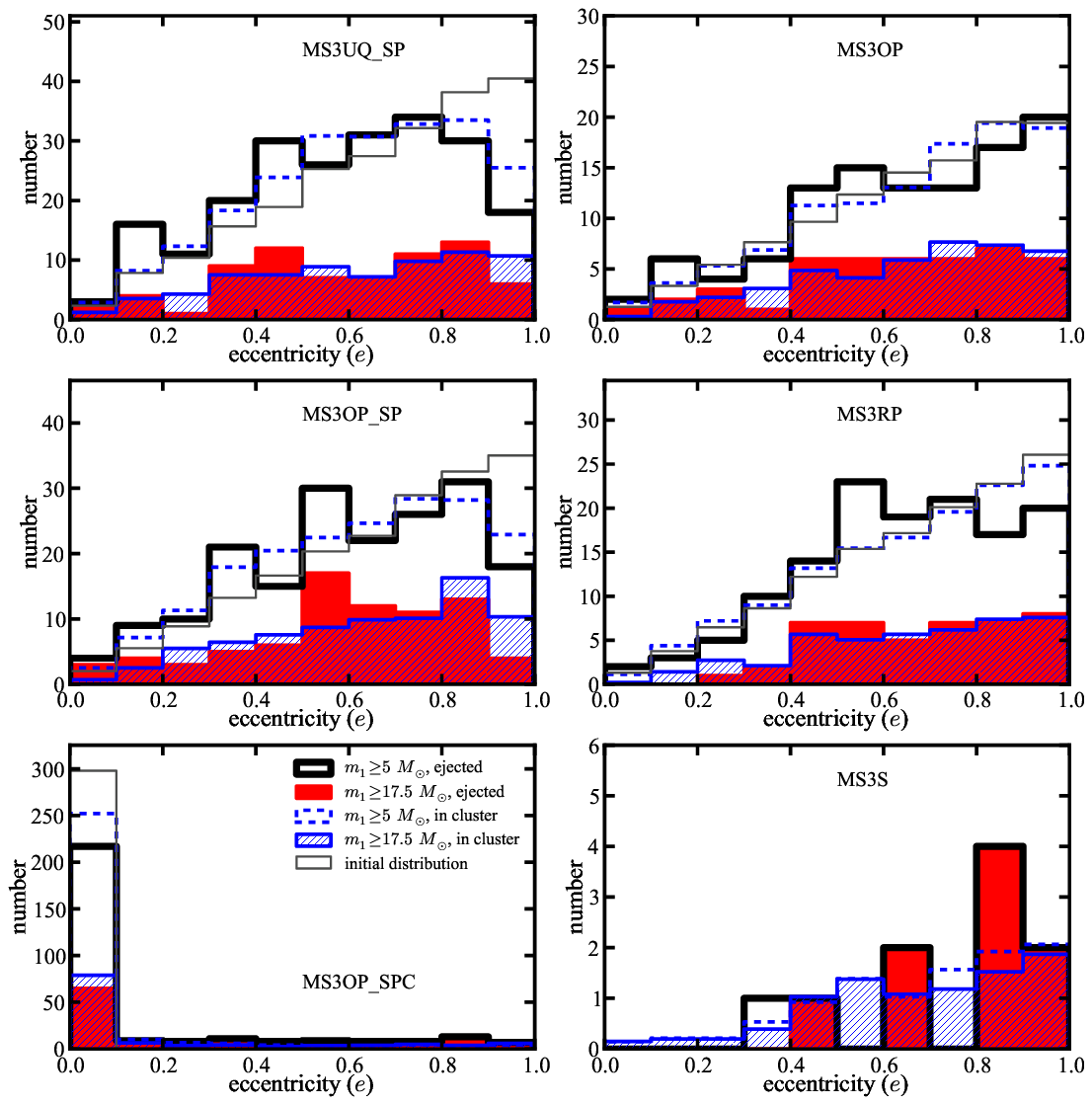}
 \caption{Same as Figs.~\ref{ejfig:binpd} and \ref{ejfig:binmr}, but for the eccentricity distributions. }
 \label{ejfig:binec}
\end{figure*}

A majority of the ejected massive (O-star) binaries (or of the inner binary systems in the case of high-order multiple systems) 
from the initially binary-rich models have a period shorter than $10^5$ ($10^4$)\,days, except in the MS3S models, 
in which all stars are initially single (Fig.~\ref{ejfig:binpd}). 
Compared to the period distribution of massive binaries that remain  
in a cluster (blue lines in the figures), the period distribution of the ejected binaries is more 
biased towards  shorter periods. 

With the thermal distribution for the initial eccentricity distribution,  the fraction of very short-period binaries 
($P/\mathrm{days} < 1$) is lower among ejected binaries than in the initial distribution in the models with 
the \citet{Set12} period distribution (MS3OP\_SP and MS3UQ\_SP). This is due to stellar collisions.
This naturally explains that the reduction in the number of very short-period massive binaries is not seen 
for the model with only circular binaries (MS3OP\_SPC).  

For the models with the Kroupa period distribution (Eq.~(\ref{eq:P_Kroupa}), the right-column panels in Fig.~\ref{ejfig:binpd}), 
the skewness to shorter periods for the period distribution of the ejected binaries is 
more prominent. The peak of the distribution, for example, appears at a shorter period 
than that of the systems in the clusters. 

The evolution of the period distribution of massive systems, particularly of those that remain in their host cluster, 
depends on the initial period distribution, which determines the dominant binary-removal mechanism as discussed in Sect.~\ref{ejsec:hiarch}. 
For the SP models, a fraction of short-period binaries is significantly smaller at 3 Myr than the initial distribution, 
which is a result of the merging of close binaries in a highly eccentric orbit. 
For the models with the Kroupa period distribution, in contrast, the binary fraction decreases for long periods, as a result of the binary disruption.

The mass-ratio distributions of the ejected systems are similar to those of the systems 
that remain in the clusters. But binaries with the lowest mass ratio ($q<0.1$) 
are slightly more deficient among the ejected binaries than in those that remain in clusters.
Unlike the other two orbital parameters discussed above, the mass-ratio distributions of massive binaries are not 
significantly altered from the initial distributions (Fig.~\ref{ejfig:binmr}), as is also the case for  
the lower mass counterparts \citep{MKO11}.

A fraction of binaries exchanged their partner, which led to a small change in the mass-ratio distribution. 
However, neither models with random pairing nor those with ordered pairing  
reproduce the observed mass-ratio distribution of massive binaries, which is an almost uniform distribution, 
as a result of the dynamical evolution. The uniform mass-ratio distribution, taking all primaries together, 
is thus the initial distribution for massive binaries ($m_{1}\geq 5\,\msun$).

All binary-rich models but one (MS3OP\_SPC) adopt the thermal distribution for the initial eccentricity 
distribution. The models that are initially biased towards short periods present a significant 
decrease of high-eccentricity ($e\geq0.9$) orbits in populations within clusters as well as in ejected populations. 
This is because stellar mergers result from close binaries with high eccentricity 
(i.e. the peri-centre distance is very small).  
There is no significant difference between the eccentricity distribution 
of the ejected systems and the systems that remain in the cluster (Fig.~\ref{ejfig:binec}). 
 
Above we briefly mentioned the dynamical evolution of massive binary populations. Although it is important 
to study how the dynamical evolution affects the binary populations
to understand the true initial binary population of massive systems formed in a cluster, 
given the observed (i.e. dynamically already evolved) distribution function,
this is beyond the scope of this study. 
A more detailed discussion on the dynamical evolution of the massive binary populations in clusters 
will appear in a future study (Oh et al. in preparation), where the true physical initial or birth distribution
functions of massive binaries will be constrained similarly to the procedure applied in \citet{PK95a,PK95b} for late-type stars.

\section{Discussion}
\label{ejsec:dis}

\begin{figure}
\resizebox{0.95\hsize}{!}{\includegraphics{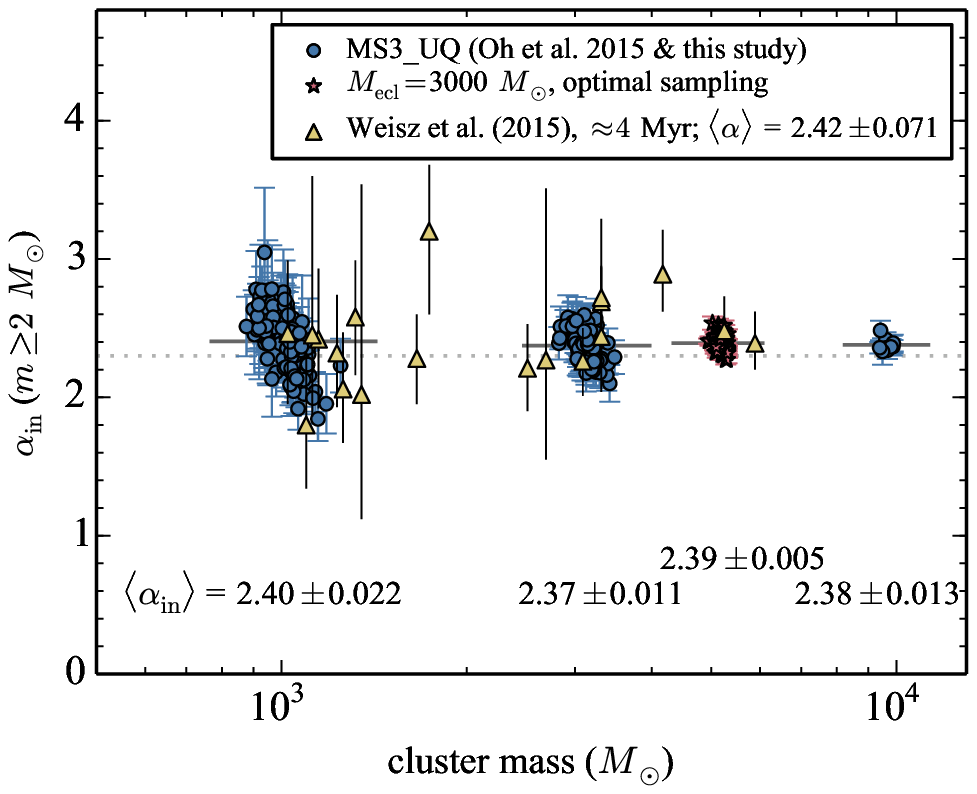}}
\caption{Present-day MFs of all stars ${\geq}2\,\msun$ for clusters with $\mcl=10^3$, $10^{3.5}$, 
and $10^{4}\,\msun$ at 3 Myr. Grey horizontal bars and numbers below the points are the average values 
for each \nbody\ model with different cluster mass. The uncertainties on the average $\ain$ values are 
standard deviations of the mean. The red star is the model with $\mcl=3000\,\msun$ 
and with initial conditions similar to the MS3UQ\_SP model, but stellar masses derived 
by optimal sampling (see text). For this model, cluster masses are shifted to higher masses by 0.25 
on the logarithmic scale to separate the model from the MS3UQ\_SP model. 
The triangles are clusters with an age of ${\approx}4$\,Myr in M31 from \citet{Weisz15}.}\label{ejfig:combimf}
\end{figure}

In Sect.~\ref{ejsec:mf} we showed that the average values of the present-day MF slopes
for all stars ${\geq}2\,\msun$ that  remain in the clusters that are efficient in ejecting massive stars 
($2.37<\ain<2.5$) are similar to the mean MF slope of young star clusters in M31 derived 
by \citet{Weisz15}, $\alpha=2.45$. 
The cluster masses in \citeauthor{Weisz15} span $10^3$--$10^4\,\msun$, while in this study we only considered  
one cluster mass, $\mcl=10^{3.5}\,\msun$, which is the middle value of their cluster mass range 
on the logarithmic scale. It is therefore necessary to check whether the agreement between 
our present-day MF and that of \citeauthor{Weisz15} only appears for the specific cluster mass 
we used in this study, 
or whether it is also there for different cluster masses that are similar to the cluster mass range of \citeauthor{Weisz15}. 
Here, we derive $\ain$ for two additional cluster masses, $\mcl=10^3$ and $10^{4}\,\msun$, 
for the most realistic model in our library, MS3UQ\_SP. We adopted the MS3UQ\_SP models 
with these two cluster masses from \citet{OKP15} and derived the present-day MF slope of all stars 
with a mass ${\geq}2\,\msun$ that remain in the cluster, $\alpin$, 
for each cluster with the same procedure as described in Sect.~\ref{ejsec:mf}. 
Additionally, we included one model similar to MS3UQ\_SP, but in which stellar masses are obtained using 
 the optimal sampling procedure in which the IMF has no Poissonian sampling scatter \citep{Ket13}.

The MF slopes are plotted as a function of cluster mass in Fig.~\ref{ejfig:combimf}. 
We note that in Fig.~\ref{ejfig:combimf} the data points of the optimal sampling model 
are shifted to a higher cluster mass (by adding 0.25 on the logarithmic scale to the cluster masses) 
to separate them from the MS3UQ\_SP models. 
The mean MF slopes of youngest clusters (only ages of ${\approx}4$\,Myr) in \citeauthor{Weisz15} is $2.42$, 
slightly smaller than that of their whole sample clusters, $2.45$.
For all three cluster masses of the MS3UQ\_SP model, the averaged present-day MFs at 3 Myr are $\ain\approx 2.4$, 
regardless of cluster mass, and this agrees very well with the mean MF of the \citeauthor{Weisz15} sample 
for ${\approx}4$\,Myr old clusters (see also the numbers in Fig.~\ref{ejfig:combimf}).
The averaged MF slope for the model with optimal sampling is $2.39$, very similar to the values of the other models. 
Our analysis implies that the canonical IMF evolves to a steeper present-day MF even in the first few Myr 
because of dynamical and stellar evolution, at least within the cluster mass range shown here, 
and that the IMF of young clusters in M31 should be the canonical IMF.
The steeper averaged MF in the observations would have resulted from the 
(dynamical and stellar) evolution of the clusters. 
However, we note that some additional degree of IMF steepening above a few solar masses could 
be present in the clusters in M31 because of a possible metallicity dependence of the IMF \citep[cf.][]{Markset12}.

The scatter of MFs of individual clusters is larger for clusters with lower mass (i.e. smaller $N$) 
as a result of the stochastic behaviour of random sampling.  
For the model with optimal sampling, the scatter of the MF slopes is smaller than that for the MS3UQ\_SP model 
with a similar cluster mass, and so are their uncertainties.
This is because there is no stochasticity in optimal sampling for the IMF since the optimal sampling produces 
only one set of stellar masses for a cluster mass. The spread of the present-day MFs then appears 
as a result of the slightly different evolution of the individual clusters.  
It should also be noted that the canonical IMF, $2.3$, the input IMF for our \nbody\ models, is reproduced 
for the averaged IMFs from these three additional models deduced with the procedure described 
in Sect.~\ref{ejsec:mf}, as is shown in all models with $\mcl=10^{3.5}\,\msun$. 
The similar steepening of the MFs at 3\,Myr for all cluster masses is not caused 
by any initial systematic biases that are due to different cluster mass.

In the previous sections, we described that the dynamical ejections of massive stars vary with 
the initial conditions of their birth cluster and its massive star population. 
To constrain the initial configurations of massive stars in a star cluster, our results
need to be compared with observations and should be able to rule out sets of initial conditions 
in the models that result in outcomes that are inconsistent with observations.  

Since individual $10^{3.5}\,\msun$ clusters eject only a small number of O-star systems, the number of ejected O-star 
systems can vary strongly from cluster to cluster. Some clusters, for instance, eject no O-star systems,   
while others loose 3--4 O-star systems. 
A cautionary remark is due when comparing our averaged results to an individual cluster. 
For the comparison, a large sample of clusters and of O stars in the field are needed. 
From the \nbody\ calculations we can find all stars ejected from a star cluster. 
In reality, a fraction of the ejected massive systems may not be traced back to their birth cluster 
and may not be recognised as ejected systems if they have travelled too far from their birth place 
or have experienced the two-step ejection process \citep{PK10}, which makes it almost impossible 
to trace the stars to their origin. 
Furthermore, O stars form in a wide range of cluster masses in a galaxy. The ejection fractions and their
properties also depend on the cluster mass \citep{OKP15}. 
Thus this study cannot represent the true observed field population or be compared to the general properties 
of the field O stars since here we only studied a single cluster mass and snapshots at 3\,Myr to make the analysis 
manageable and to pave the way towards even more comprehensive work. 
Ultimately, such work should include the cluster mass function in terms of a full-scale dynamical 
population synthesis approach (see \citealt{MK11} for more details and also \citealt{OKP15}).

It is expected that \emph{Gaia} will deliver vast data on the kinematics of stars in the Galaxy,  
including the O-star population. Comprehensive \nbody\ studies on the evolution of young star clusters 
formed in the Milky-Way-like galaxy incorporating the cluster mass function are thus required to interpret the data thoroughly.

\section{Summary}
\label{ejsec:sum} 
We investigated the effects of initial conditions of star clusters on the dynamical ejections 
of massive stars from moderately massive ($\mcl=10^{3.5}\,\msun$) young star clusters
by means of direct \nbody\ calculations with diverse initial conditions. 
The ejection fraction of massive systems is most sensitive to the initial density 
of the cluster (i.e. the initial size in this study because we used the same mass for all clusters) compared to
other properties. 
But the binary population is also an important factor for the ejection efficiency. The ejection fraction 
is higher, for example, when massive binaries are composed of massive components, and 
the properties of the ejected systems, such as their velocities and multiplicity,  
strongly depend on the properties of the initial binary population. 

The mass function of ejected stars is highly top-heavy because ejections with increasing stellar mass 
becomes more efficient. 
This tendency with stellar mass also alters the mass function of stars in a cluster 
so that it becomes steeper (top-light) 
than the IMF. The observed stellar mass functions in young (but more than a few Myr old) clusters may 
therefore deviate from their IMF \citep{PK06,BK12}. 
This may be evident in the MF slopes of young star clusters in M31 
for which the MFs are homogeneously derived with high-resolution observations by \citet{Weisz15}. The mean MF 
for the clusters is steeper than the canonical IMF, but is in excellent agreement 
with the average present-day (at 3\,Myr) MFs for our \nbody\ models with the most realistic initial conditions 
and with different cluster masses (Sects.~\ref{ejsec:mf} and \ref{ejsec:dis}).   
We stress that the IMFs of young clusters in M31 in \citet{Weisz15} therefore do appear to be indistinguishable 
from the canonical IMF and that the stellar MF of clusters can be altered even within the first few Myr of 
cluster evolution through dynamical ejections and collisions of massive stars, and through stellar evolution.

We showed that high-order multiple systems containing O stars can also be ejected readily.  
The ejected systems have a lower multiplicity fraction, especially high-order multiplicity, 
than those that remain in the cluster. 
The period distributions of the ejected binary systems are biased 
to shorter periods than those for systems that remain in the cluster. 
The mass ratio and eccentricity distributions of the ejected systems are similar 
to those of the systems that remain in the cluster. 
The evolved mass-ratio distribution of massive stars almost preserves the shape of the initial distribution.
We thus emphasize that the birth mass-ratio distribution of massive (particularly O star) binaries 
must be close to a uniform distribution as in observations of O-star binaries (\citealt{Set12,Ket14};
cf. B-star binaries, \citealt{ST02}).

The ejection fraction and the properties of ejected systems depend on the initial conditions 
of the clusters and their massive population. 
Applying our results to a large survey of the kinematics of massive stars in the Galactic field, 
as will be possible with \emph{Gaia}, for instance, will help to constrain the birth configuration of massive stars and of the star clusters 
in which they form. Further research using \nbody\ models is required to achieve a full-scale dynamical 
population synthesis to be applied to the \emph{Gaia} data.  

\begin{acknowledgements}
We sincerely thank Sverre Aarseth for freely distributing his \nbody\ codes and for his comments.  
We also thank Dan Weisz for kindly providing the data on MF slopes of young star clusters in M31 from \citet{Weisz15}.
SO was partially supported for this research through a stipend from the International
Max Planck Research School (IMPRS) for Astronomy and Astrophysics at the
Universities of Bonn and Cologne, and through stipends from the SPODYR group.
\end{acknowledgements}

\bibliographystyle{aa}
\bibliography{reference}

\appendix
\section{Additional figures}
\subsection{system mass versus velocity of the ejected systems for all models with $\rhi\leq 0.3$\,pc. }
\label{ejsec:apvel}
The velocity and system mass of ejected systems, same as Fig.~\ref{ejfig:mv}, are plotted in Fig.~\ref{aejfig:mv} 
for all models with $\rhi\leq 0.3$\,pc.

	\begin{figure*}[hbtp]
	 \centering
\resizebox{\hsize}{!}{\includegraphics{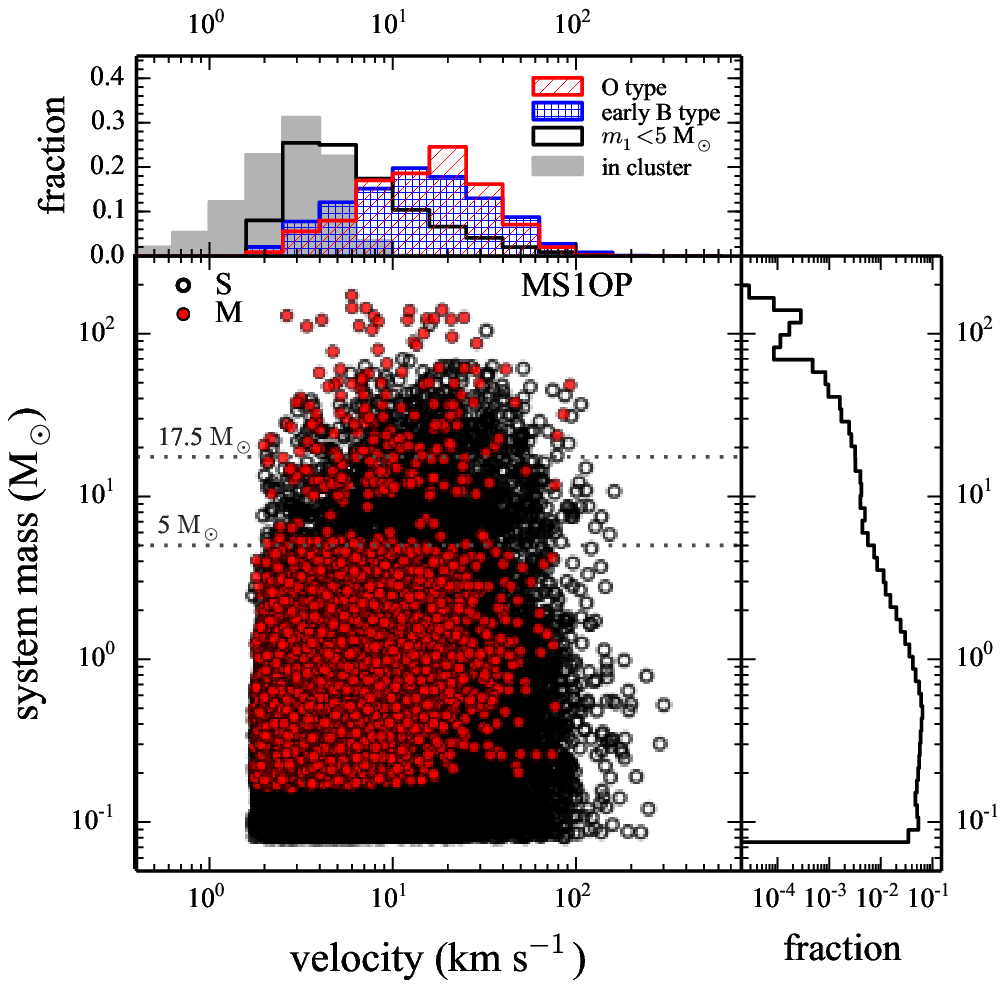}
	 \includegraphics{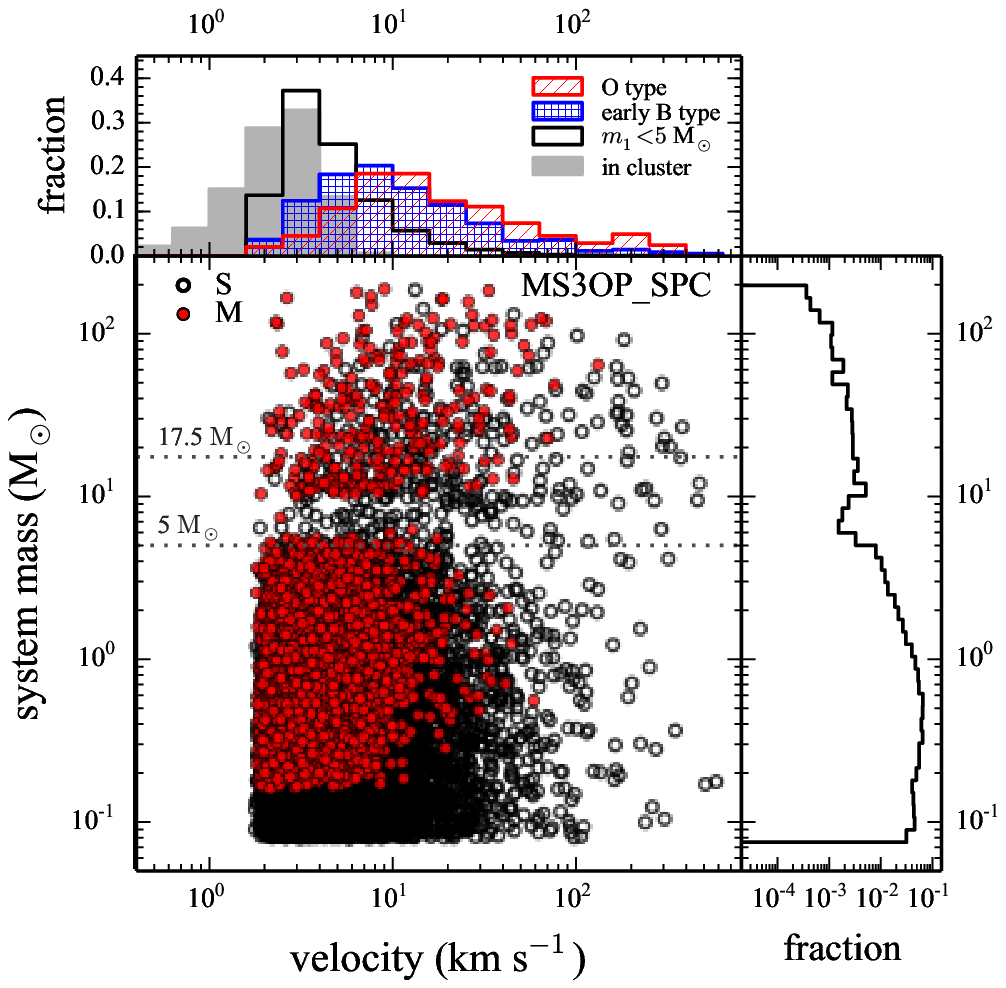}}\\
\resizebox{\hsize}{!}{	 \includegraphics{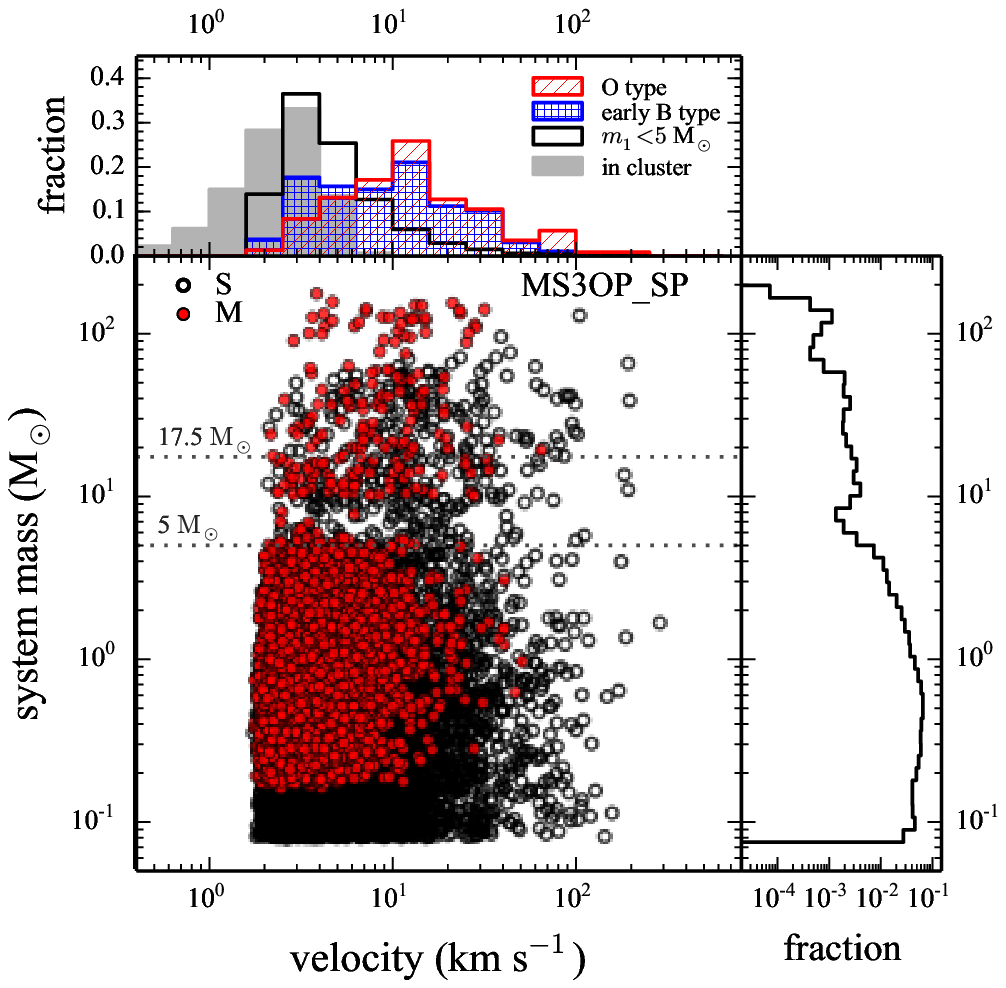}
	 \includegraphics{mvfig3rhMS3UQ_SP.eps}}
	 \caption{Same as Fig.~\ref{ejfig:mv}, but for all models with $\rhi\leq0.3$\,pc.}
	 \label{aejfig:mv}
	\end{figure*}

\setcounter{figure}{\value{figure}-1} 	
	\begin{figure*}[tbp]
	 \centering
\resizebox{\hsize}{!}{	 \includegraphics{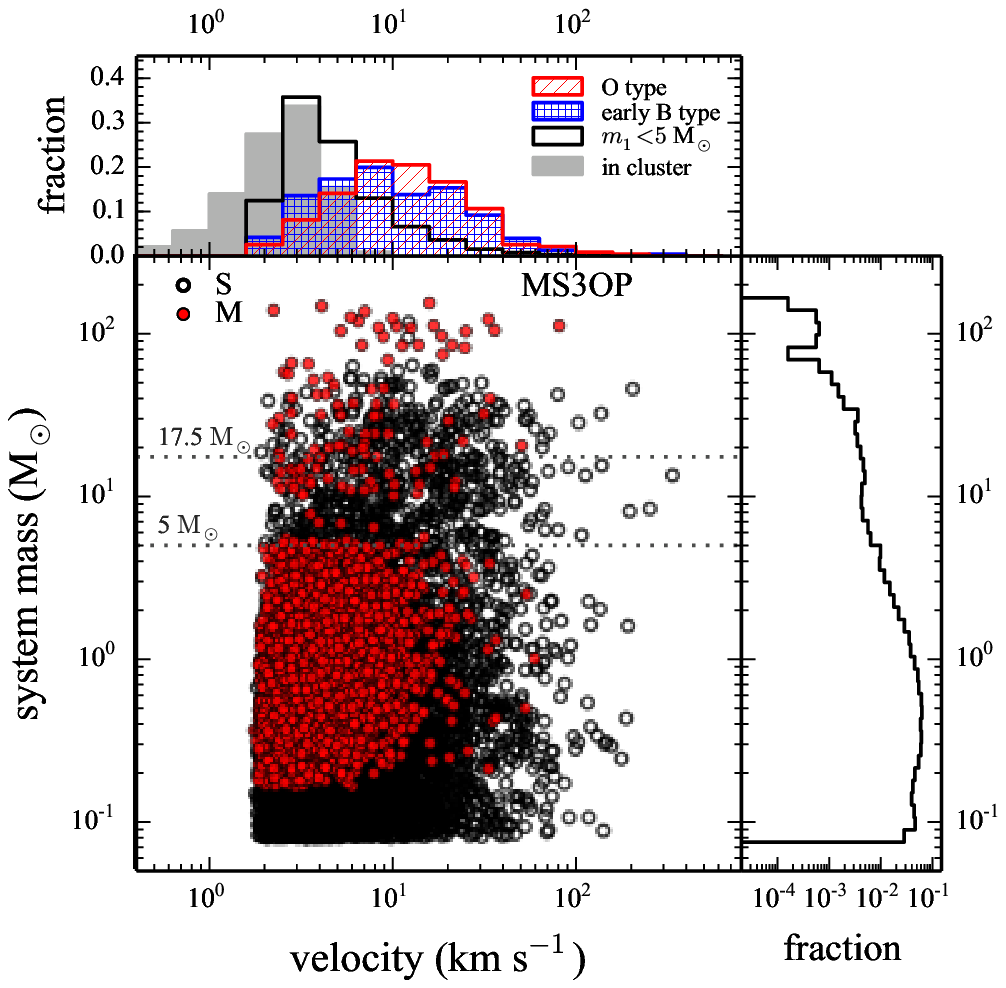}
	 \includegraphics{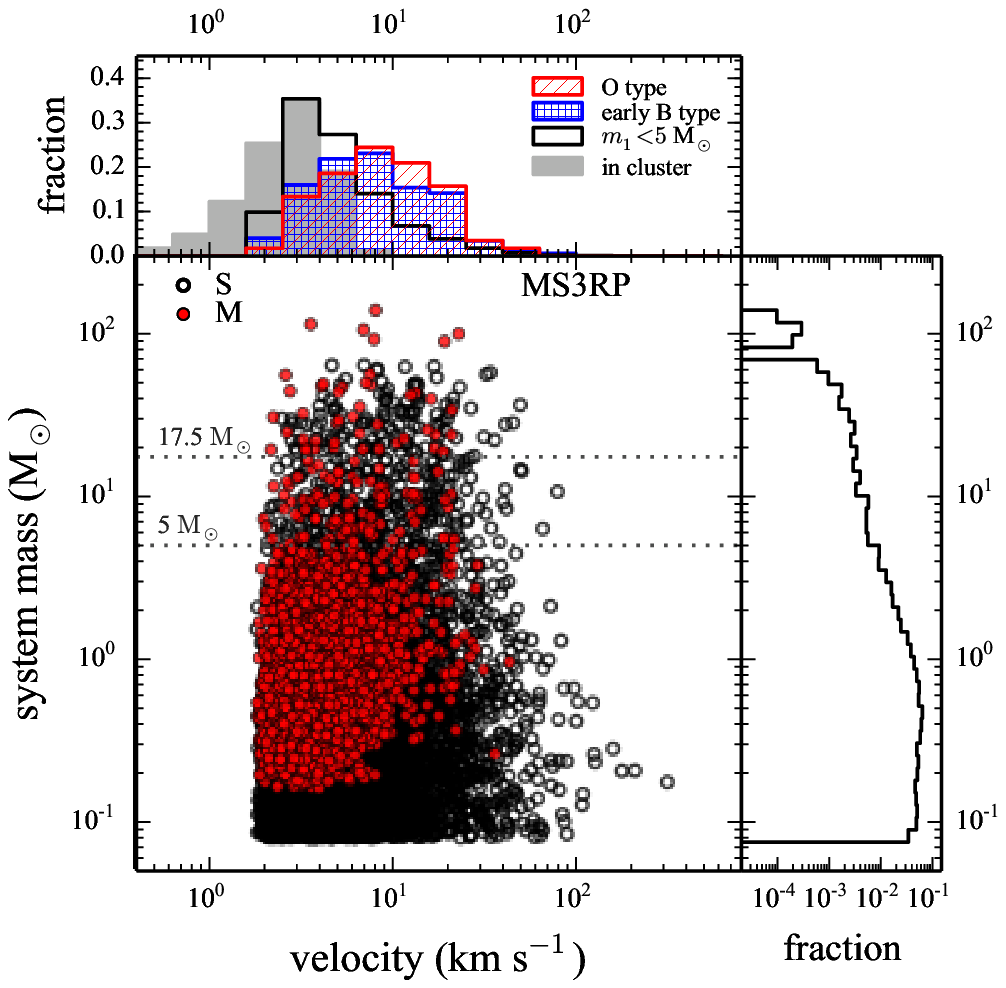}}
\resizebox{\hsize}{!}{	 \includegraphics{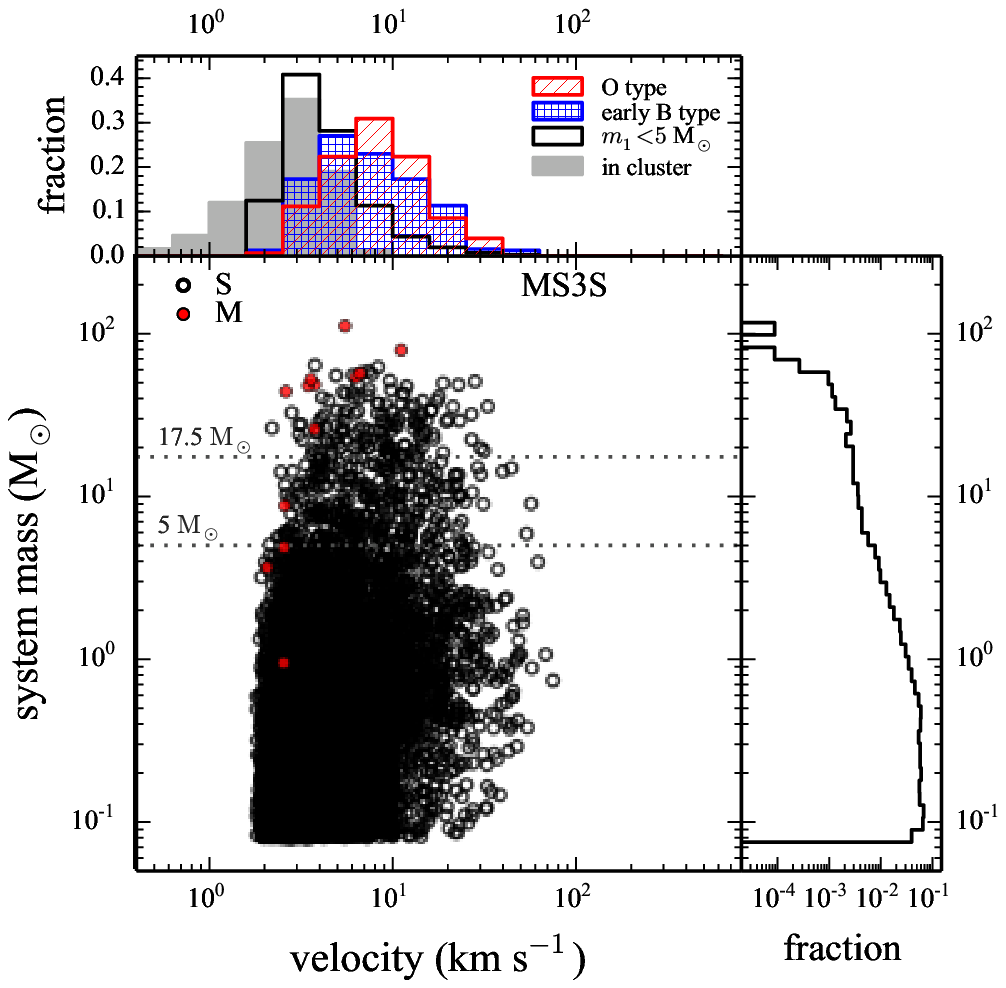}
	 \includegraphics{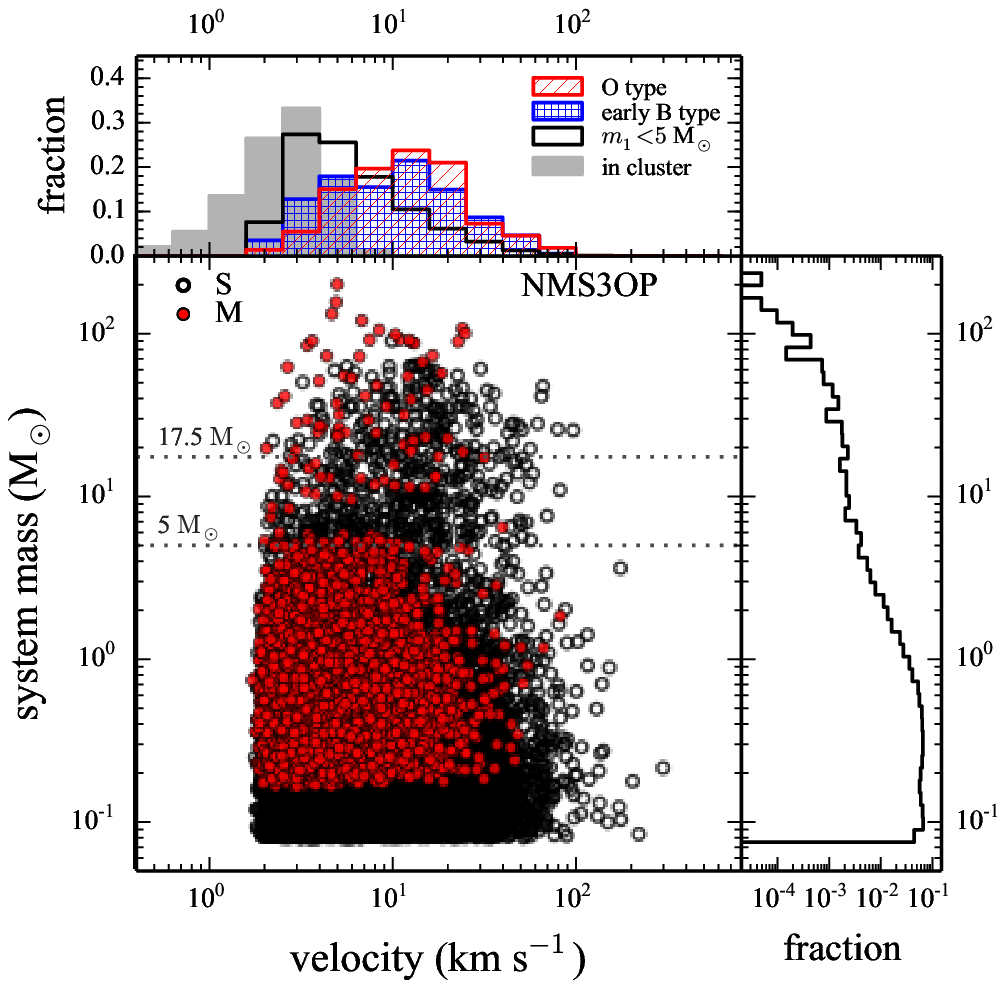}}
	 \caption{(Continued)}
	\end{figure*}

\setcounter{figure}{\value{figure}-1} 	
	\begin{figure*}[tbp] 
\resizebox{\hsize}{!}{	 \includegraphics{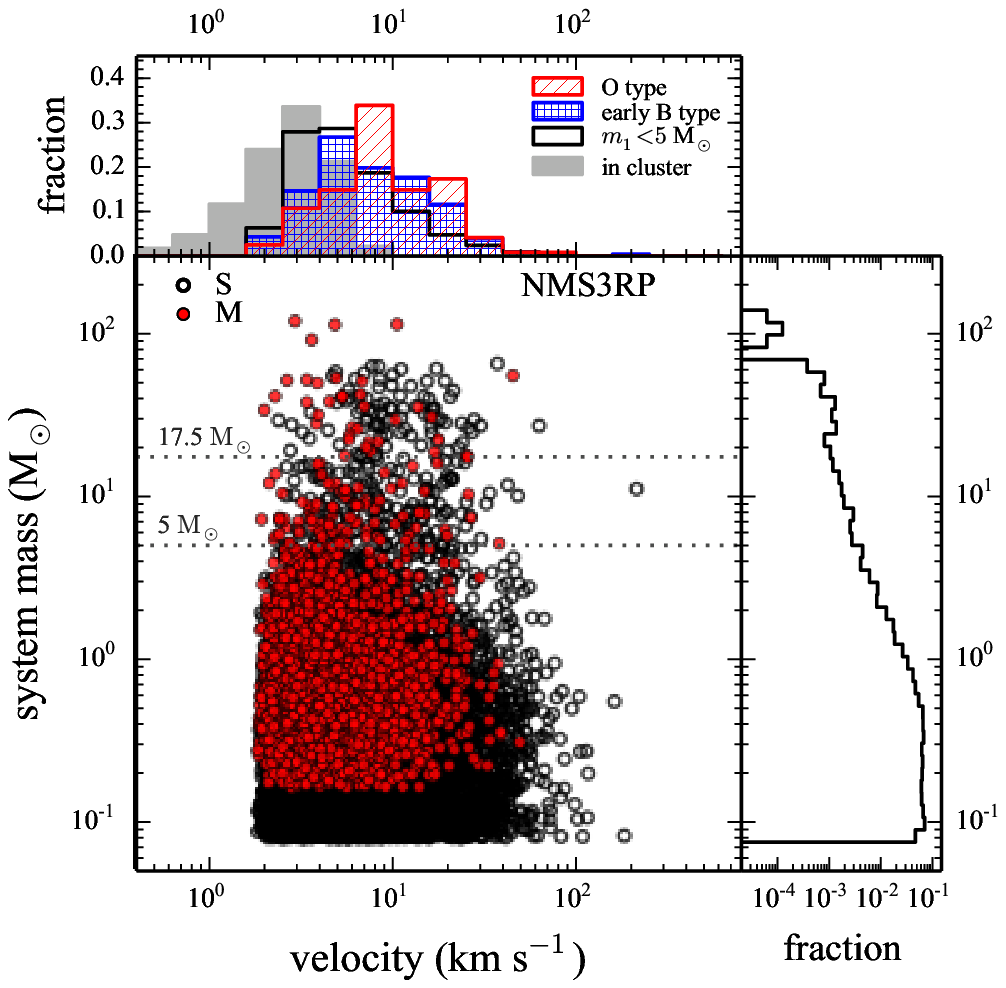}
	 \includegraphics{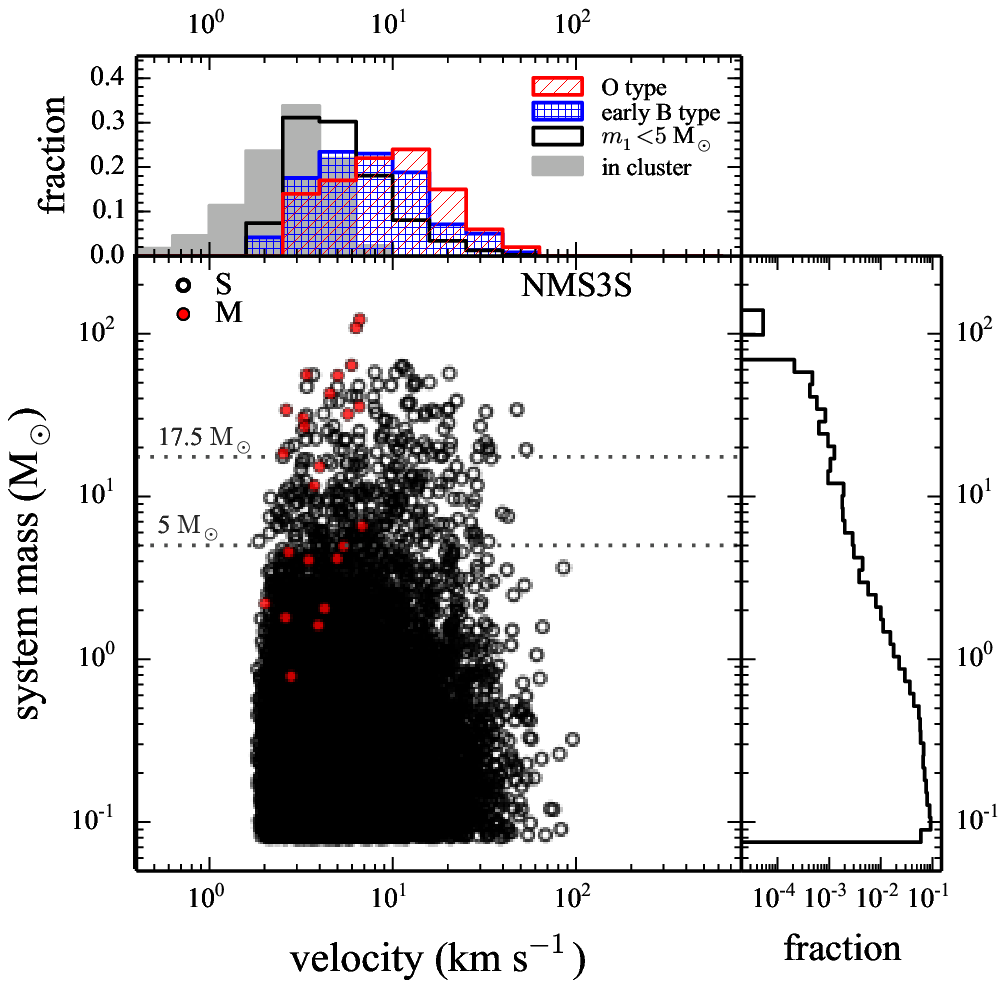}}
	 \caption{(Continued)}
	\end{figure*}

\subsection{$\tej$ and the core radius}\label{ejsec:aptejrc}
The distributions of $\tej$ and core radius as a function of time, same as Fig.~\ref{ejfig:tauej1}, 
are plotted in Fig.~\ref{aejfig:tej} for all models with $\rhi\leq 0.3$\,pc.

	\begin{figure*}[hbtp]
	 \centering
\resizebox{\hsize}{!}{	 \includegraphics{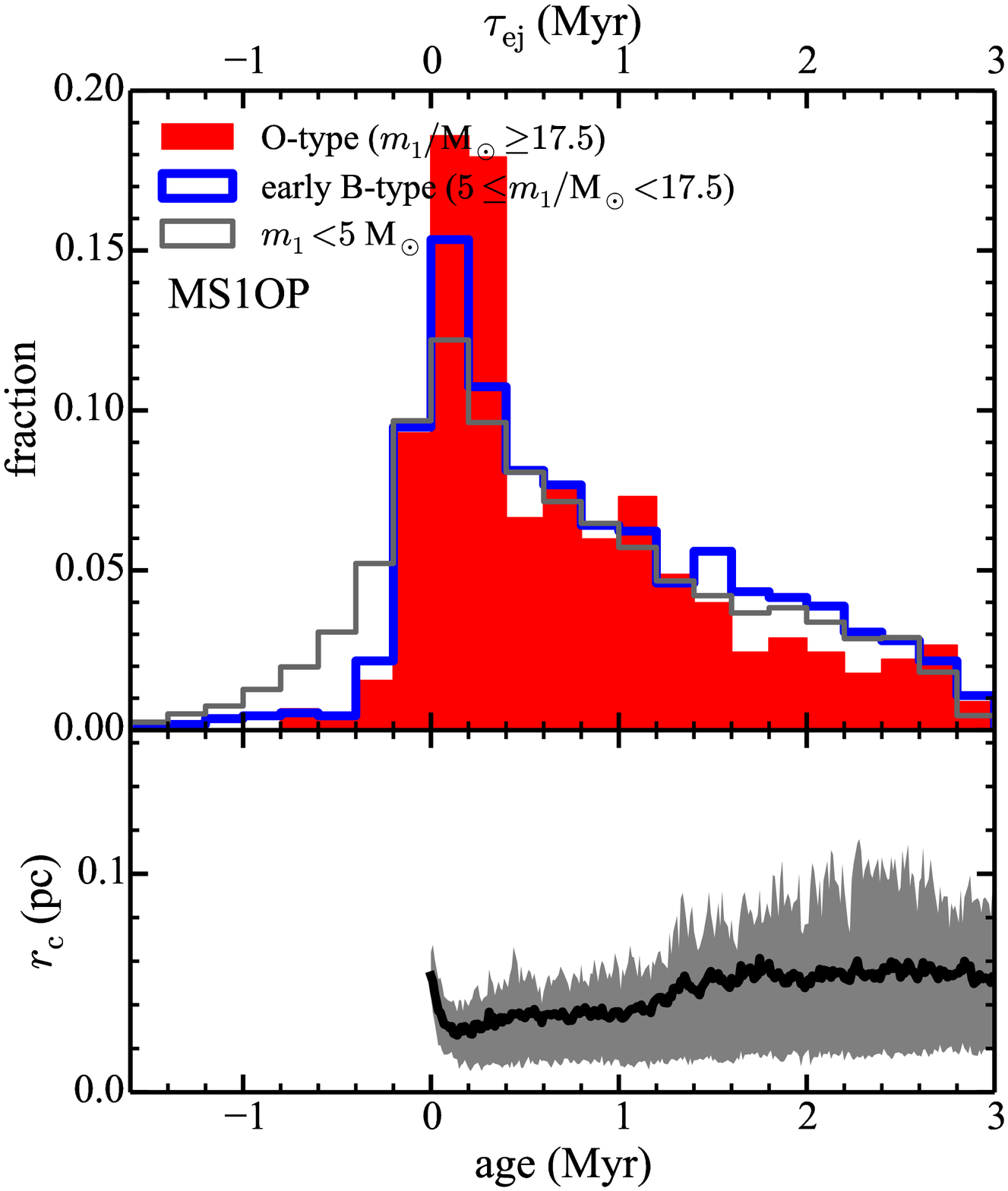}
	 \includegraphics{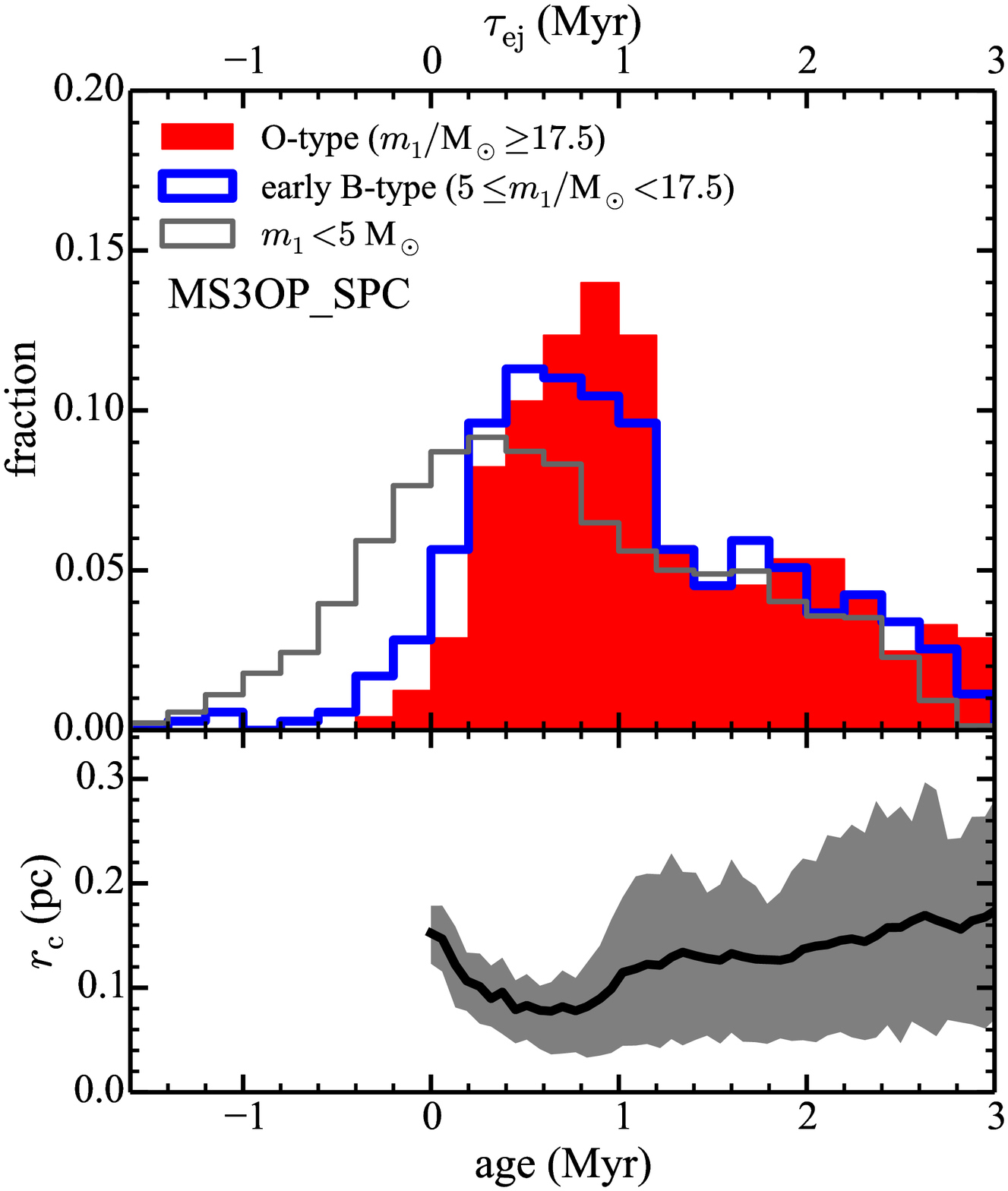}}\\
\resizebox{\hsize}{!}{	 \includegraphics{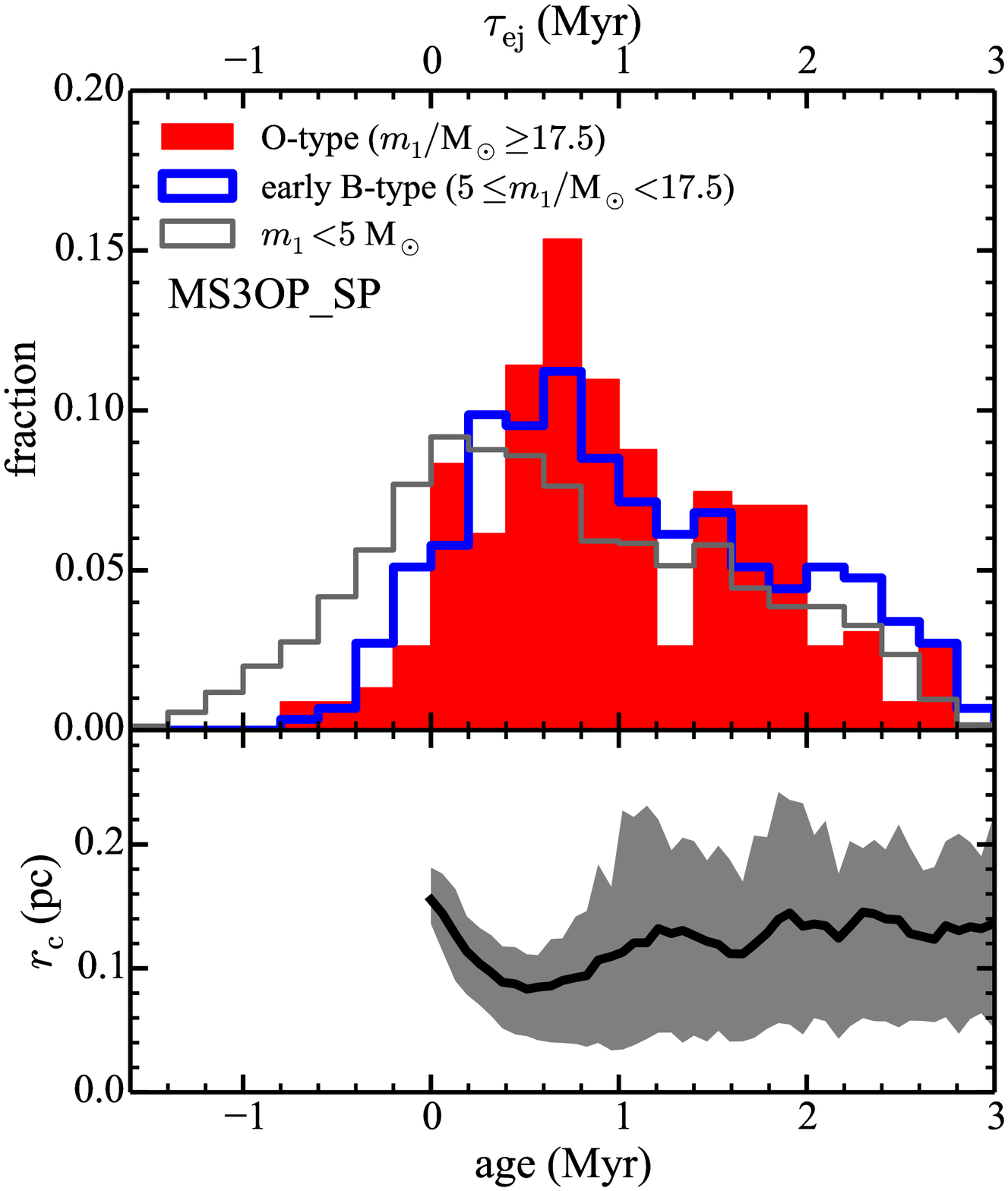}
	 \includegraphics{tauej_MS3UQ_SP.eps}}
	 \caption{Same as Fig.~\ref{ejfig:tauej1}, but for all models with $\rhi\leq0.3$\,pc.}
	 \label{aejfig:tej}
	\end{figure*}
   \setcounter{figure}{\value{figure}-1} 	
	\begin{figure*}[tbp]
	 \centering
\resizebox{\hsize}{!}{	 \includegraphics{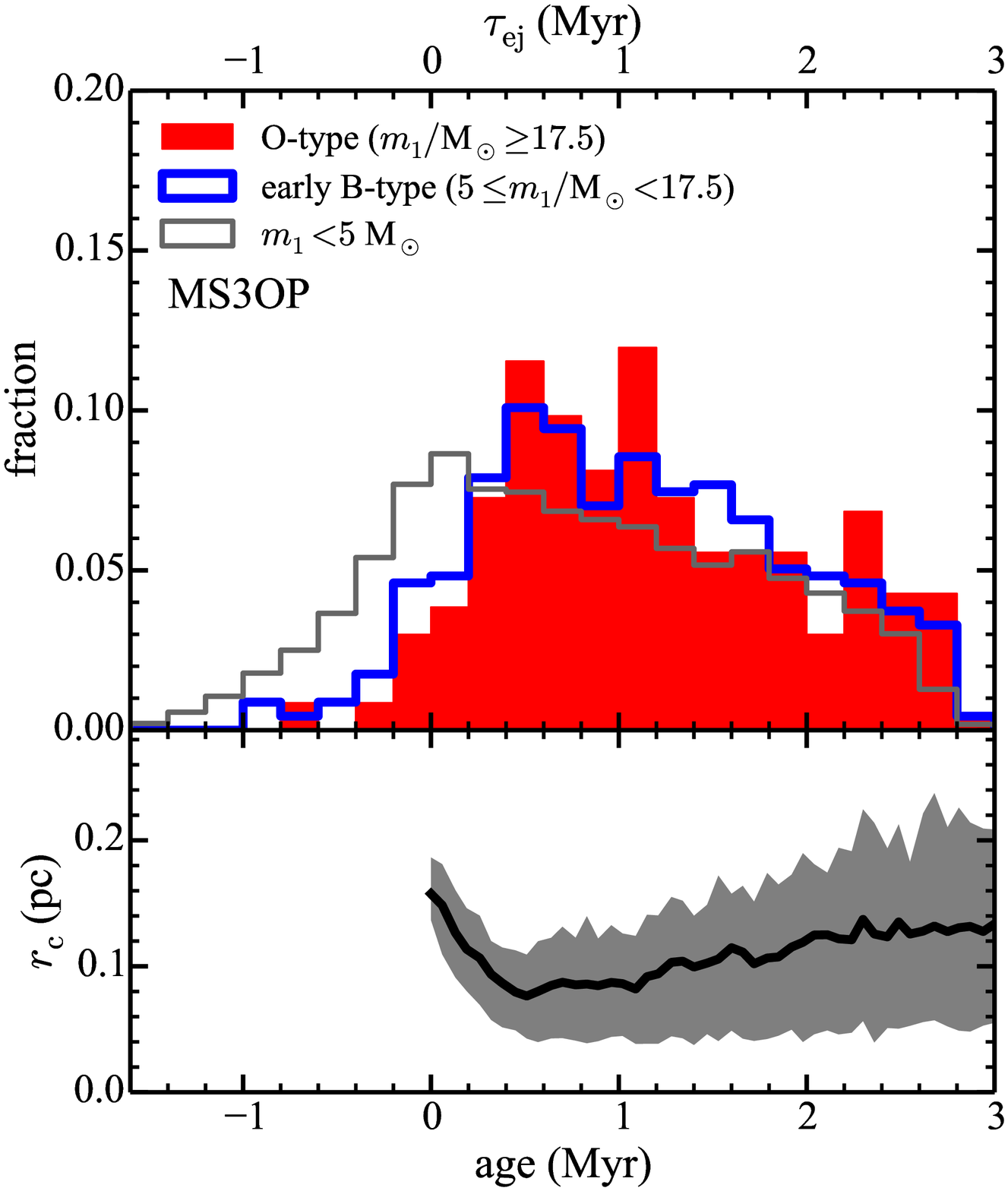}
	 \includegraphics{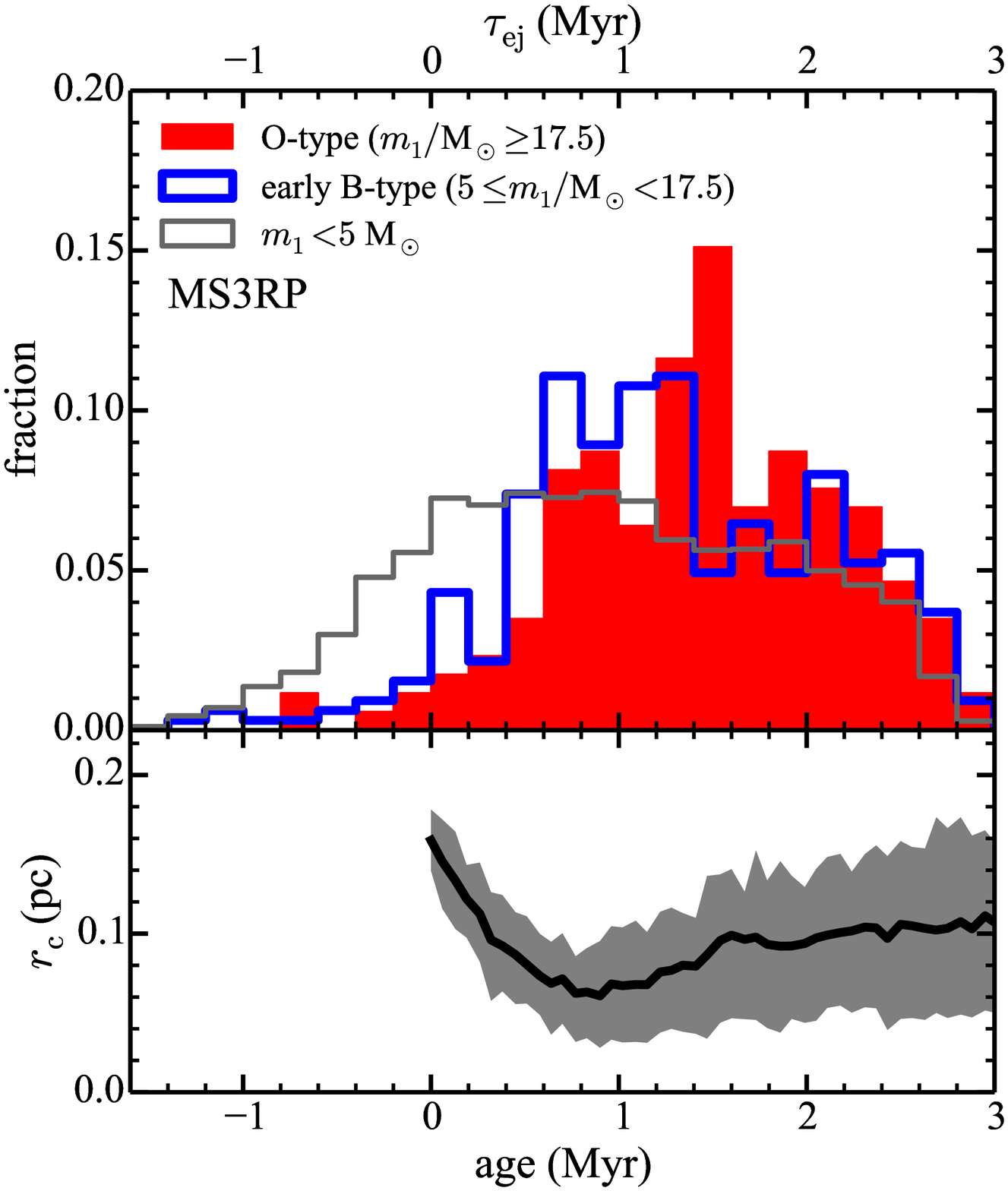}}\\
\resizebox{\hsize}{!}{	 \includegraphics{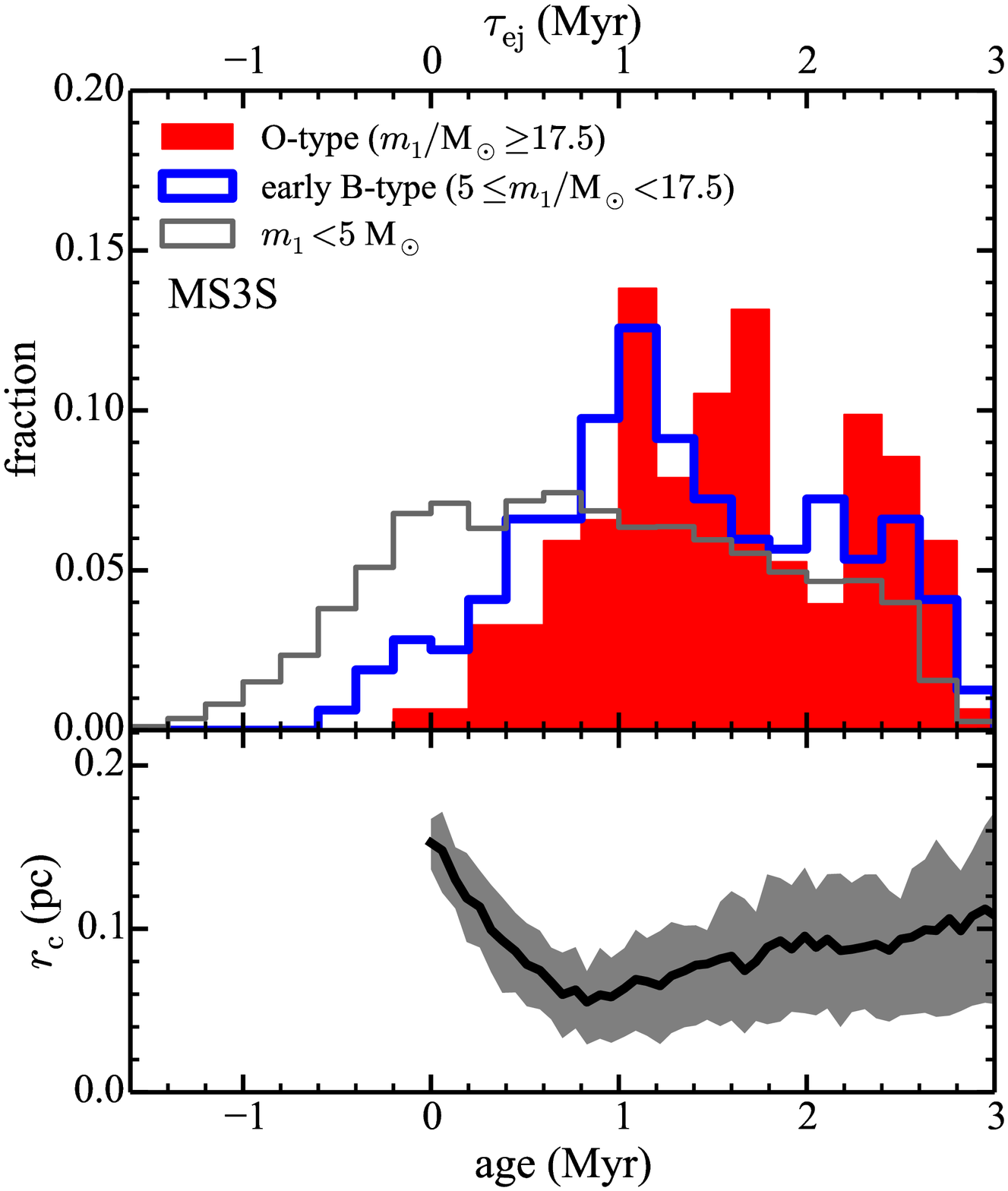}
	 \includegraphics{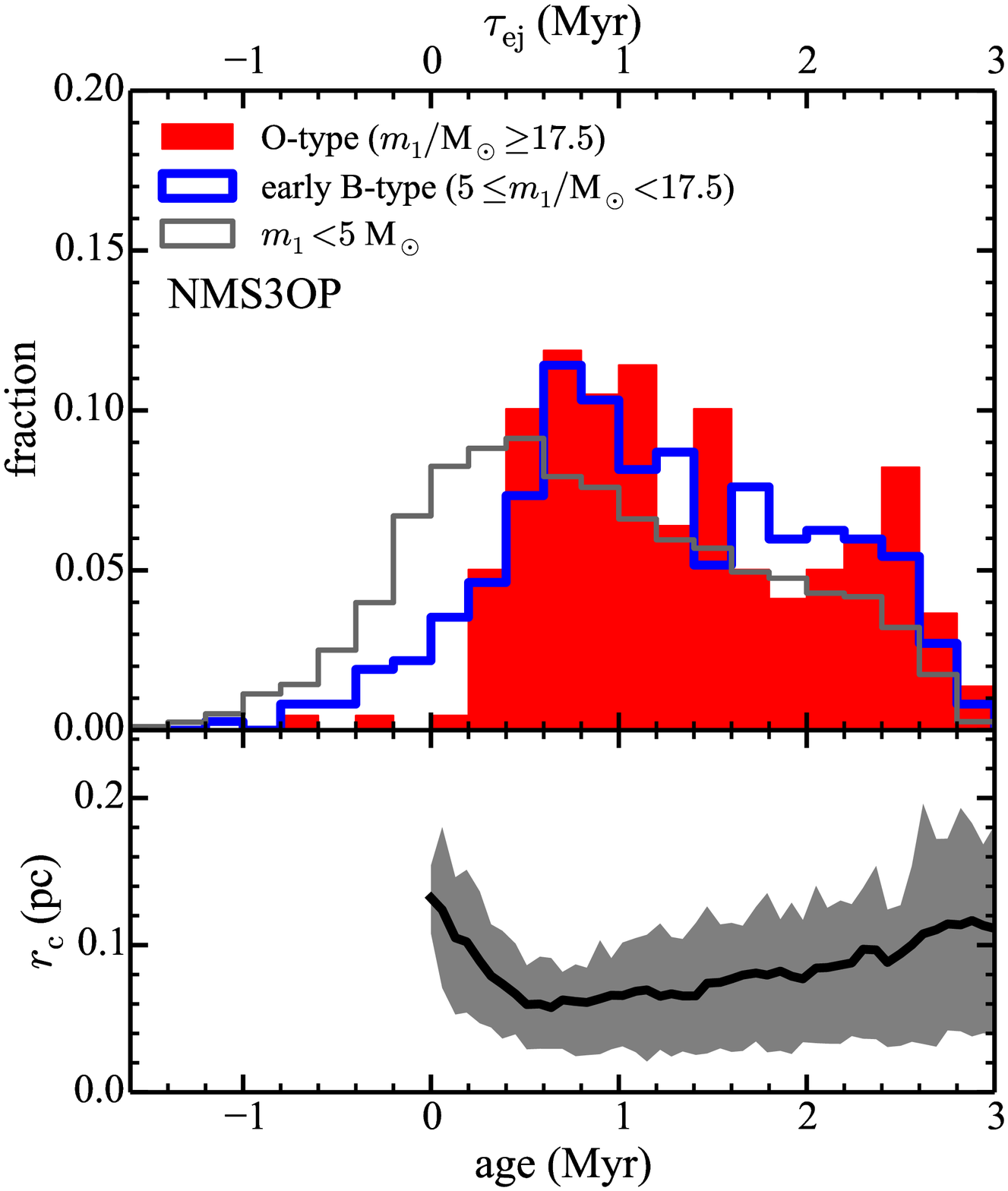}}
	 \caption{(Continued)}
	\end{figure*} 
	\setcounter{figure}{\value{figure}-1} 	

	\begin{figure*}[tbp] 
	 \centering
\resizebox{\hsize}{!}{	 \includegraphics{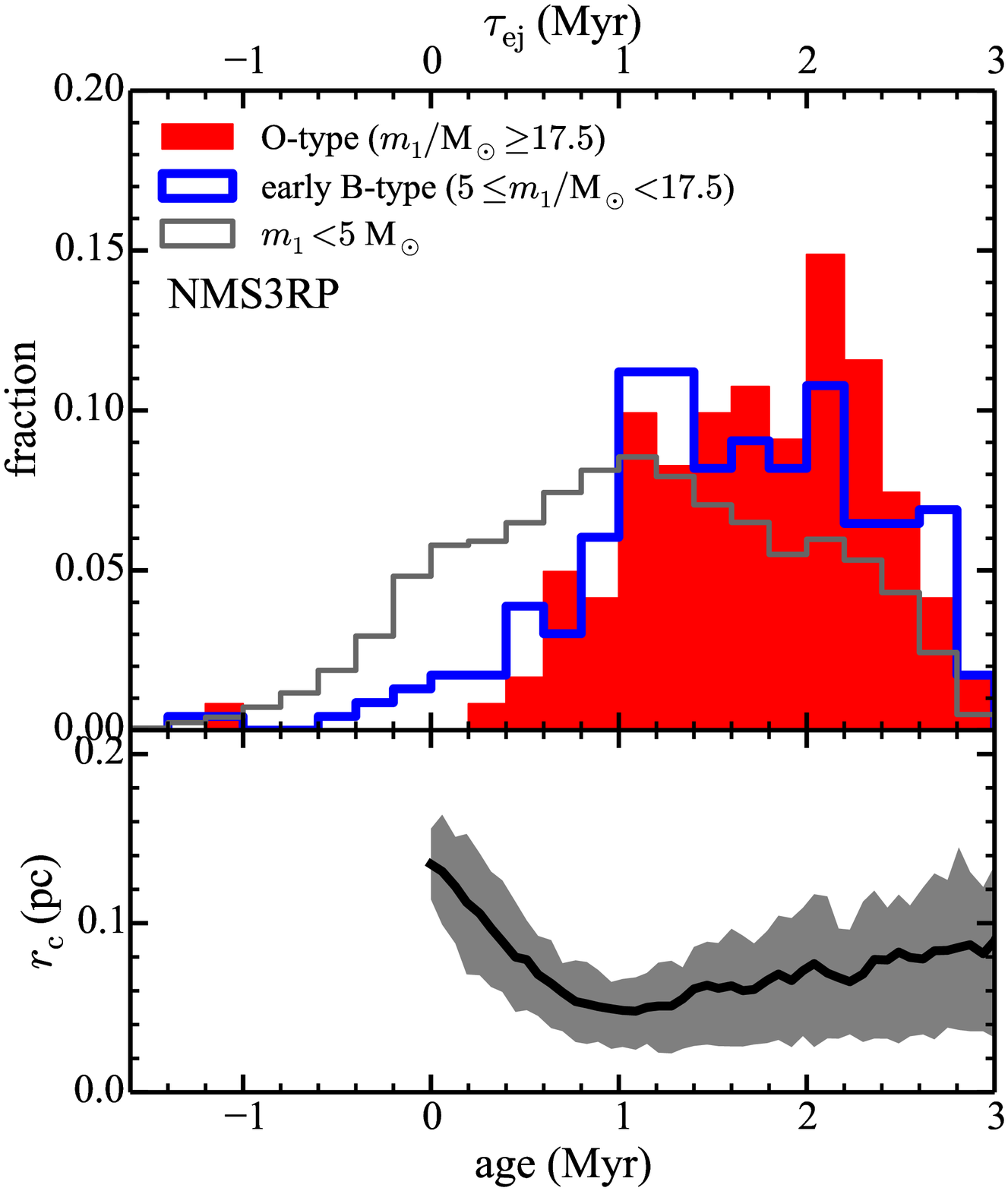}
	 \includegraphics{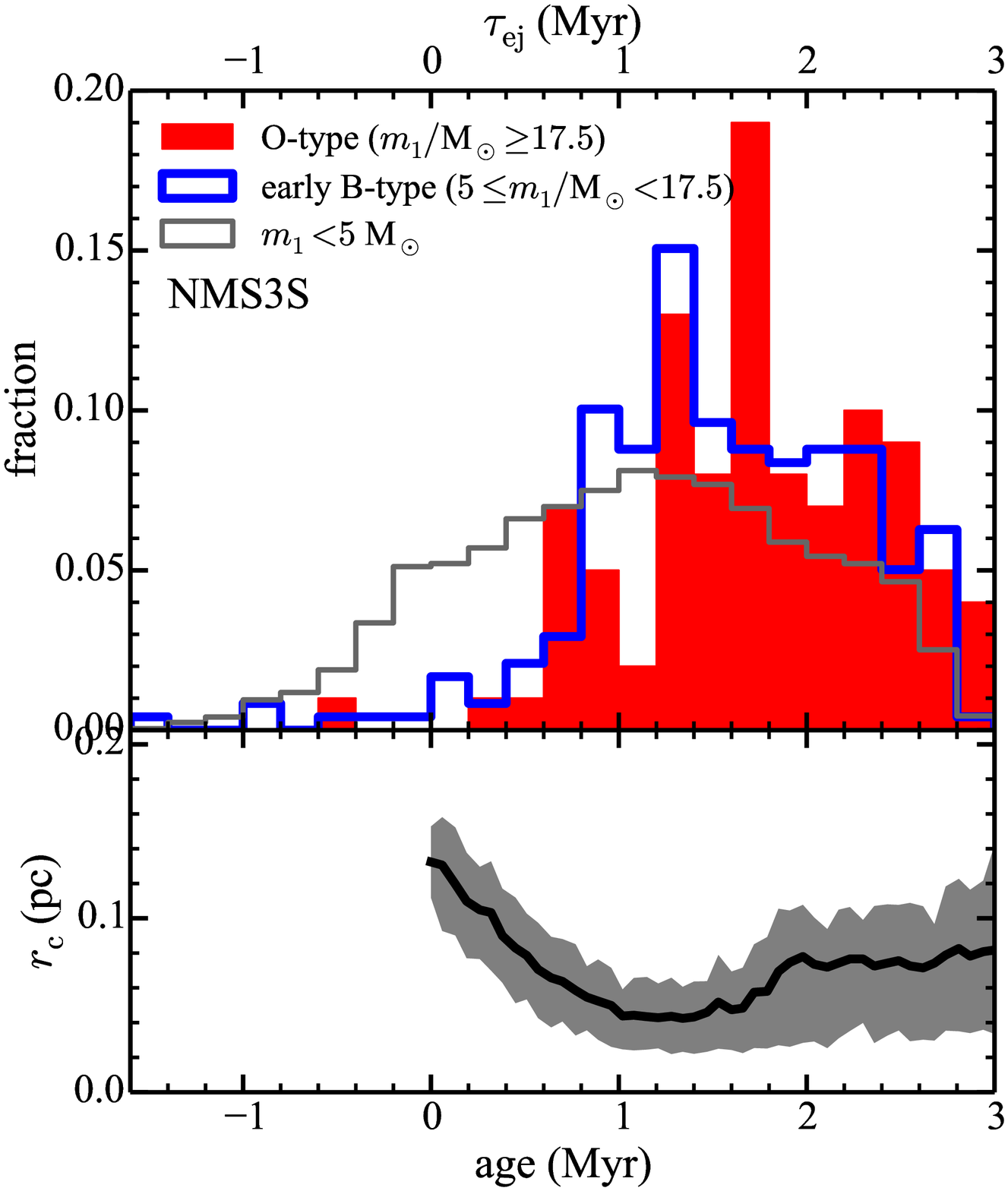}}
	 \caption{(Continued)}
	\end{figure*}

\subsection{$\tej$ versus velocity}\label{ejsec:aptejvel}
Heat maps of $\tej$ versus velocity are plotted in Fig.~\ref{aejfig:tejv} for all models with $\rhi\leq 0.3$\,pc. 
The systems with a negative $\tej$ mostly have a low velocity, implying deceleration by the potential 
of their birth cluster.

  \begin{figure*}[hbtp]
   \centering
\resizebox{\hsize}{!}{	 \includegraphics{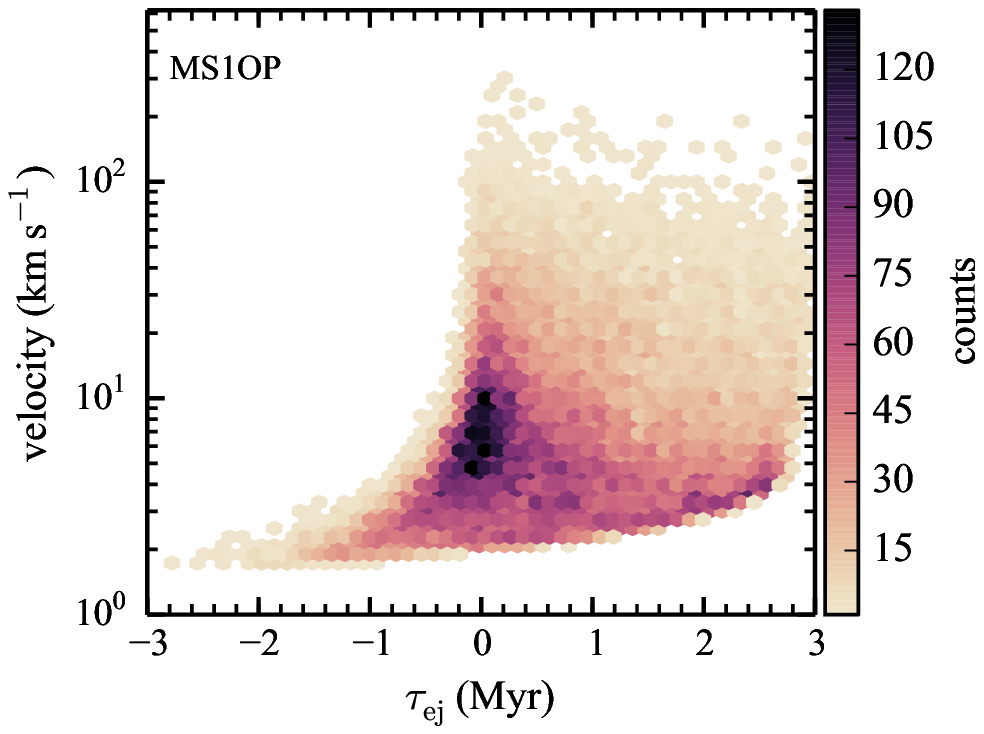}
	 \includegraphics{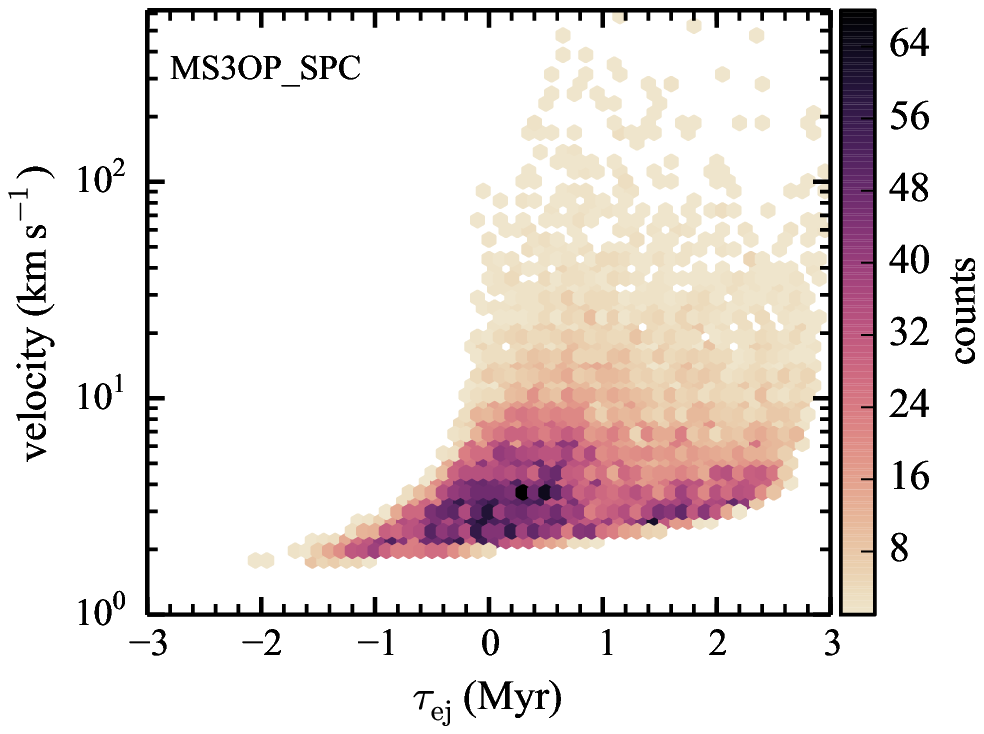}}\\
\resizebox{\hsize}{!}{	 \includegraphics{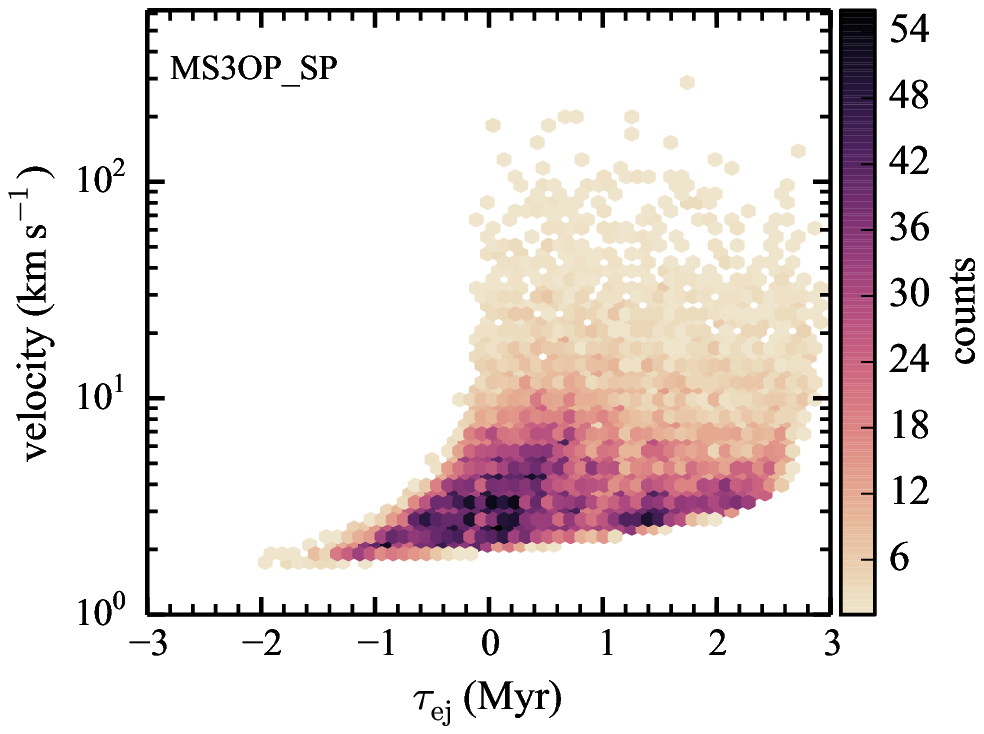}
	 \includegraphics{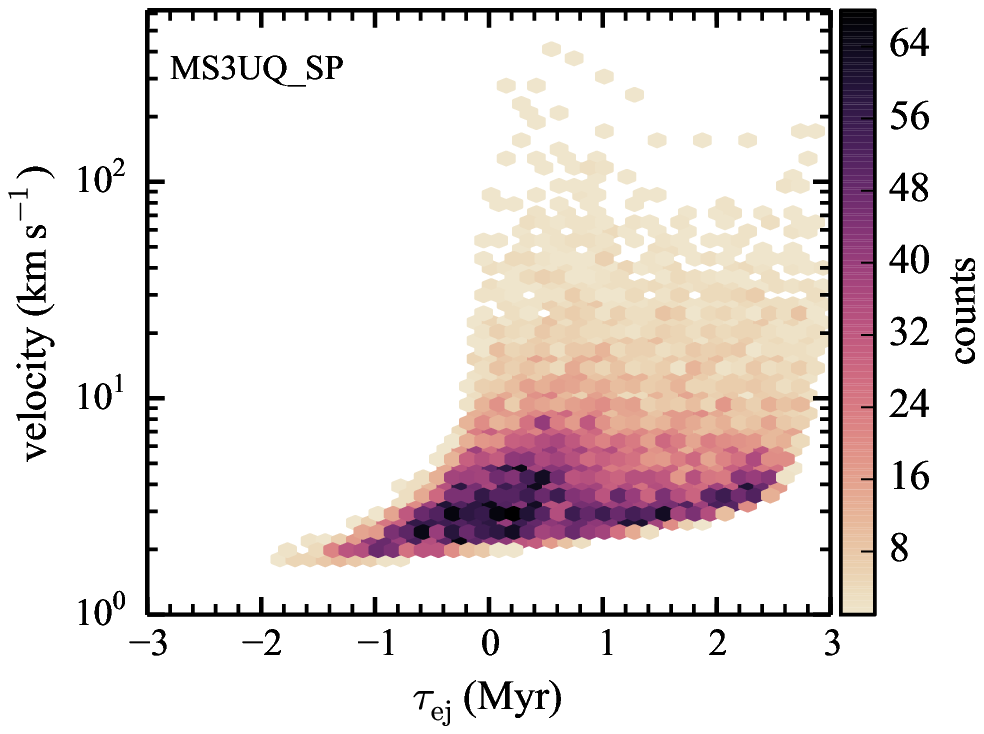}}\\
\resizebox{\hsize}{!}{	 \includegraphics{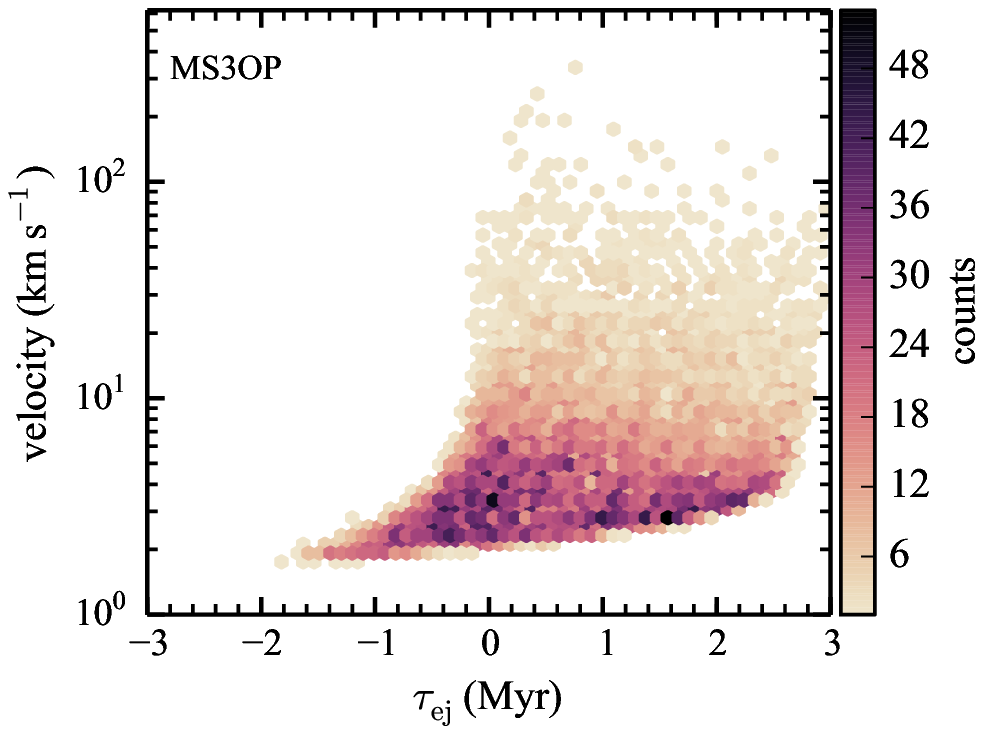}
	 \includegraphics{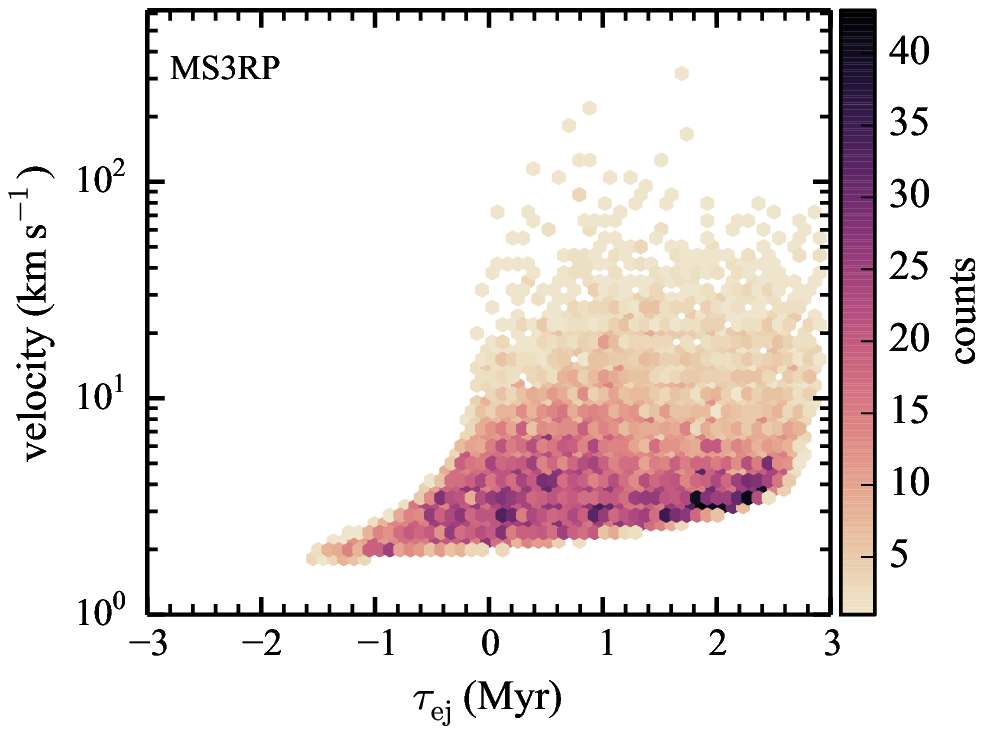}}
	 \caption{Heat maps of $\tej$ versus velocity for all models with $\rhi\leq0.3$\,pc.}
	 \label{aejfig:tejv}
	\end{figure*}
	\setcounter{figure}{\value{figure}-1} 	
	\begin{figure*} 
	 \centering
\resizebox{\hsize}{!}{	 \includegraphics{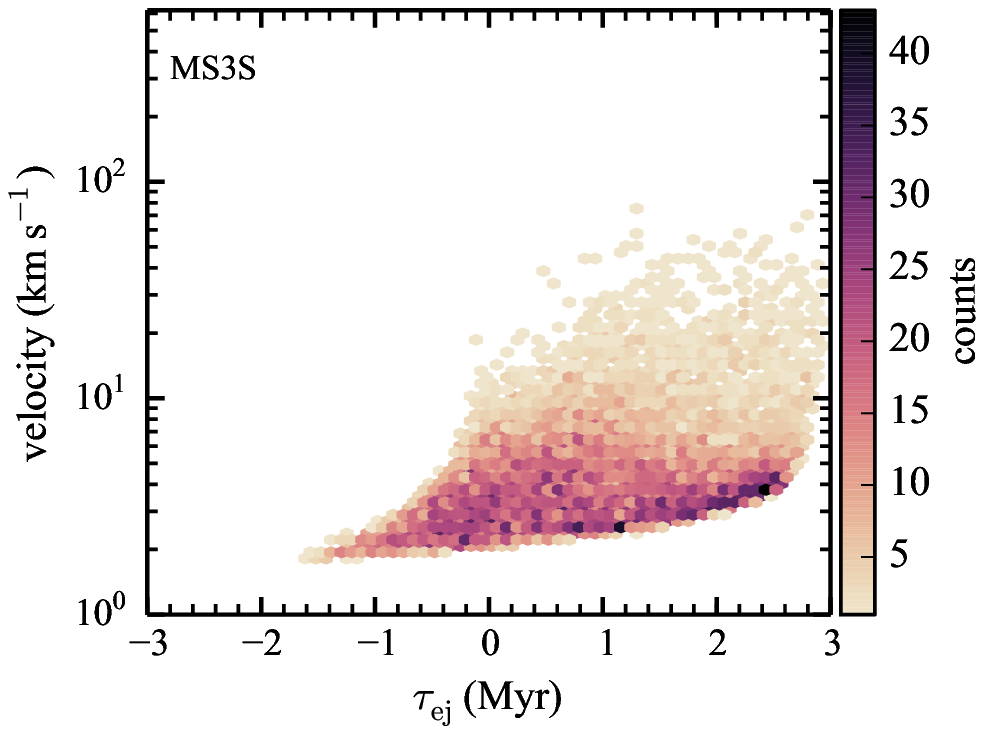}
	 \includegraphics{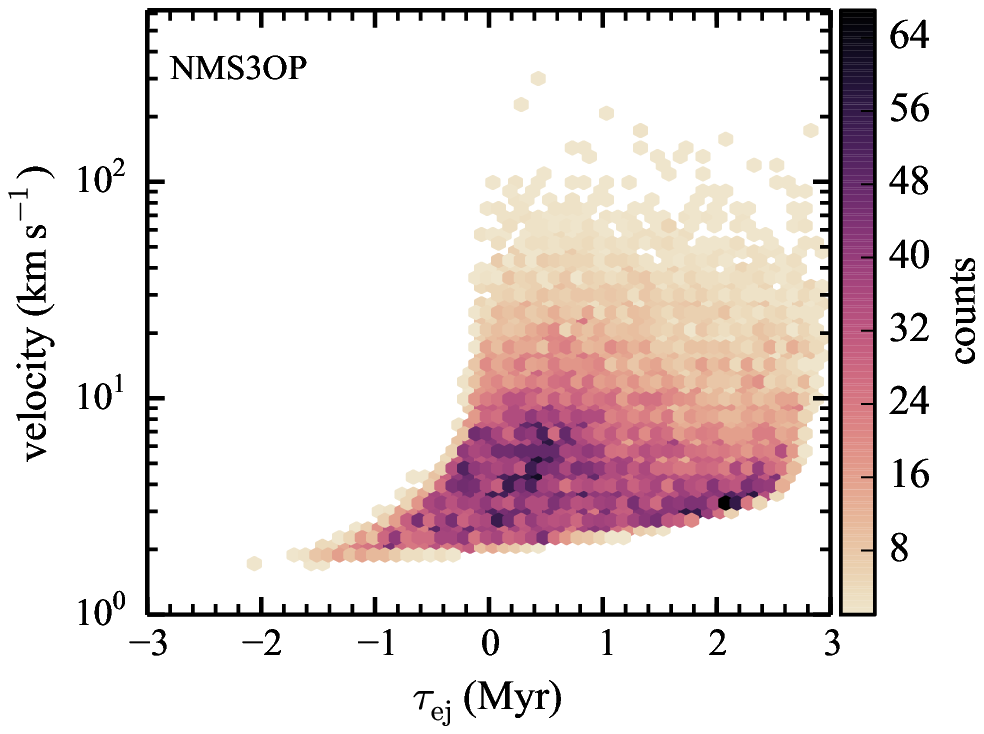}}\\
\resizebox{\hsize}{!}{	 \includegraphics{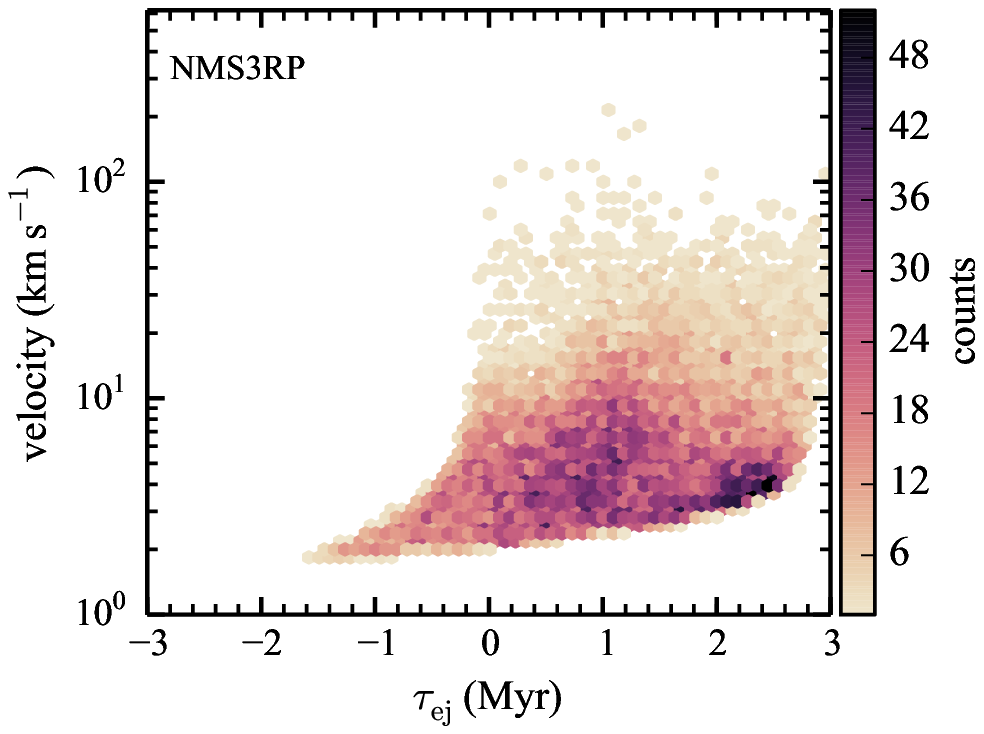}
	 \includegraphics{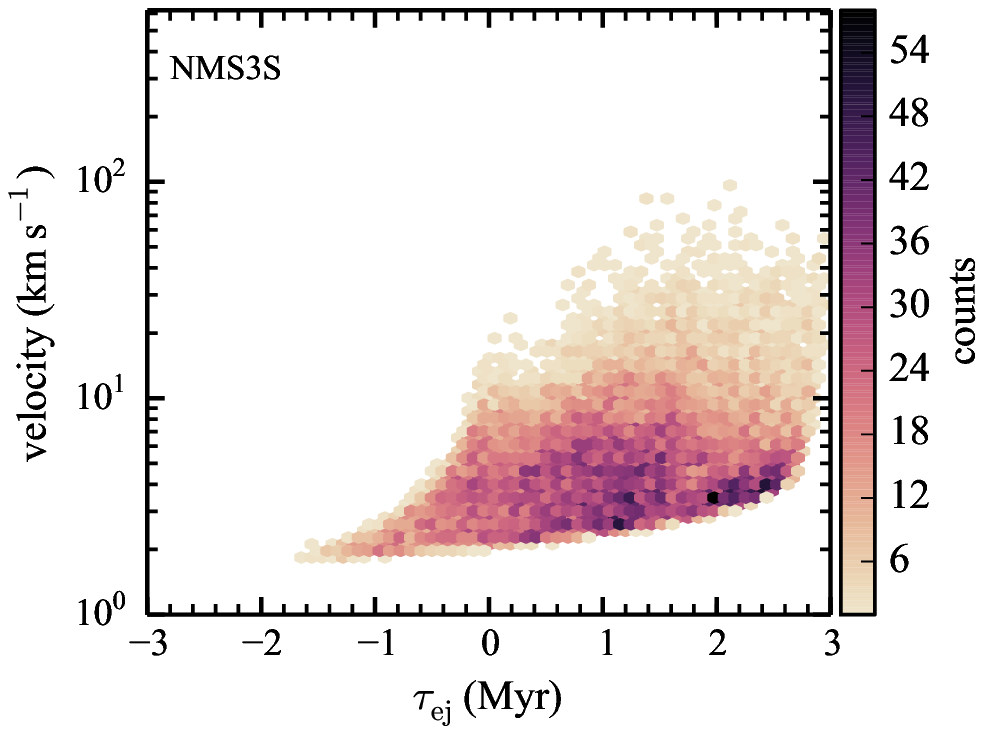}}
     \caption{(Continued)}
  \end{figure*}

\label{lastpage}
\end{document}